\documentclass[sigconf,screen]{acmart}

\acmSubmissionID{715}
    
\usepackage{physics}
\usepackage{acronym}

\usepackage{multirow}
\usepackage{graphicx}
\usepackage{float}
\usepackage{rotating}
\usepackage[skip=1pt]{caption}
\usepackage{subcaption}
\usepackage{enumitem}

\AtBeginDocument{%
  }

\acrodef{AWRF}{attention-weighted rank fairness}
\acrodef{NDCG}{normalized discounted cumulative gain}
\acrodef{EUR}{Exposed Utility Ratio}
\acrodef{RUR}{Realized Utility Ratio}
\acrodef{DP}{Demographic Parity}
\acrodef{IAA}{Inequality of Amortized Attention}
\acrodef{EEL}{Expected Exposure Loss }
\acrodef{EER}{Expected Exposure Relevance}
\acrodef{EED}{Expected Exposure Disparity}
\acrodef{logDP}{Demographic Parity}
\acrodef{logEUR}{Exposed Utility Ratio}
\acrodef{logRUR}{Realized Utility Ratio}

\acrodef{NBR}{next basket recommendation}

\copyrightyear{2025}
\acmYear{2025}
\setcopyright{cc}
\setcctype{by}
\acmConference[RecSys '25]{Proceedings of the Nineteenth ACM Conference on Recommender Systems}{September 22--26, 2025}{Prague, Czech Republic}
\acmBooktitle{Proceedings of the Nineteenth ACM Conference on Recommender Systems (RecSys '25), September 22--26, 2025, Prague, Czech Republic}\acmDOI{10.1145/3705328.3748169}
\acmISBN{979-8-4007-1364-4/2025/09}

\begin{document}

\title[A Reproducibility Study of Product-side Fairness in Bundle Recommendation]{A Reproducibility Study of Product-side Fairness\\ in Bundle Recommendation}

\author{Huy-Son Nguyen\authornotemark{}}
\authornote{Both authors contributed equally to this research.}
\orcid{0009-0006-4616-0976}
\affiliation{%
  \institution{Delft University of Technology}
  \city{Delft}
  \country{The Netherlands}
}
\email{H.S.Nguyen@tudelft.nl}

\author{Yuanna Liu\authornotemark{}}
\authornotemark[1] 
\orcid{0000-0002-9868-6578}
\affiliation{
  \institution{University of Amsterdam}
  \city{Amsterdam}
  \country{The Netherlands}
}
\email{y.liu8@uva.nl}

\author{Masoud Mansoury}
\orcid{0000-0002-9938-0212}
\affiliation{%
  \institution{Delft University of Technology}
  \city{Delft}
  \country{The Netherlands}
}
\email{m.mansoury@tudelft.nl}

\author{Mohammad Aliannejadi}
\orcid{0000-0002-9447-4172}
\affiliation{
  \institution{University of Amsterdam}
  \city{Amsterdam}
  \country{The Netherlands}
}
\email{m.aliannejadi@uva.nl}

\author{Alan Hanjalic}
\orcid{0000-0002-5771-2549}
\affiliation{%
  \institution{Delft University of Technology}
  \city{Delft}
  \country{The Netherlands}
}
\email{a.hanjalic@tudelft.nl}

\author{Maarten de Rijke}
\orcid{0000-0002-1086-0202}
\affiliation{
  \institution{University of Amsterdam}
  \city{Amsterdam}
  \country{The Netherlands}
}
\email{M.deRijke@uva.nl}

\begin{abstract}


Recommender systems are known to exhibit fairness issues, particularly on the product side, where products and their associated suppliers receive unequal exposure in recommended results. While this problem has been widely studied in traditional recommendation settings, its implications for bundle recommendation (BR) remain largely unexplored. This emerging task introduces additional complexity: recommendations are generated at the bundle level, yet user satisfaction and product (or supplier) exposure depend on both the bundle and the individual items it contains. Existing fairness frameworks and metrics designed for traditional recommender systems may not directly translate to this multi-layered setting. In this paper, we conduct a comprehensive reproducibility study of product-side fairness in BR across three real-world datasets using four state-of-the-art BR methods. We analyze exposure disparities at both the bundle and item levels using multiple fairness metrics, uncovering important patterns. Our results show that exposure patterns differ notably between bundles and items, revealing the need for fairness interventions that go beyond bundle-level assumptions. We also find that fairness assessments vary considerably depending on the metric used, reinforcing the need for multi-faceted evaluation. Furthermore, user behavior plays a critical role: when users interact more frequently with bundles than with individual items, BR systems tend to yield fairer exposure distributions across both levels. Overall, our findings offer actionable insights for building fairer bundle recommender systems and establish a vital foundation for future research in this emerging domain.

\end{abstract}



\maketitle

\section{Introduction}

Fairness in recommender systems has become an increasingly important area of research~\cite{deldjoo2024fairness,jin2023survey,wang2023survey,li2023fairness,mansoury2020feedback}. These systems often serve multiple stakeholders -- such as users, products, providers (or suppliers), and platforms -- each with different expectations and needs~\cite{abdollahpouri2020multistakeholder}. Although recommender systems are typically optimized to deliver accurate and personalized results to users, fairness requires that products and providers are equitably represented in these outputs.

On the product side, exposure is a key concern for fairness. The visibility of a product on a recommendation list can shape users' opinions (e.g., in \textit{news}~\cite{qi2022profairrec,wu2021fairness}), influence opportunities (e.g., in \textit{job matching}~\cite{li2023afairness,islam2021debiasing}), affect preferences (e.g., in \textit{music}~\cite{mehrotra2018towards}), and drive economic outcomes (e.g., in \textit{e-commerce}~\cite{dash2022fairir}). Product-side fairness aims to ensure that products receive equal exposure or exposure  proportional to their utility,
so that suppliers or products have a fair chance of being seen and selected by users~\cite{singh2018fairness,mansoury2022understanding}. 

Prior research has made significant progress in measuring and addressing product-side fairness across a variety of recommendation tasks. For example,~\citet{raj2022measuring} studied fairness in top-$K$ recommendation by modeling item exposure based on user browsing behavior. Similarly,~\citet{li2024we} explored fairness in next-basket recommendation, focusing on how repeated and exploratory recommendation patterns influence product exposure.

Despite these advances, the issue of fairness remains largely unexplored in the context of bundle recommendation (BR), which recommends sets of items grouped together as meaningful packages. BR is increasingly common in diverse domains such as e-commerce (e.g., \textit{fashion}~\cite{ding2023computational}, \textit{electronic kits}~\cite{sun2024revisiting}, \textit{online games}~\cite{deng2020personalized}), digital media (e.g., \textit{curated playlists}~\cite{sun2024survey}), services (e.g., \textit{meals}~\cite{li2024mealrec+}), and even the medical field (e.g., \textit{drug packages}~\cite{zheng2023interaction}). Unlike single-item recommendation, bundle recommendation must account not only for the relevance of individual items but also for their compatibility as a collection, introducing more complex relationships between users, bundles, and items~\cite{sun2024revisiting,sun2024survey,nguyen2024bundle}. This added complexity raises critical new challenges to fairness. Specifically, ensuring fair exposure in BR requires attention not just to which bundles are recommended, but also to how items within those bundles are presented, both of which may be influenced by biases in historical interaction data.

In this paper, we conduct the first comprehensive study of product-side fairness in bundle recommendation. Our goal is to assess how fairly current BR methods allocate exposure to both bundles and the individual items they contain, and to understand how exposure disparities arise and propagate through these systems. 

We frame our investigation around the following questions:

\begin{enumerate}[label=\textbf{(RQ\arabic*)},leftmargin=*]
    \item To what extent does popularity bias in historical user interactions lead to unfair exposure (i.e., unequal representation) at both the bundle and item levels in BR methods?
    \item How fairly do BR methods allocate exposure to bundles of varying popularity, and how does this affect the exposure of items within those bundles?
    \item How does variation in users' interaction preferences, such as a tendency to engage more with bundles versus individual items, impact fairness outcomes in BR scenario?
\end{enumerate}

\noindent%
To answer these questions, we implement a thorough empirical study using four state-of-the-art BR methods across three benchmark datasets from diverse domains: books, music playlists, and fashion outfits. We employ fairness evaluation protocols based on six widely adopted exposure fairness metrics~\cite{raj2022measuring,li2024we,liu2024measuring}, measuring both bundle- and item-level disparities.
In addition, we devise a novel approach to investigate the impact of user preferences on bundle- and item-level fairness through a user grouping strategy. Thereby, this study offers nuanced views of how fairness manifests in BR systems.


\section{Related Work}

This section describes relevant studies on bundle recommendation and fairness issues, which serve as the background of our work.

\subsection{Bundle Recommendation}

In bundle recommendation, a variety of methods have been developed to improve the accuracy of recommending pre-defined item sets to users. 
An early approach of \citet{chen2019matching}, named DAM, emphasizes the necessity of affiliated items within bundles by jointly optimizing user-item and user-bundle interactions through attention mechanisms and multi-task learning.
Using advances in graph neural networks~\cite{gao2023survey,he2020lightgcn,KipfW17}, BGCN~\cite {chang2020bundle} surpasses DAM~\cite{chen2019matching} by unfolding user preferences into an item view and bundle view, adopting graph convolutional networks~\cite{KipfW17} to encode each type of preference separately.
\citet{deng2020personalized} employ BundleNet, performing message passing on the user-bundle-item tripartite graph.  

With the growth of multi-view architectures, the adoption of contrastive learning has addressed challenges related to the inconsistency and synergy of information propagation between different views in bundle recommendation~\citep{zhao2022multi,ma2022crosscbr,nguyen2024bundle,ma2024multicbr}. 
CrossCBR~\cite{ma2022crosscbr}, a multi-view learning approach with LightGCNs~\cite{he2020lightgcn}, achieves notable advancements by using cross-view contrastive losses to capture cooperative signals from two distinct views.
Building on CrossCBR~\cite{ma2022crosscbr}, MultiCBR~\cite{ma2024multicbr} restructures the learning process by reversing the sequence of fusion and contrast operations. Instead of relying on a quadratic number of cross-view contrastive losses, MultiCBR~\cite{ma2024multicbr} simplifies the framework by employing only two self-supervised contrastive losses, enhancing both its effectiveness and efficiency.
Recent approaches based on distillation~\cite{ren2023distillation} or synergy optimization~\cite{kim2024towards} have achieved state-of-the-art performance.

Investigation of intricate item connections that affect bundle tactics and user preferences has led to several in-depth studies on bundle recommendation~\cite{zhao2022multi,sun2024revisiting,nguyen2024bundle,wei2023strategy,sun2024survey}.
First, \citet{zhao2022multi} introduces the intention disentanglement technique by modeling multiple latent features from both local and global views, using contrastive loss to align these views. This architecture faces performance limitations due to its hyperparameter sensitivity and computational complexity.
BundleGT~\cite{wei2023strategy} leverages hierarchical graph transformer networks to capture strategy-aware representations by modeling latent associations at the bundle and user levels.
To enhance bundle representation learning at the item-level, EBRec~\cite{du2023enhancing} augments user-item interactions by exploiting user-bundle-item correlations generated through pre-trained models.
BunCa~\cite{nguyen2024bundle} harnesses item-level causation within user preferences and bundle construction, employing a refined attention operation on item-item graphs to better capture asymmetric relationships among items in real-world scenarios.
Beside modeling item relations in set recommendation via graphs~\cite{nguyen2023hhmc,nguyen2024bundle,wei2023strategy}, some generative approaches, such as BRIDGE~\cite{bui2024bridge} and DisCo~\cite{bui2025personalized}, have been further developed to produce new appropriate bundles aligning with user preferences.  

These advancements listed above~\cite{wei2023strategy,nguyen2024bundle,zhao2022multi} uncover user preferences underlying inferred bundling strategies, but have not inspected the impact of product-side fairness on the final prediction.
Our study reproduces prominent BR models~\cite{wei2023strategy,nguyen2024bundle,ma2022crosscbr,ma2024multicbr} within a unified experimental framework and concentrates on investigating item-level and bundle-level fairness based on their outcomes. Our scrutiny can also be extended to enhance current approaches in bundle construction~\cite{ma2024leveraging,sun2024revisiting}, or bundle generation~\cite{sun2024adaptive,chang2021bundle} tasks. 

\vspace{-4mm}
\subsection{Fairness in Recommender Systems}

Information access systems, such as search engines and recommendation systems, exhibit characteristics of multi-stakeholder, ranking-based displays, centrality of personalization, and user responses in real applications~\cite{ekstrand2022fairness}. These factors pose challenges in evaluating and optimizing fairness. In recent years, the research community has shown a growing interest in fairness within recommender systems, focusing on fairness definitions~\cite{beutel2019fairness,yang2017measuring}, evaluation metrics~\cite{zehlike2017fa,sapiezynski2019quantifying}, and optimization algorithms~\cite{biega2018equity,mansoury2024mitigating} to foster a fair and healthy recommendation ecosystem. 

Fairness of recommender systems can be divided into user fairness and product fairness~\cite{wang2023survey}. Product fairness investigates whether products are allocated a fair distribution of exposure by being recommended based on various fairness principles, such as equal opportunity and statistical parity~\cite{raj2022measuring}. On the user side, user fairness measures the deviation of recommendation performance among different user groups~\cite{wang2021user}. Specifically, users can be grouped by the level of activity or the consumption of popular items~\cite{rahmani2022experiments}. To optimize product fairness and user fairness in a recommender system, fairness algorithms, including in-processing methods~\cite{xu2025bridging} and post-processing methods~\cite{naghiaei2022cpfair}, have been proposed to enhance fairness at different stages of the recommendation algorithm pipeline. A recent open-source toolkit, FairDiverse~\cite{xu2025fairdiverse}, collects and develops multiple fairness-aware algorithms in search and recommendation. 

Apart from the general recommendation scenarios, fairness is also explored in some special domains, such as \ac{NBR}, where users exhibit both repetitive and exploratory purchase behaviors~\cite{li2023next}. \citet{liu2024measuring} re-consider item fairness metrics to assess representative \ac{NBR} methods and test the robustness of these metrics. \citet{li2024we} discover a shortcut to achieving better fairness and accuracy performance in \ac{NBR} and proposes a fine-grained evaluation paradigm. \citet{liu2025repeat} jointly optimize item fairness and repeat bias for \ac{NBR} methods through mixed-integer linear programming. These works show that fairness behaves differently in different recommendation scenarios, which may impact the evaluation design and optimization strategies. 

Fairness in bundle recommendation remains unexplored. Bundle recommendation algorithms directly recommend bundle lists to users, and indirectly influence the exposure of items within the predicted bundles. Therefore, our work aims to measure the bundle-level and item-level fairness of BR methods and investigate their potential correlation. To the best of our knowledge, we are the first to study product-side fairness in bundle recommendation.  

\vspace{-2mm}
\section{Methodology}

This section outlines our experimental setup used to evaluate exposure at both the bundle and item levels.

\subsection{Problem Formulation}
Given a set of users $\mathcal{U} = \{u_1, u_2, \ldots, u_{|\mathcal{U}|}\}$, a set of bundles $\mathcal{B} = \{b_1, b_2, \ldots, b_{|\mathcal{B}|}\}$, and a set of items $\mathcal{I} = \{i_1, i_2, \ldots, i_{|\mathcal{I}|}\}$, the BR task represents user-bundle interactions, user-item interactions, and bundle-item affiliations through three binary matrices: $X \in \{0,1\}^{|\mathcal{U}| \times |\mathcal{B}|}$, $Y \in \{0,1\}^{|\mathcal{U}| \times |\mathcal{I}|}$, and $Z \in \{0,1\}^{|\mathcal{B}| \times |\mathcal{I}|}$ as inputs to train BR models. In these matrices, an entry of $1$ indicates an observed connection between the corresponding user-bundle, user-item, or bundle-item pair, while $0$ denotes the absence of such an interaction.
The objective of the BR task is to predict accurately unobserved user-bundle interactions by leveraging the implicit feedback contained within $X$, $Y$, and $Z$ matrices.
Specifically, for each user $u \in \mathcal{U}$, a list of top-$K$ bundles can be inferred and ranked through user-specific relevance probability $\hat y(b|u)$, where $b \in \mathcal{B}$.

\paragraph{Fairness approach}
Our study evaluates fairness at both the item-level and bundle-level across recommended bundles for all users using fairness metrics described in Section~\ref{sss:fair_metric}. Each user receives a list of bundles $L$, as a recommendation, where each bundle $b \in L$ contains a set of items. We aim to evaluate whether the bundles and items obtained by these BR algorithms are fairly exposed, and to investigate the correlation between bundle-level fairness and item-level fairness in the BR scenario. Table~\ref{notation} presents the notation used in this paper.


\begin{table}[t]
\vspace{-2mm}
\caption{Notation used in the paper.}
\label{notation}
\centering
\begin{tabular}{@{}ll@{} }
\toprule
$X$ & binary user-bundle interaction matrix \\
$Y$ & binary user-item interaction matrix \\
$Z$ & binary bundle-item affiliation matrix \\
$L$  & ranked list of $K$ bundles \\ 
$L(b)$  & rank of bundle $b$ in $L$      \\
$y(b|u)$  & relevance of $b$ to $u$      \\ 
$y(i|i\in b)$  & relevance of item $i$ in bundle $b$      \\ 
$G^{+}$ & popular group \\
$G^{-}$ & unpopular group \\
$\vb{a}_L$ & exposure vector for bundles in $L$ \\ 
$\vb{a}_L(b)$ & exposure of $b$ in $L$ \\ 
$\vb{a}_L(i|i\in b)$ & exposure of item $i$ in bundle $b$ \\ 
$G(L)$  & group alignment matrix for bundles in $L$ \\
$\mathbf{\epsilon}_L$ & the exposure of groups in $L$ $(G(L)^T \vb{a}_L)$\\
\bottomrule
\end{tabular}
\vspace*{-4mm}
\end{table}


\subsection{Datasets}

Following~\cite{chang2020bundle,ma2022crosscbr,nguyen2024bundle,ma2024multicbr,du2023enhancing}, we use three benchmark datasets from various sectors with the same training ratio to evaluate BR models: Youshu\footnote{\url{https://www.yousuu.com/}}, NetEase\footnote{\url{https://music.163.com/}}, and iFashion~\cite{chen2019pog}.

The Youshu dataset involves predicting which book lists (bundles) users are likely to purchase.
In the NetEase dataset, individual tracks represent items, and user-generated playlists serve as the target bundles for music recommendation.
The iFashion dataset constructs outfits as bundles, which are combinations of clothing and accessories, to attract user engagement in a fashion recommendation scenario.
The key statistics of the three datasets are summarized in Table \ref{tab:data}.
These datasets cover diverse application domains and show varying interaction densities and bundle sizes, ensuring a general evaluation.

Three types of historical interaction are used as model inputs: \textit{user-item}, \textit{user-bundle}, and \textit{bundle-item}.
Following~\cite{ma2022crosscbr}, datasets are randomly split into training, validation, and test sets on user-bundle interactions for each user by a $7:1:2$ ratio. 


Notably, the \textit{c-score} metric, as reported in~\cite{ren2023distillation}, measures the consistency between collaborative user-user pairs via interaction matrices $X$ and $Y$. On the iFashion dataset, it demonstrates high consistency, indicating that users with similar item-level preferences also tend to have similar bundle-level preferences. In contrast, on the Youshu and NetEase datasets, the low \textit{c-score} values suggest that in large-bundle datasets, there is weaker alignment between user-bundle and user-item collaborative relationships.


\begin{table*}[ht]
    \caption{Statistics of the Youshu, NetEase, and iFashion datasets.}
    \label{tab:data}
    \centering
    \begin{tabular}{l cccccccccccc}
        \toprule
        \multirow{2}{*}{\textbf{Dataset}} & \multirow{2}{*}{$|\mathcal{U}|$} & \multirow{2}{*}{$|\mathcal{I}|$} & \multirow{2}{*}{$|\mathcal{B}|$} & \multirow{2}{*}{$\mathcal{U}$--${\mathcal{I}}$} & \multirow{2}{*}{$\mathcal{U}$--${\mathcal{B}}$} & \multirow{2}{*}{\#Avg.I/B} & \multirow{2}{*}{$\mathcal{U}$--${\mathcal{B}}$ Dens.} & \multirow{2}{*}{\textit{c--score}} &&  \multicolumn{3}{c}{User groups} \\ \cline{11-13}

         &  &  &  &  &  &  &  &  && $|g_1|$ &$|g_2|$ & $|g_3|$ \\
        \midrule
        Youshu           & \phantom{0}8,039   & \phantom{0}32,770  & \phantom{0}4,771     & \phantom{1,}138,515     & \phantom{0,0}51,377        & 37.03  &  0.13\% & 0.0812 && \phantom{0,}677 & \phantom{0,}135 & \phantom{0}7,194 \\
        NetEase          & 18,528  & 123,628 & 22,864    & 1,128,065   & \phantom{0,}302,303       & 77.80   & 0.07\% & 0.0599 && \phantom{0,}498 & \phantom{0}240 & 17,790 \\
        iFashion         & 53,897  & \phantom{0}42,563  & 27,694    & 2,290,645   & 1,679,708     & \phantom{0}3.86 &  0.07\% & 0.2917 && 5,565 & 3,397 & 44,935   \\
        \bottomrule
    \end{tabular}
\vspace*{-4mm}
\end{table*}



\subsection{Evaluation Protocols}

\subsubsection{Utility ranking metrics}
Following common evaluation practices in bundle recommendation~\cite{sun2024survey,ma2022crosscbr,nguyen2024bundle,du2023enhancing}, we use recall at $K$ ($R@K$) and normalized discounted cumulative gain at $K$ ($N@K$) to assess the utility of BR methods. 
$R@K$ computes the proportion of test bundles that appear within the top-$K$ ranked prediction, while $N@K$ evaluates the ranking quality by emphasizing the placement of relevant bundles at higher positions in the predicted list. 

\subsubsection{Fairness ranking metrics}
\label{sss:fair_metric} Following~\cite{liu2024measuring}, we measure fairness using a popular bundle group and an unpopular bundle group over distributions of ranked bundle lists among all users. Since fair exposure is unlikely to be satisfied in a single list in the recommendation task~\cite{raj2022measuring}, we exclude exposure fairness metrics designed for single ranking, such as \ac{AWRF}~\cite{sapiezynski2019quantifying}. Assume $\pi(L|u)$ is a user-dependent distribution and $\rho(u)$ is a distribution over all users, then $\rho(u)\pi(L|u)$ is the distribution of overall rankings among all the users. $\epsilon_L = G(L)^T\vb{a}_L$ is the group exposure of a single ranking; its expectation $\epsilon_\pi = E_{\pi\rho}[\epsilon_L]$ is the group exposure among all the rankings. We select six representative fair ranking metrics covering two categories of fairness principles: 

\textbf{Equal opportunity} takes into account the utility of the ranked lists, and requires that exposure should be proportional to relevance. The \ac{EUR}~\cite{singh2018fairness} uses a ratio-based metric to measure how much the exposure of each group deviates from the ideal of being proportional to its utility $Y(G)$:
\begin{equation}
    \text{EUR} = \frac{\mathbf{\epsilon_\pi} \left( G^+ \right) / Y \left( G^+ \right)}{\mathbf{\epsilon_\pi} \left( G^- \right) / Y \left( G^- \right)}.
\end{equation}
The \ac{RUR}~\cite{singh2018fairness} further ensures that click-through rates for each group $\mathbf{\Gamma}(G)$ should be proportional to utility:   
\begin{equation}
    \text{RUR} = \frac{\mathbf{\Gamma} \left( G^+ \right) / Y \left( G^+ \right)}{\mathbf{\Gamma} \left( G^- \right) / Y \left( G^- \right)}.
\end{equation}
The \ac{EEL}~\cite{diaz2020evaluating} measures the Euclidean distance between expected exposure $\epsilon_{\pi}$ and target exposure $\epsilon^*$ among groups; similarly, the \ac{EER}~\cite{diaz2020evaluating} measures the dot product of expected exposure and target exposure: 
\begin{equation}
\text{EEL} = \lVert \mathbf{\epsilon_\pi} - \mathbf{\epsilon}^\ast \rVert_2^2, \\\ 
 \text{EER} = 2\mathbf{\epsilon_\pi}^T \mathbf{\epsilon}^\ast
\end{equation}

\textbf{Statistical parity} ensures equal exposure across groups without considering utility. In particular, the \ac{EED}~\cite{diaz2020evaluating} quantifies the disparities in how overall exposure is distributed among groups:    
\begin{equation}
    \text{EED} = \lVert \mathbf{\epsilon_\pi} \rVert_2^2.
\end{equation}
\ac{DP}~\cite{singh2018fairness} calculates the ratio of average exposure obtained by the two groups: 
\begin{equation}
    \text{DP} = \mathbf{\epsilon_\pi} \left( G^+ \right) / \mathbf{\epsilon_\pi} \left( G^- \right).
\end{equation}
Following~\cite{raj2022measuring}, we take logDP, logEUR, and logRUR in the experiments to deal with the empty-group case. We use the Geometric browsing model~\cite{biega2018equity} to compute exposure for all fair ranking metrics, i.e., $\gamma(1-\gamma)^{L(d)-1}$, where $\gamma$ is the patience parameter. 

\paragraph{Item-level fairness evaluation} We measure item-level fairness between a popular item group and an unpopular item group. First, we obtain the exposure of bundles within a list via the browsing model. Then, each item within a bundle shares equal average exposure of the bundle exposure, i.e., item exposure $\vb{a}_L(i | i \in b) = \vb{a}_L(b) / |b|$, where $|b|$ is the number of items within bundle $b$. 
For the test data, the relevance of item $i$ within a bundle is: $y(i|i \in b) = y(b|u) / |b|$. In this way, we can use the above fair ranking metrics to measure the item-level fairness. 

\subsection{Bundle Recommendation Models}

To reproduce BR baselines, we investigate the following representative BR methods with open-source materials, including multi-view learning, graph neural network, and data augmentation approaches. 
Four well-known BR algorithms are used to facilitate our evaluations of utility and fairness. 

\begin{itemize}[leftmargin=*]
    \item \textbf{CrossCBR} \cite{ma2022crosscbr} uses contrastive learning between two distinct views (\textit{user-item view} and \textit{user-bundle view}) to exploit their cooperation. In each view, a corresponding bipartite graph (\textit{user-item graph} or \textit{user-bundle graph}) is constructed and then propagated using LightGCN operation~\cite{he2020lightgcn}. Aligning separately learned views enables each view to distill augmented data from the other, resulting in mutual enhancement. CrossCBR improves the self-discrimination of representations by increasing the dispersion among different users and bundles.
    
    \item \textbf{MultiCBR} \cite{ma2024multicbr} is a multi-view learning framework that jointly captures user-bundle, user-item, and bundle-item interactions. Built on CrossCBR~\cite{ma2022crosscbr}, it uses bundle-item affiliations to enhance sparse bundle representations. By adopting an ``\emph{early fusion - late contrast}'' method, MultiCBR models both cross-view and ego-view preferences for improved user modeling, and reduces computational cost by requiring only two contrastive losses instead of a quadratic number like CrossCBR.
    
    \item \textbf{EBRec} \cite{du2023enhancing} learns item-level bundle representations through two key modules, which incorporate high-order B-U-I (bundle-user-item) correlations to capture richer collaborative signals. Moreover, it enhances B-U-I correlations by augmenting observed user-item interactions with synthetic data generated by pre-trained models, further improving representation quality.

    \item \textbf{BunCa} \cite{nguyen2024bundle} is a causation-aware multi-view learning framework for BR that captures asymmetric item interconnection via bundle construction and user preferences to enhance the ultimate representation of bundles/users. Contrastive learning is inherited from prior work~\cite{ma2022crosscbr,ma2024multicbr} to enhance representation alignment and discrimination across views.

\end{itemize}

\vspace{-2mm}
\subsection{Implementation Details}


To examine group fairness using the fair ranking metrics from Section~\ref{sss:fair_metric}, we create a popularity-based bundle and item partition, which is determined by the frequency of purchases across all historical user interactions.
Following~\cite{ge2021towards,liu2024measuring}, highly interacted bundles making up roughly $20\%$ of interactions are categorized as the popular group ($G^+$), while the remaining $80\%$ constitute the unpopular group ($G^-$). Particularly, the interaction frequency for each item is determined to compute item popularity by first reconstructing the user-item matrix, $Y^\prime$, as follows:
\begin{equation}\label{eq:y_prime}
    Y^\prime=(X \times Z) + Y
\end{equation}
then we normalize the matrix by dividing its entries by the total number of interactions. Given $Y^\prime$, we follow the same procedure as~\cite{ge2021towards,liu2024measuring} (also described above for bundles) to compute item popularity. This way of computing item popularity takes into account the impact of users' interactions with bundles.

\begin{figure*}[t!]
    \centering
    \begin{subfigure}[b]{.9\textwidth}
        \includegraphics[width=0.19\textwidth]{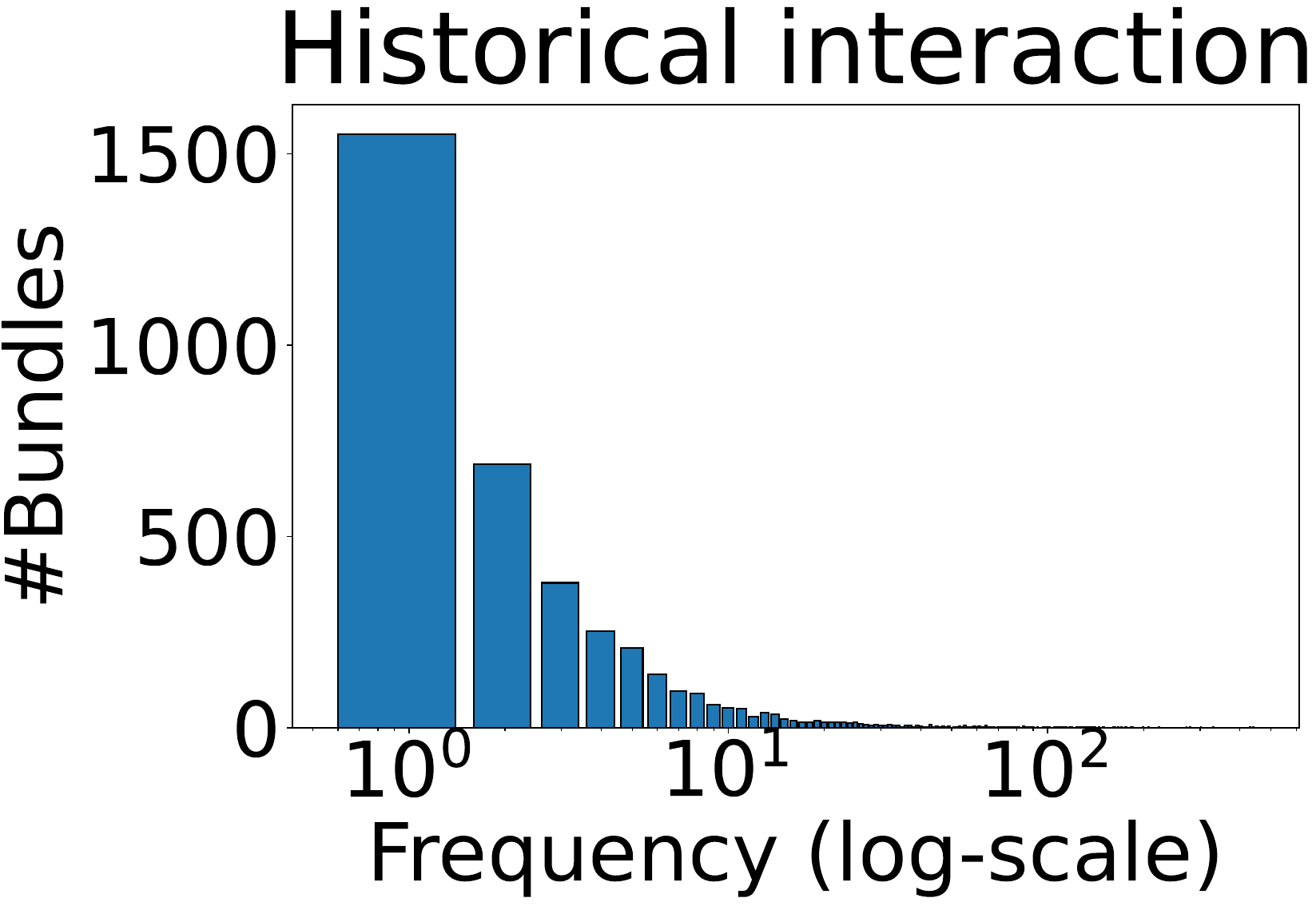}
        \includegraphics[width=0.19\textwidth]{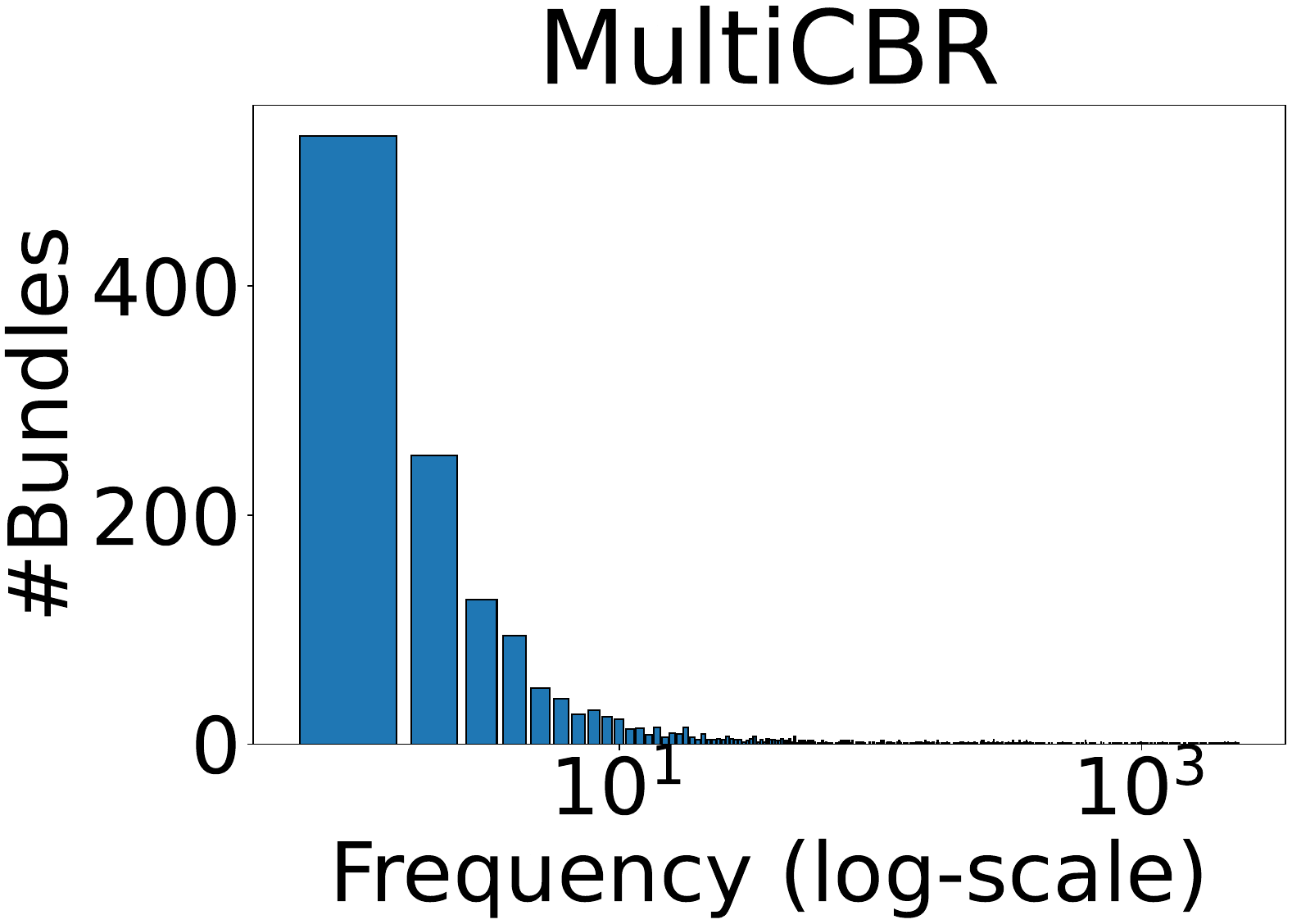}
        \includegraphics[width=0.19\textwidth]{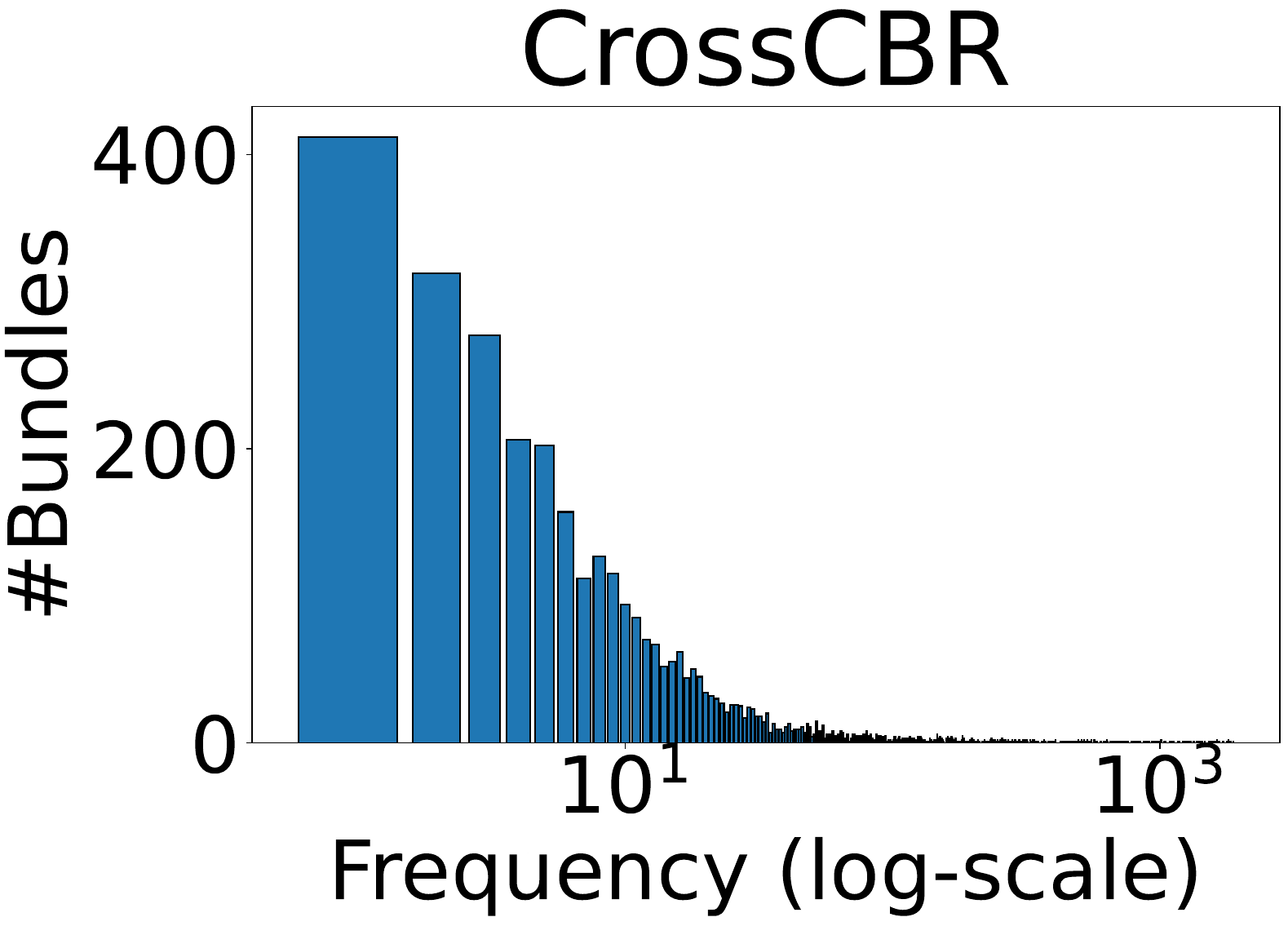}
        \includegraphics[width=0.19\textwidth]{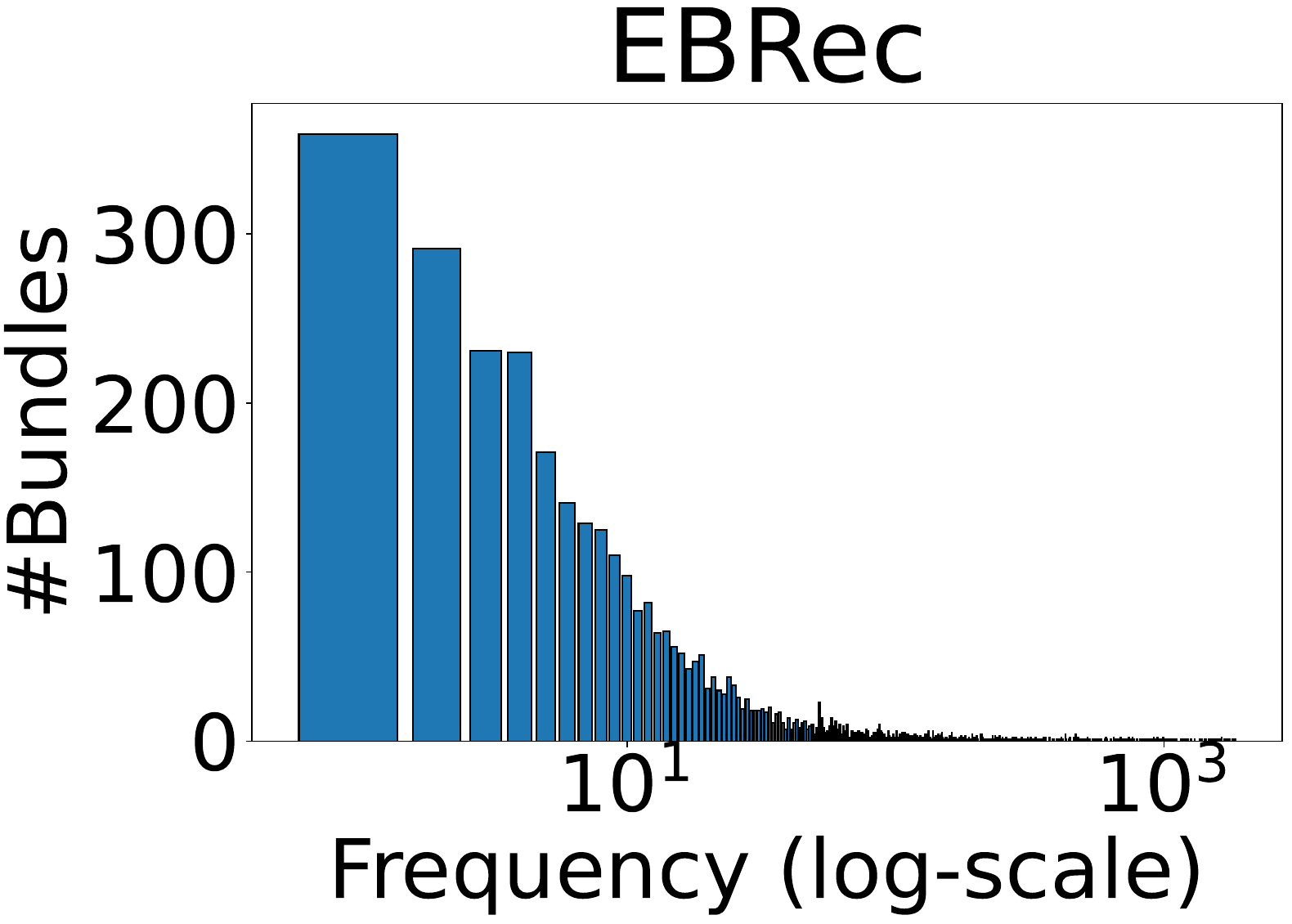}
        \includegraphics[width=0.19\textwidth]{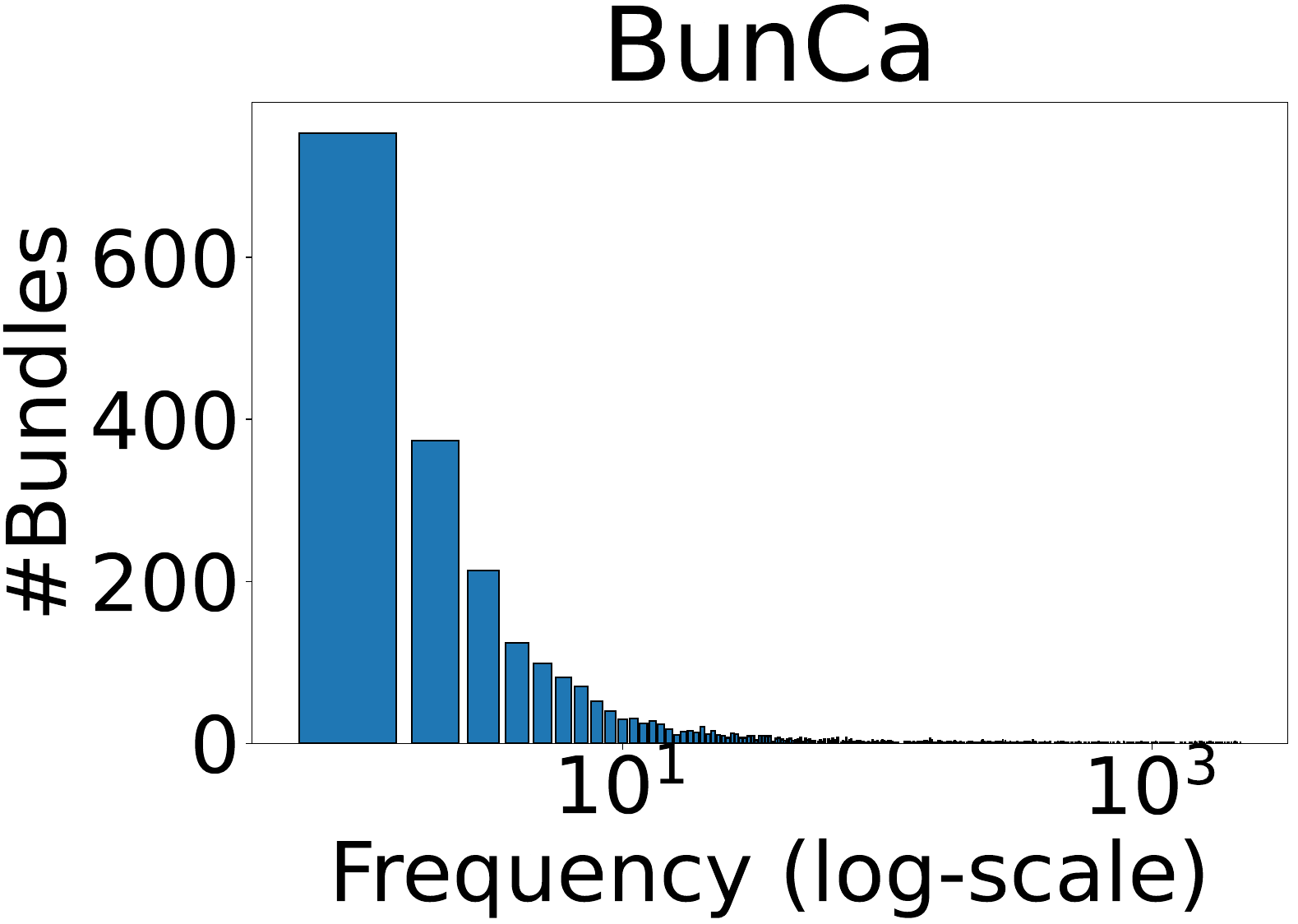}
        \vspace*{-2mm}
        \caption{Bundle distribution}\label{}
    \end{subfigure}
    \begin{subfigure}[b]{.9\textwidth}
        \includegraphics[width=0.19\textwidth]{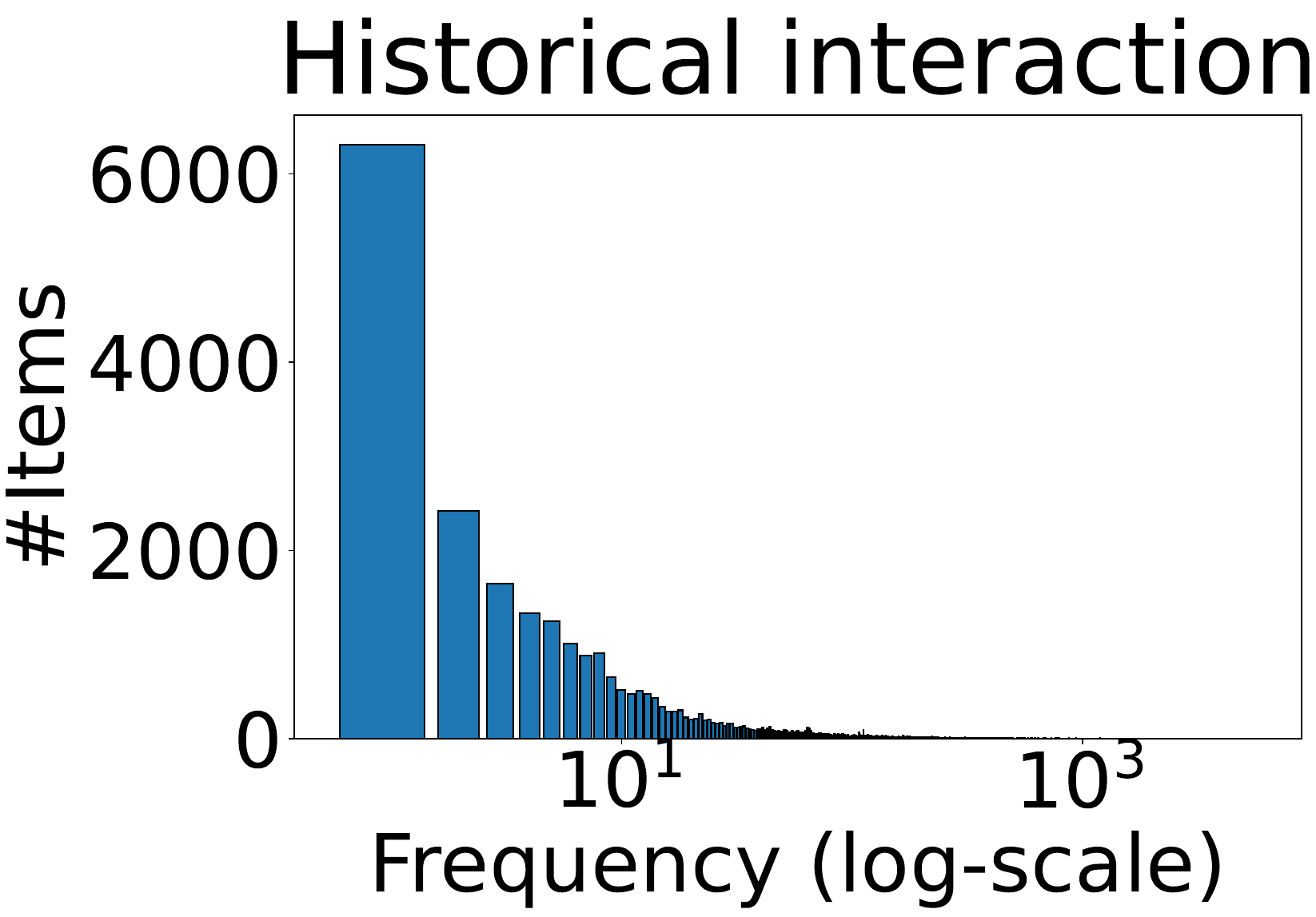}
        \includegraphics[width=0.19\textwidth]{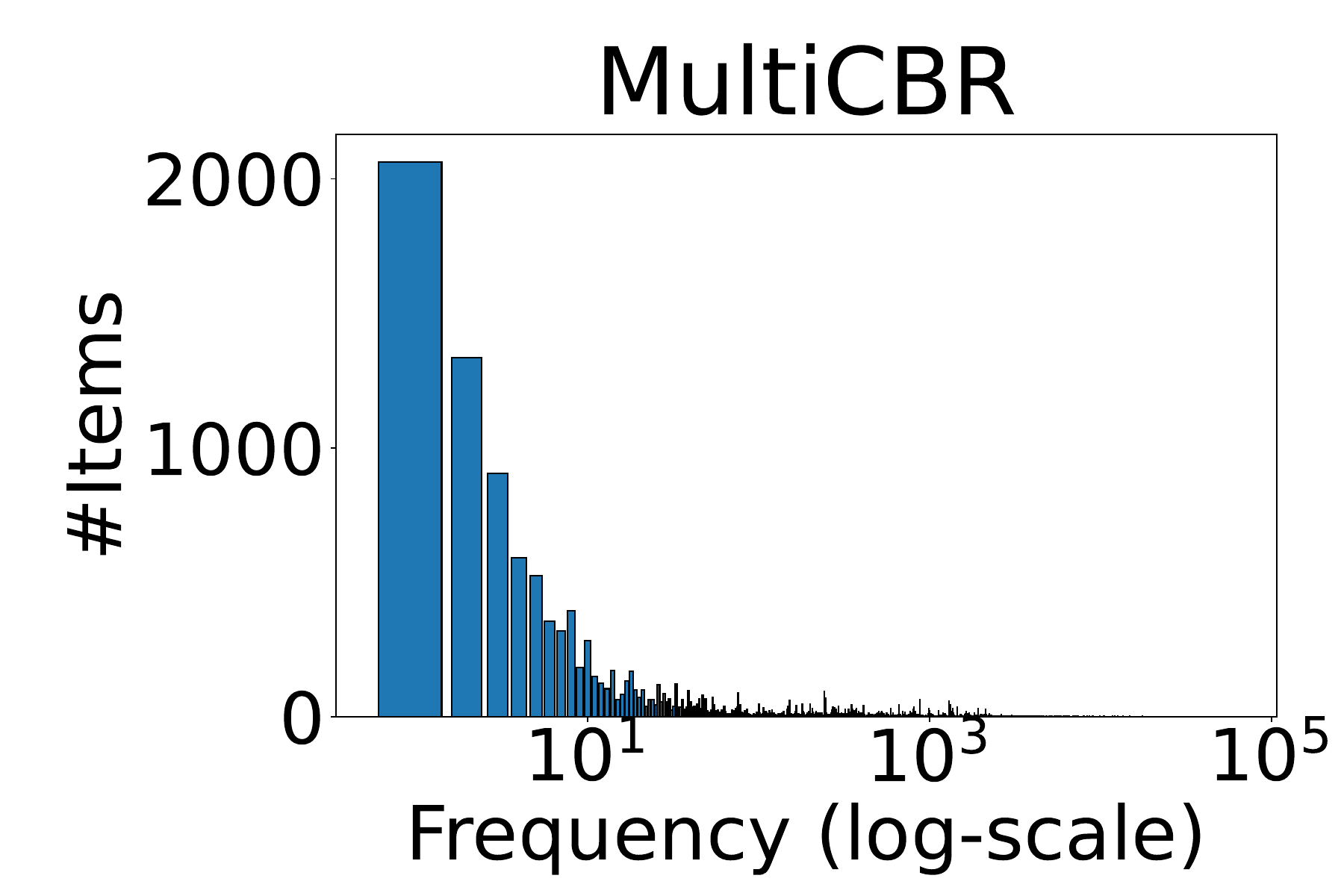}
        \includegraphics[width=0.19\textwidth]{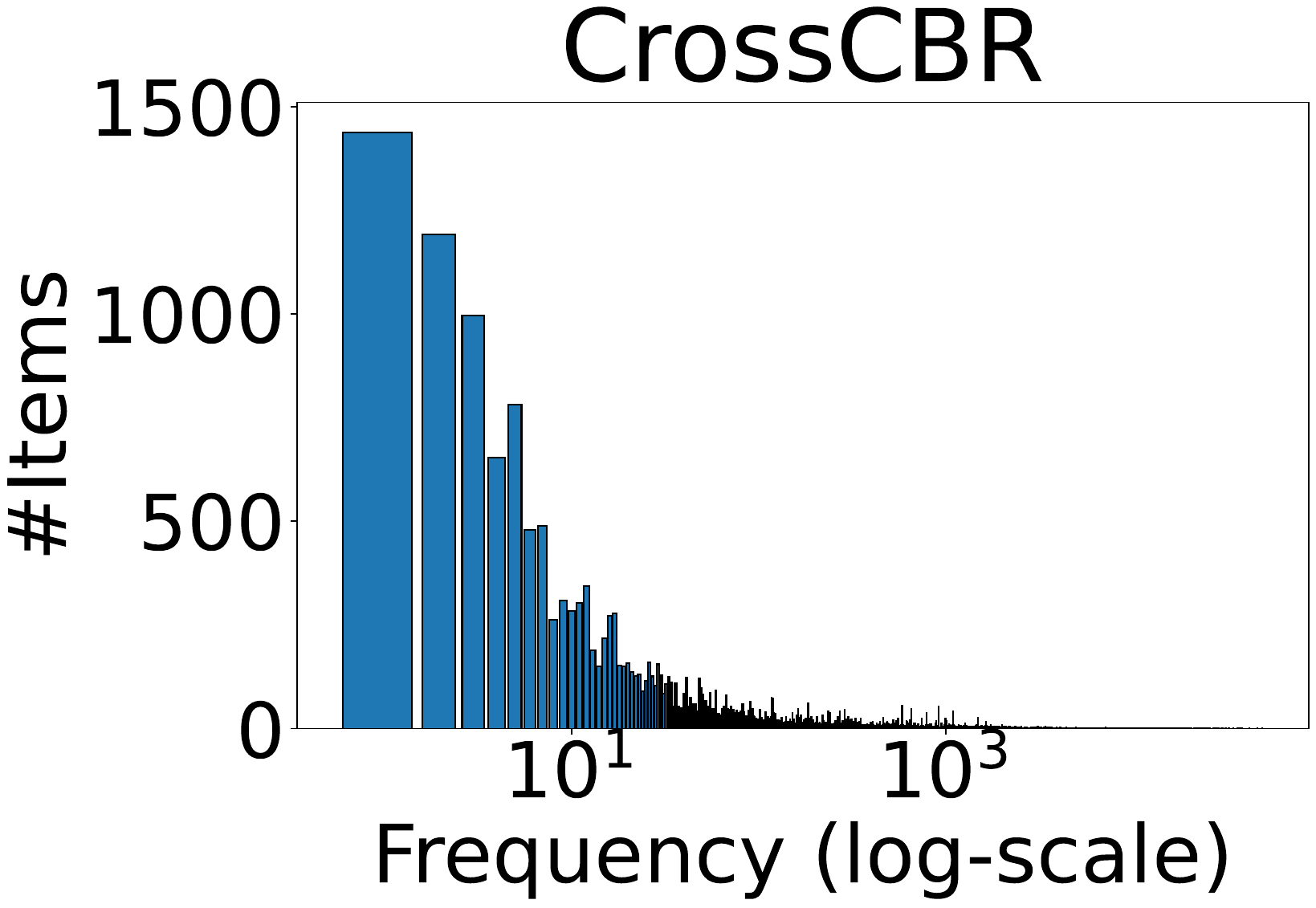}
        \includegraphics[width=0.19\textwidth]{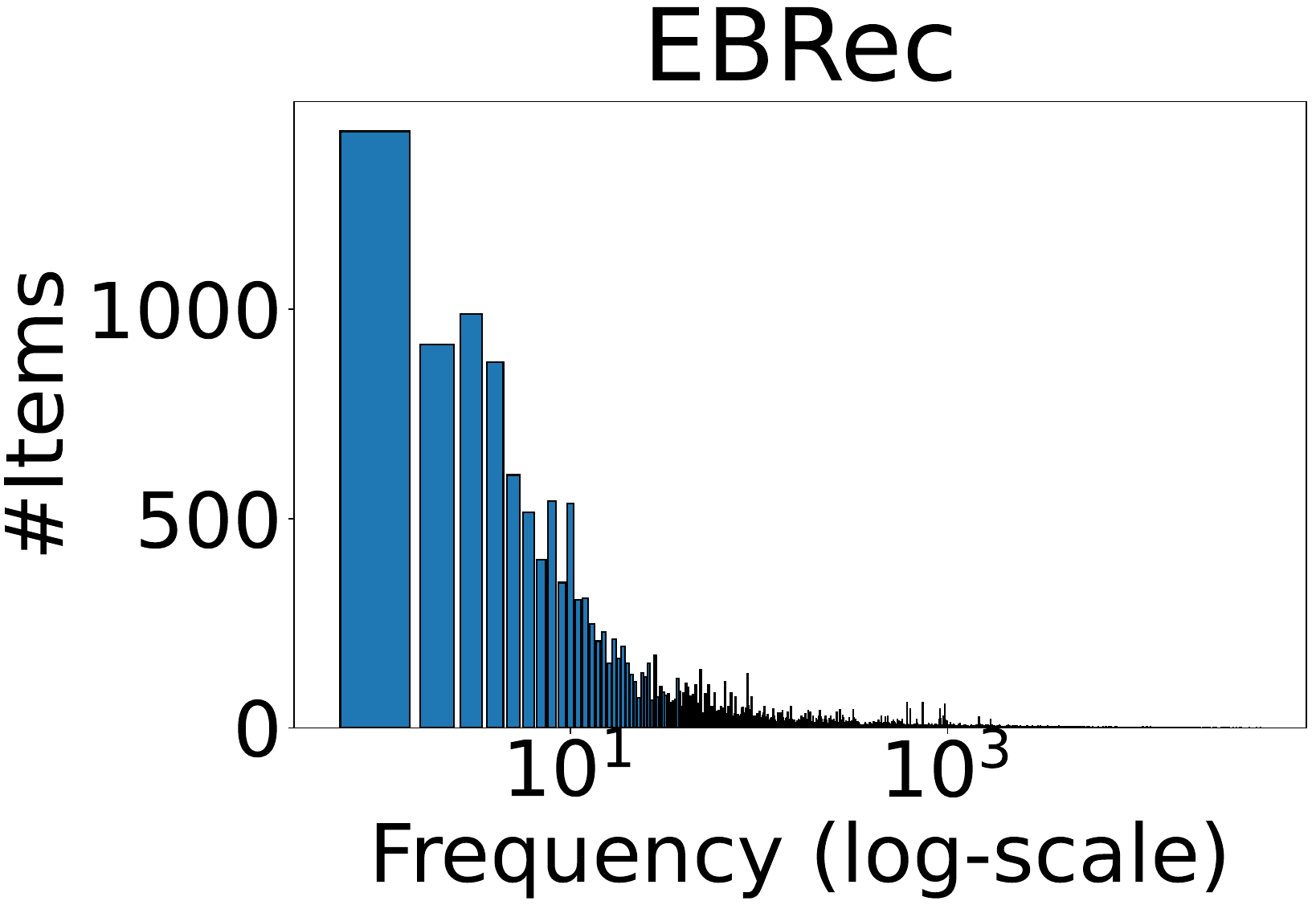}
        \includegraphics[width=0.19\textwidth]{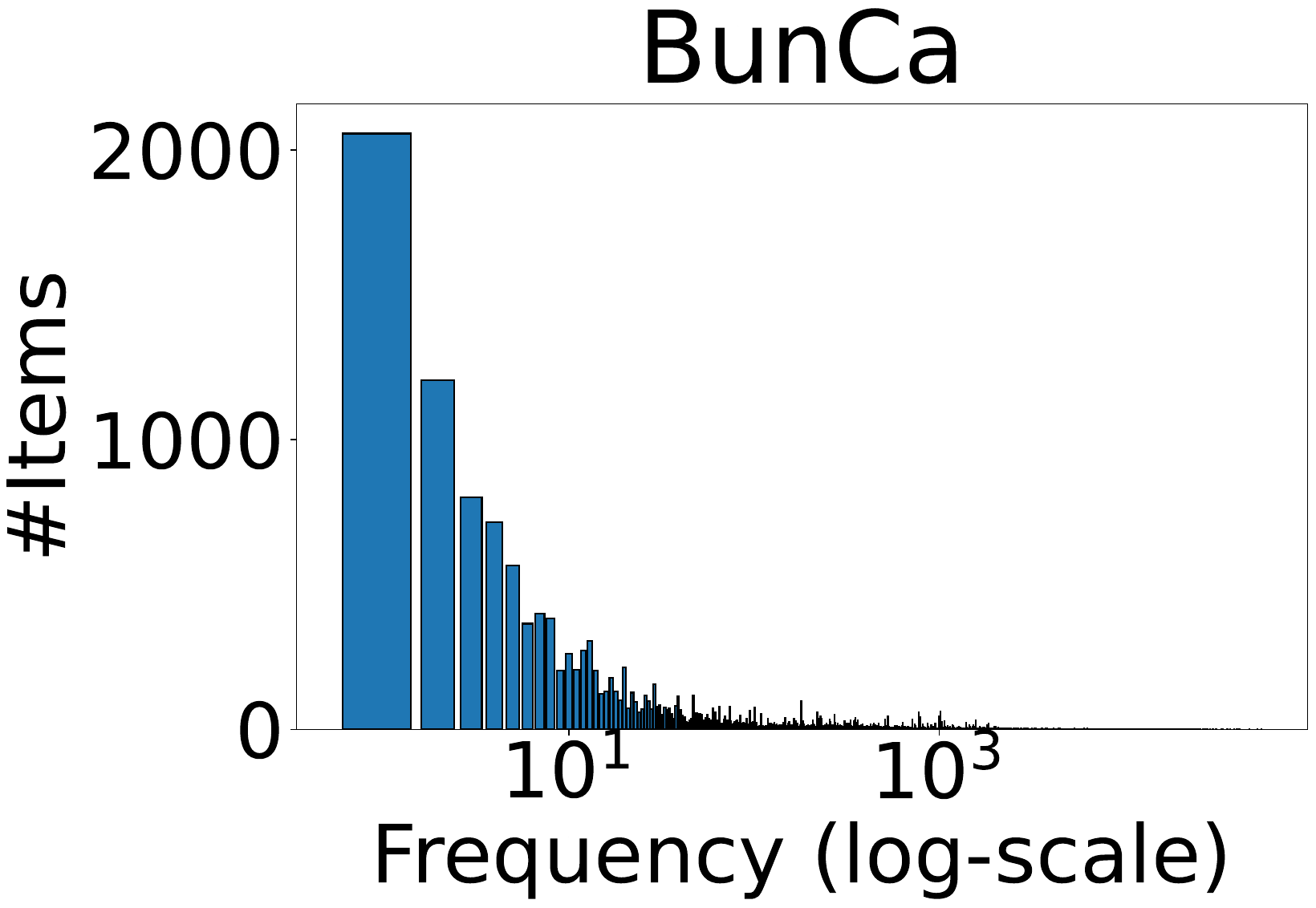}
        \vspace*{-2mm}
        \caption{Item distribution}\label{}
    \end{subfigure}
\caption{Distribution of bundle and item frequency in historical interaction and BR outcomes on Youshu dataset.}\label{fig:dist_youshu}
\Description{Distribution of bundle and item frequency on Youshu.}
\vspace*{-1.5mm}
\end{figure*}
\begin{figure*}[t!]
    \centering
    \begin{subfigure}[b]{.9\textwidth}
        \includegraphics[width=0.19\textwidth]{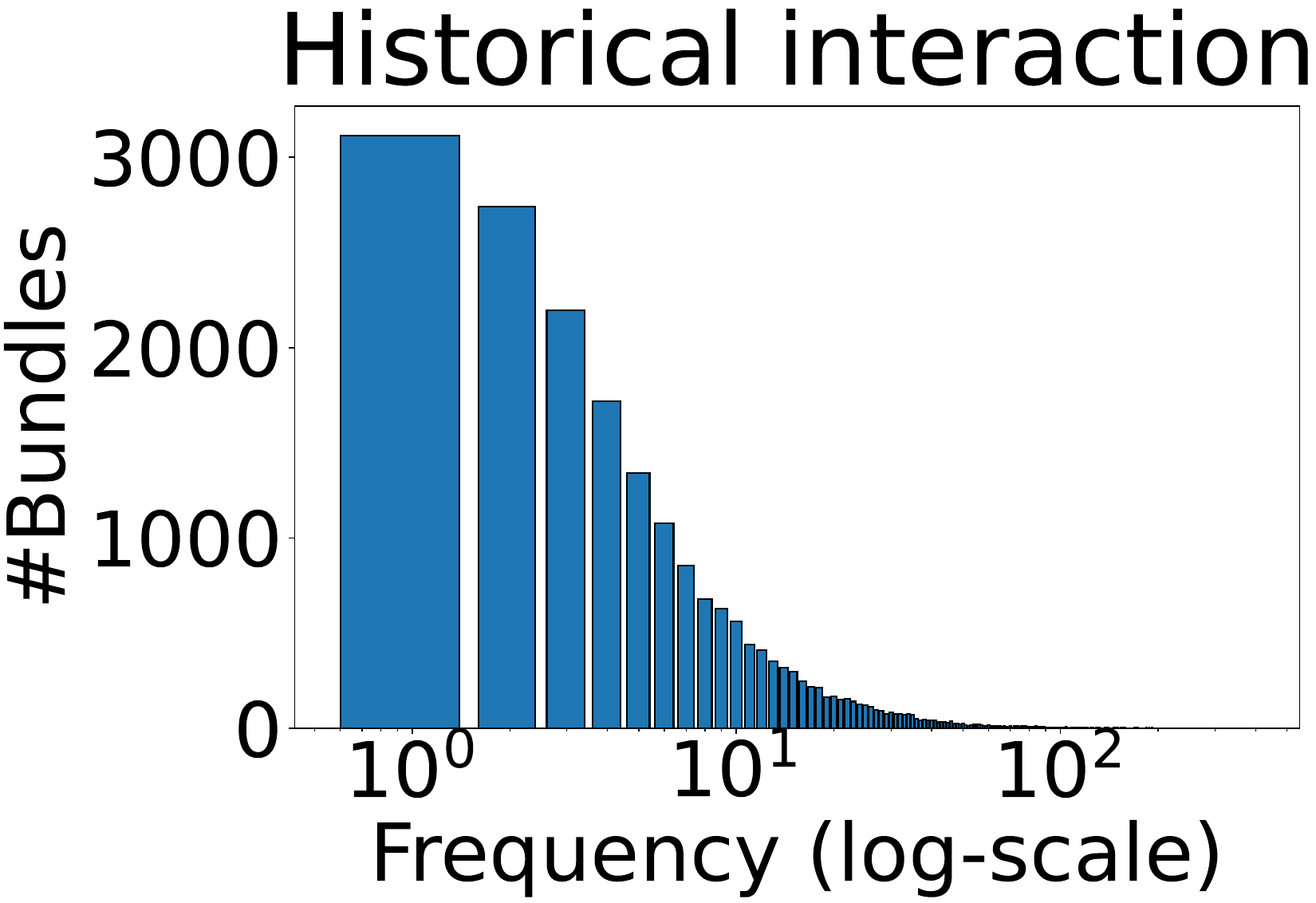}
        \includegraphics[width=0.19\textwidth]{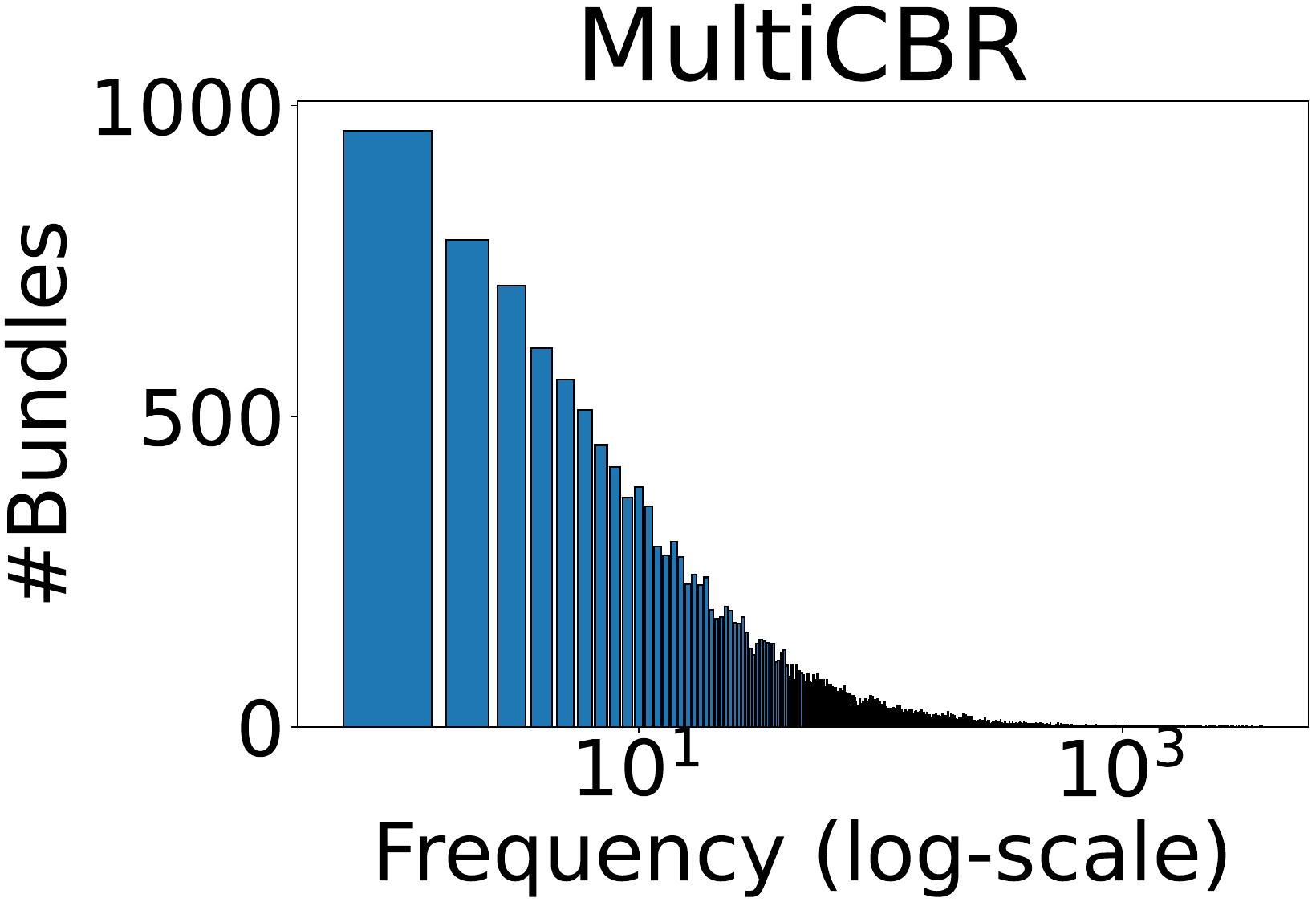}
        \includegraphics[width=0.19\textwidth]{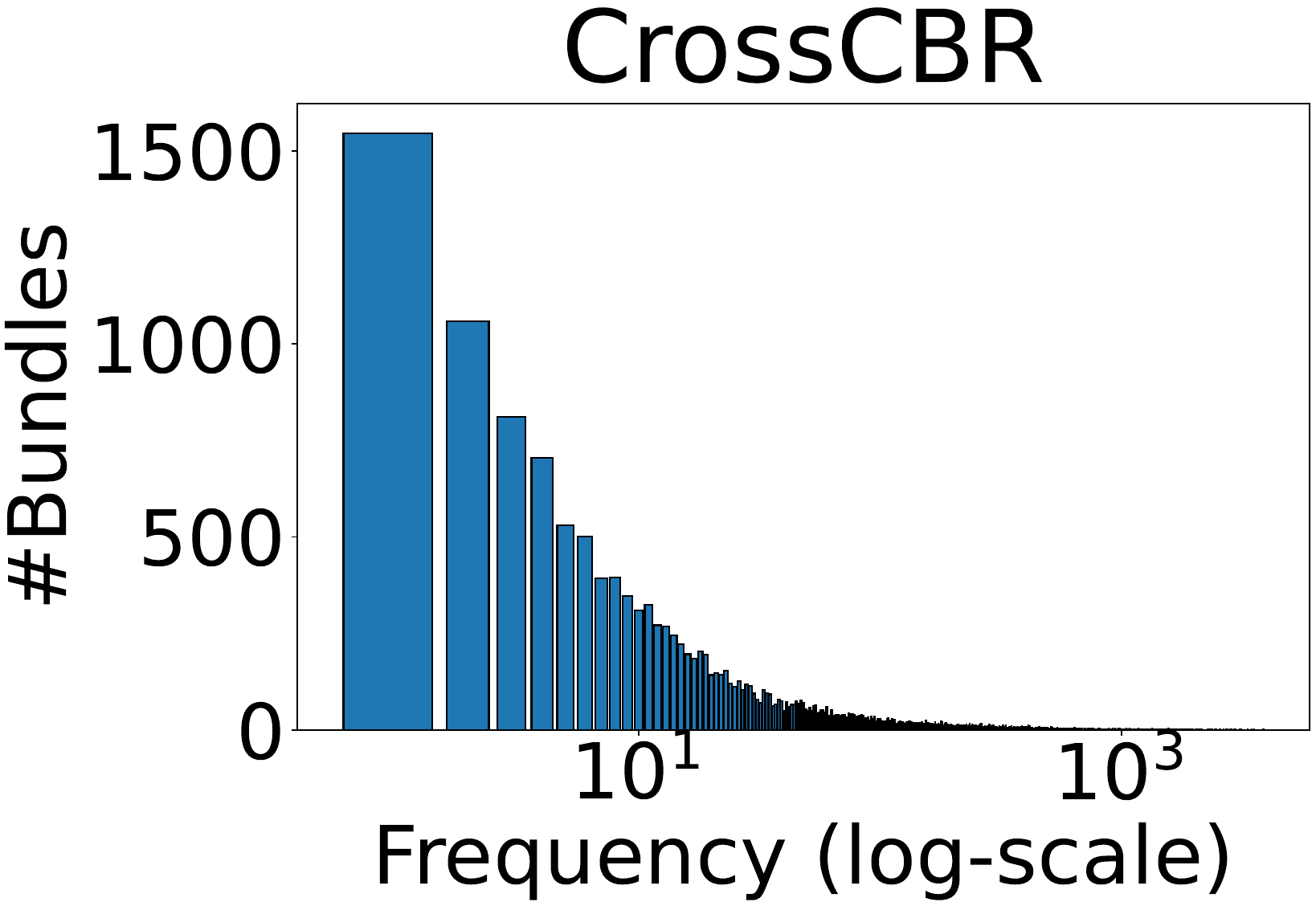}
        \includegraphics[width=0.19\textwidth]{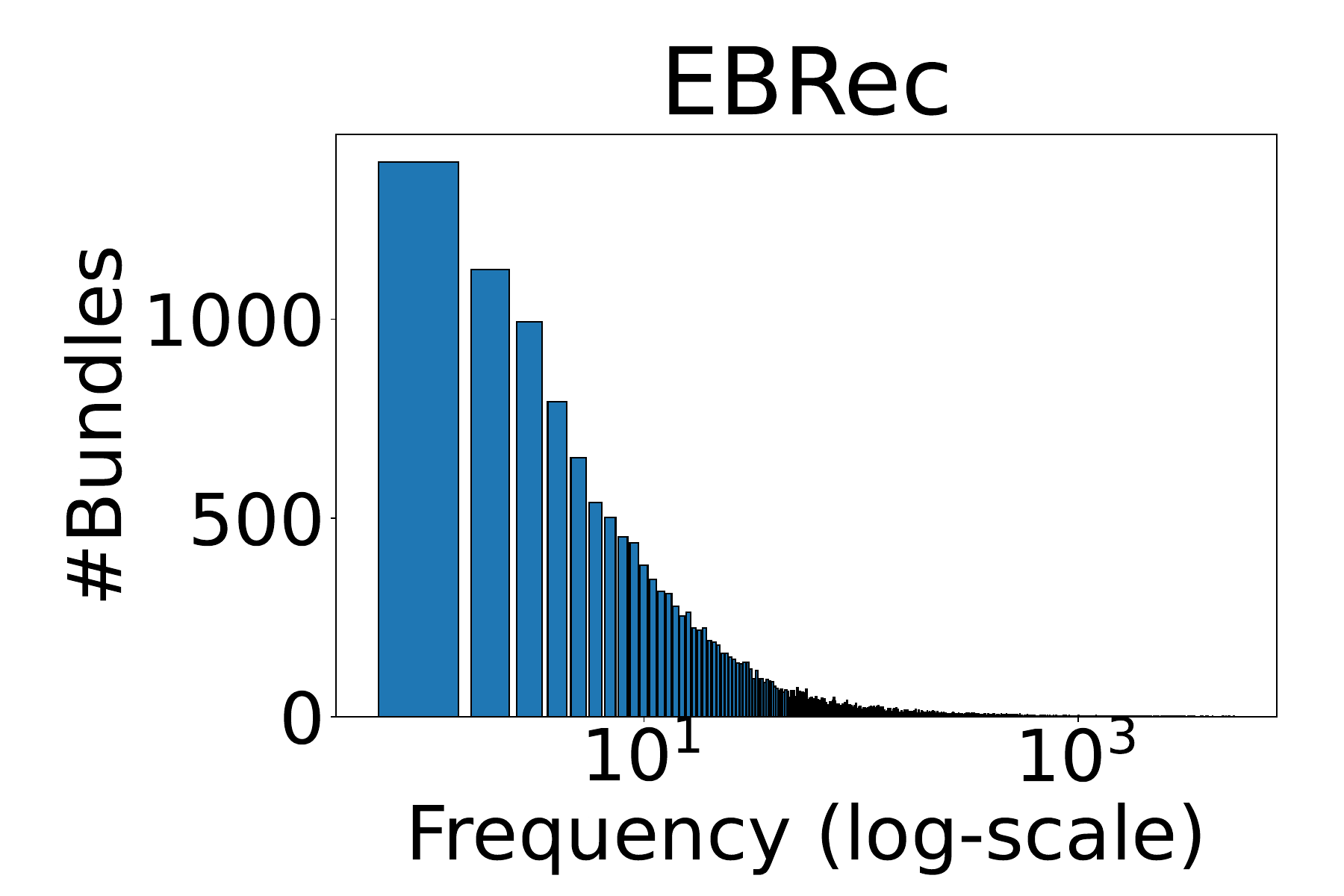}
        \includegraphics[width=0.19\textwidth]{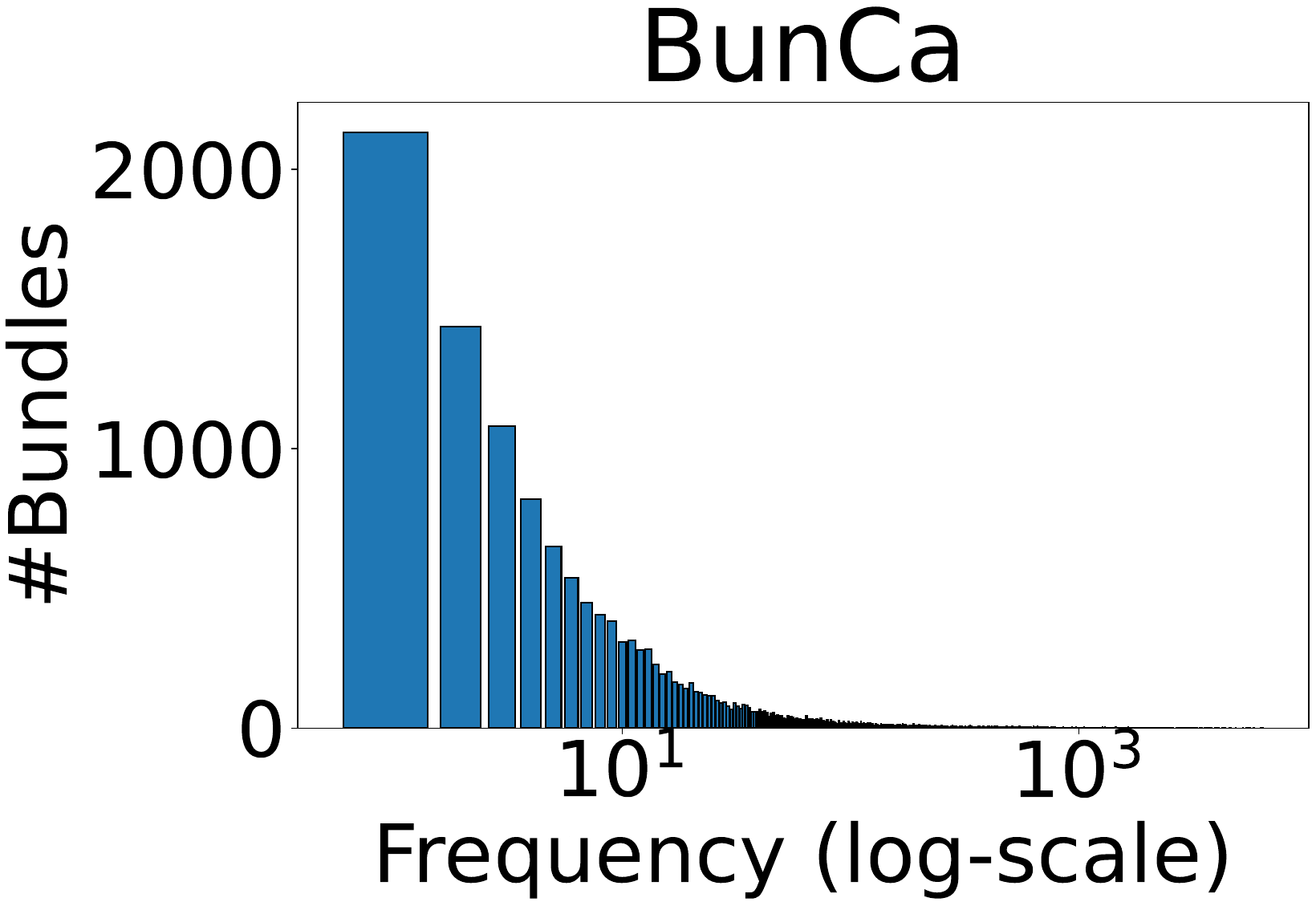}
        \vspace*{-2mm}
        \caption{Bundle distribution}\label{}
    \end{subfigure}
    \begin{subfigure}[b]{.9\textwidth}
        \includegraphics[width=0.19\textwidth]{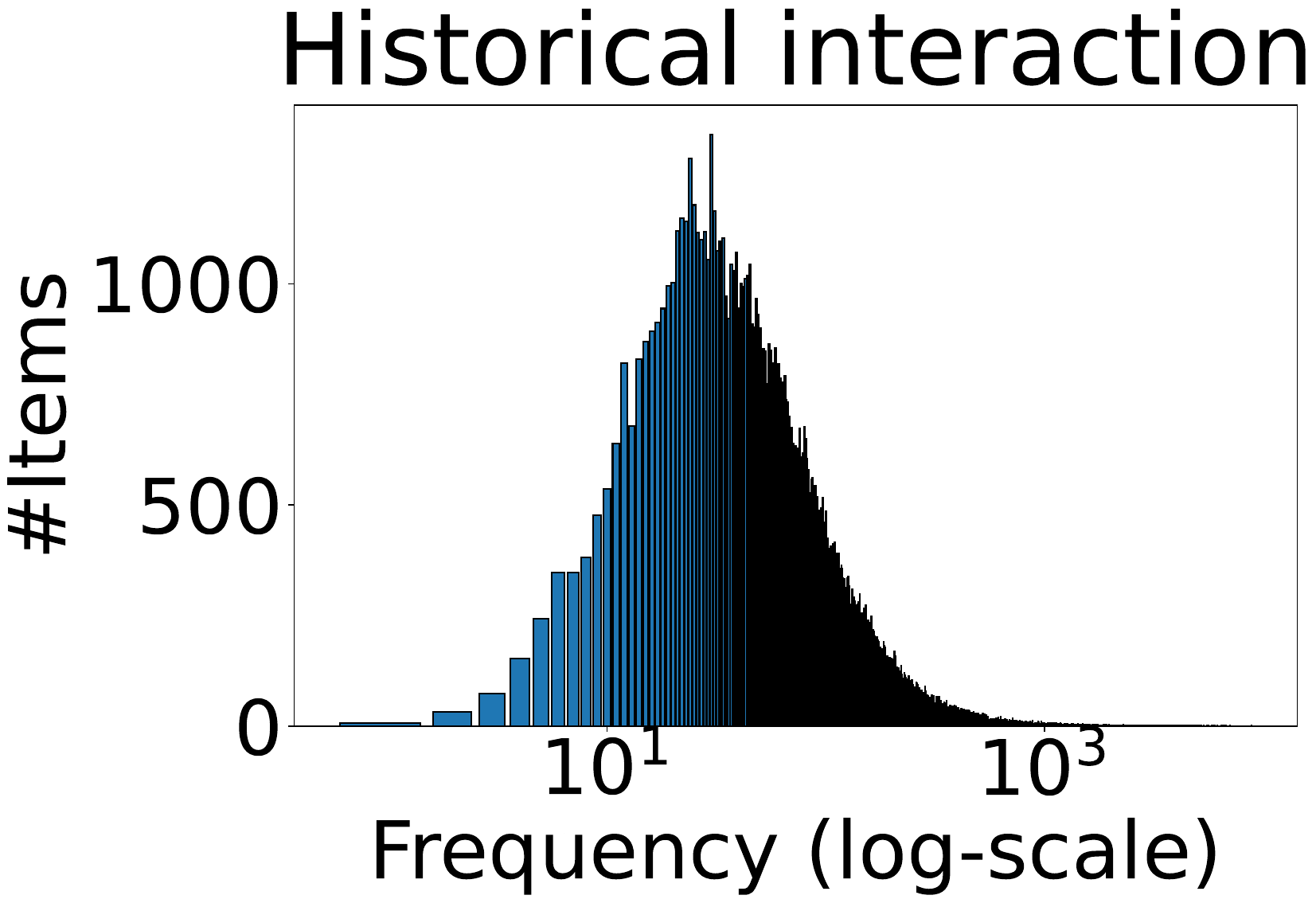}
        \includegraphics[width=0.19\textwidth]{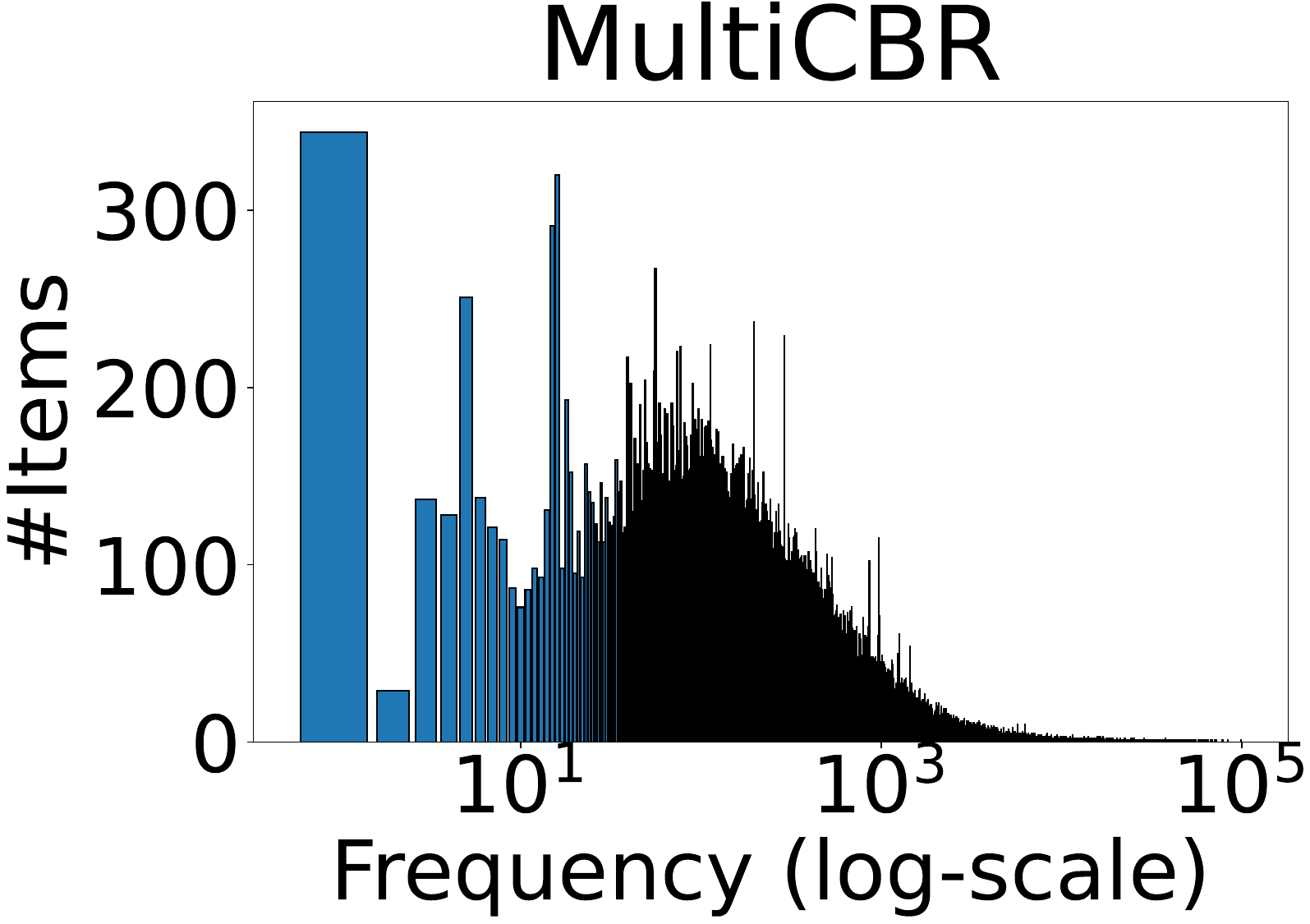}
        \includegraphics[width=0.19\textwidth]{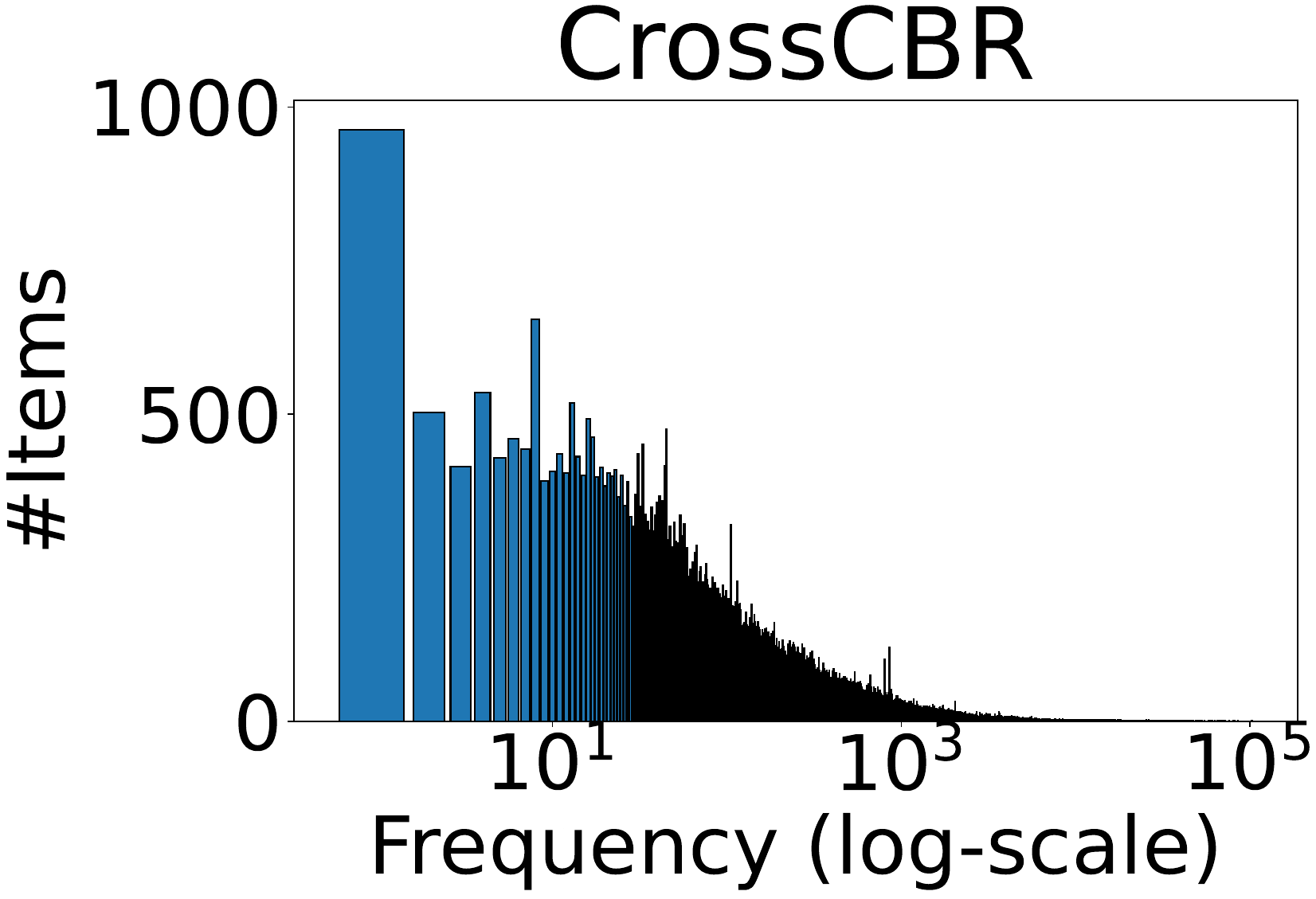}
        \includegraphics[width=0.19\textwidth]{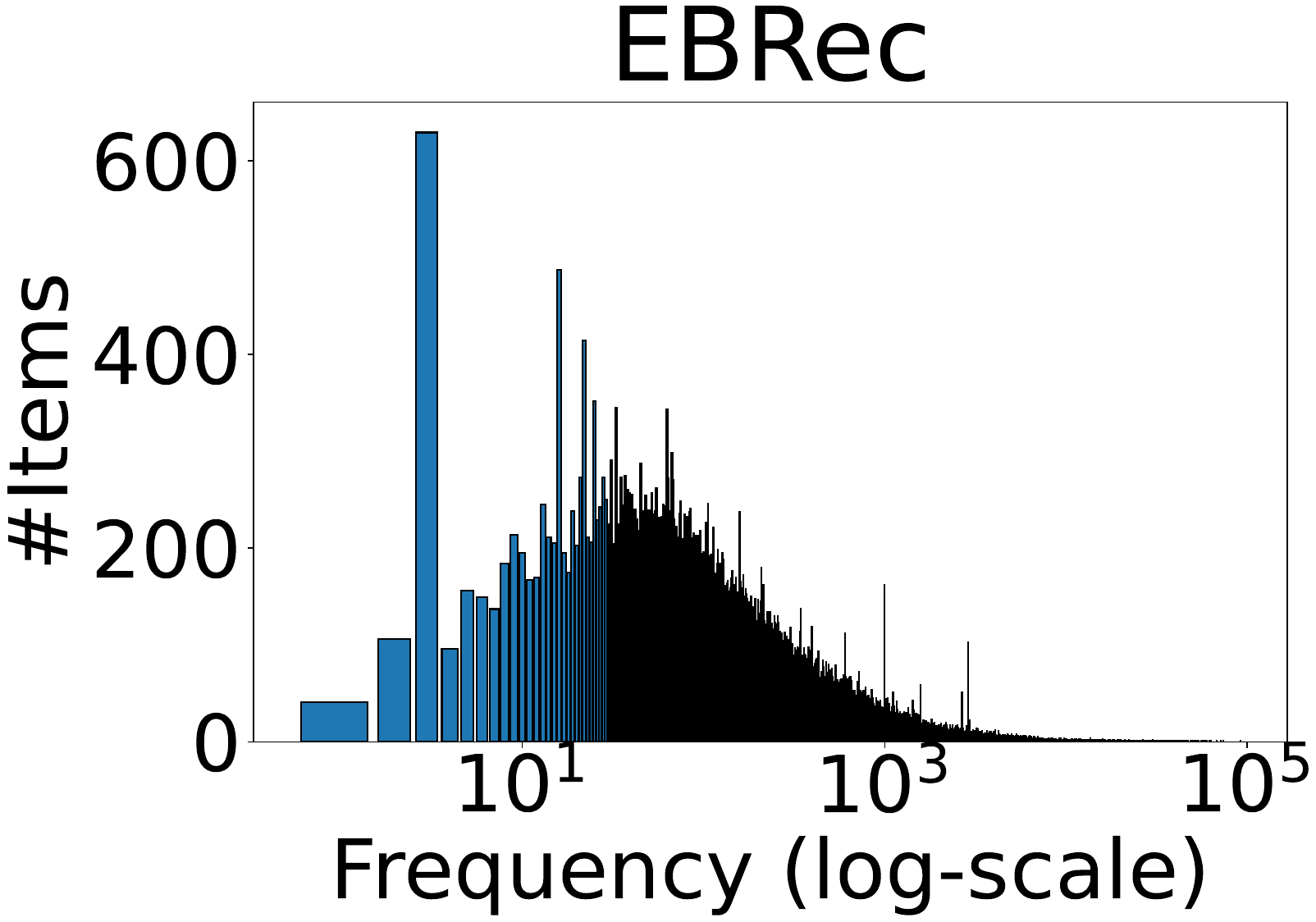}
        \includegraphics[width=0.19\textwidth]{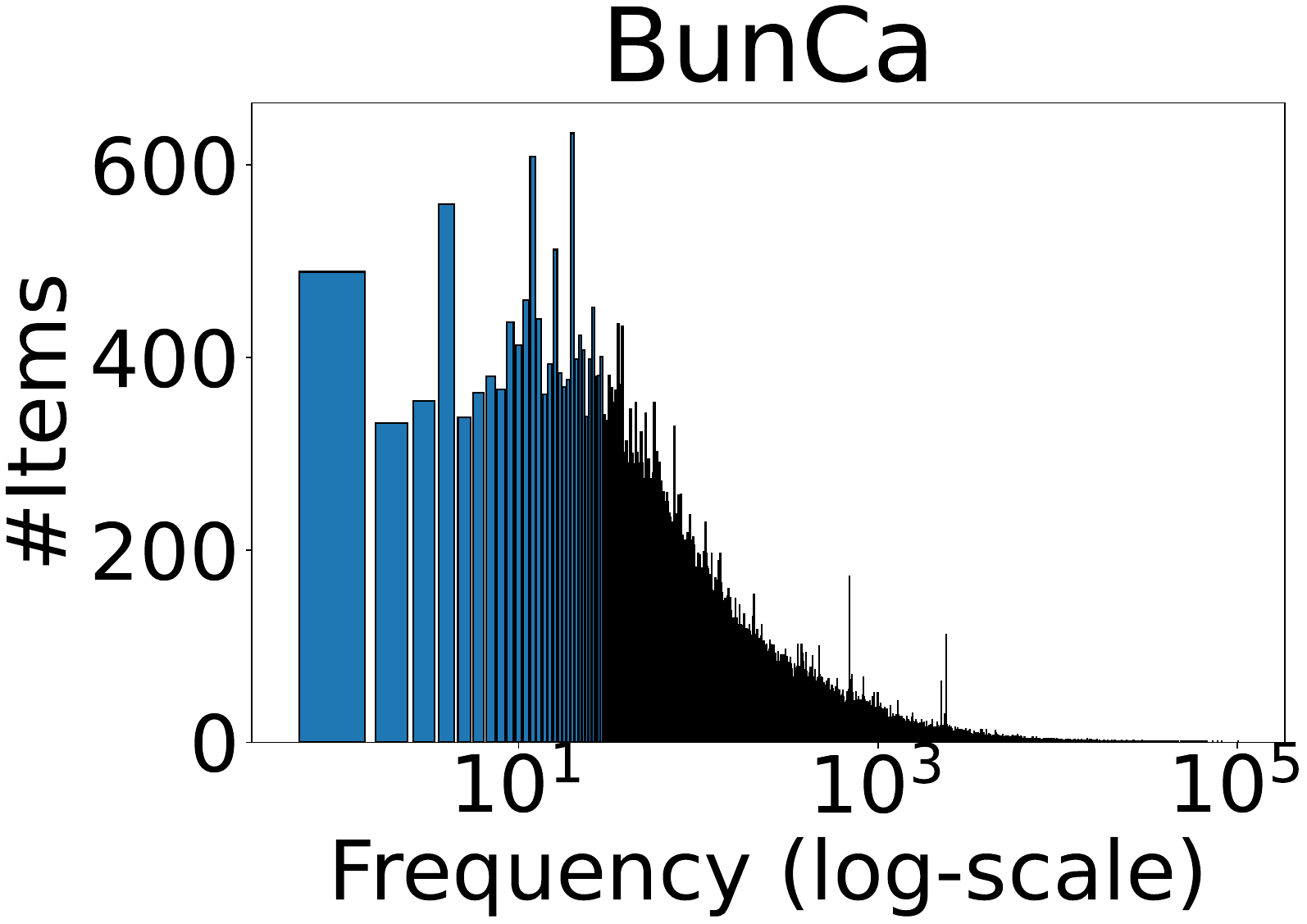}
        \vspace*{-2mm}
        \caption{Item distribution}\label{}
    \end{subfigure}
\caption{Distribution of bundle and item frequency in historical interaction and BR outcomes on NetEase dataset.}\label{fig:dist_netease}
\Description{Distribution of bundle and item frequency on NetEase.}
\vspace*{-1.5mm}
\end{figure*}
\begin{figure*}[t!]
    \centering
    \begin{subfigure}[b]{.9\textwidth}
        \includegraphics[width=0.19\textwidth]{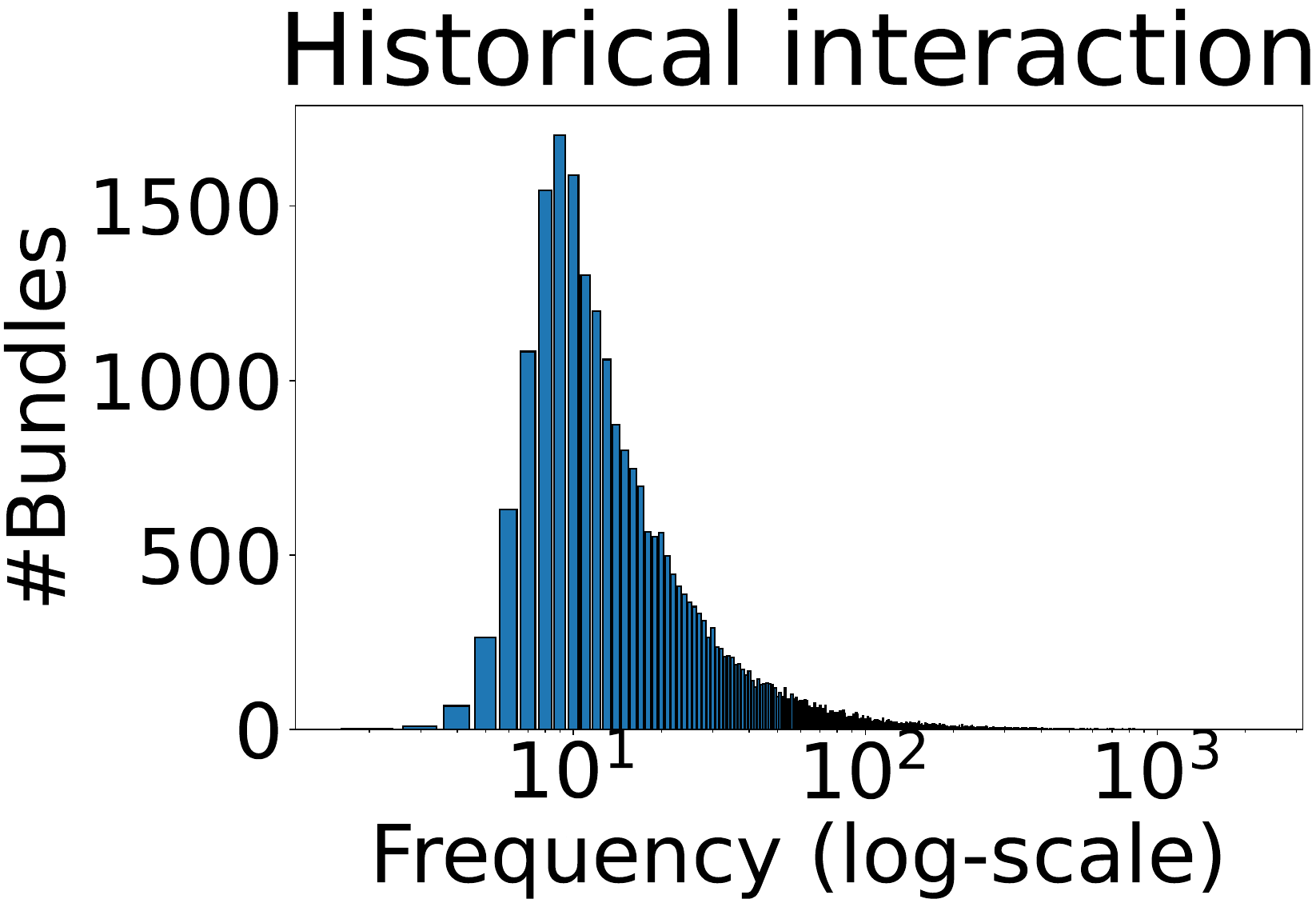}
        \includegraphics[width=0.19\textwidth]{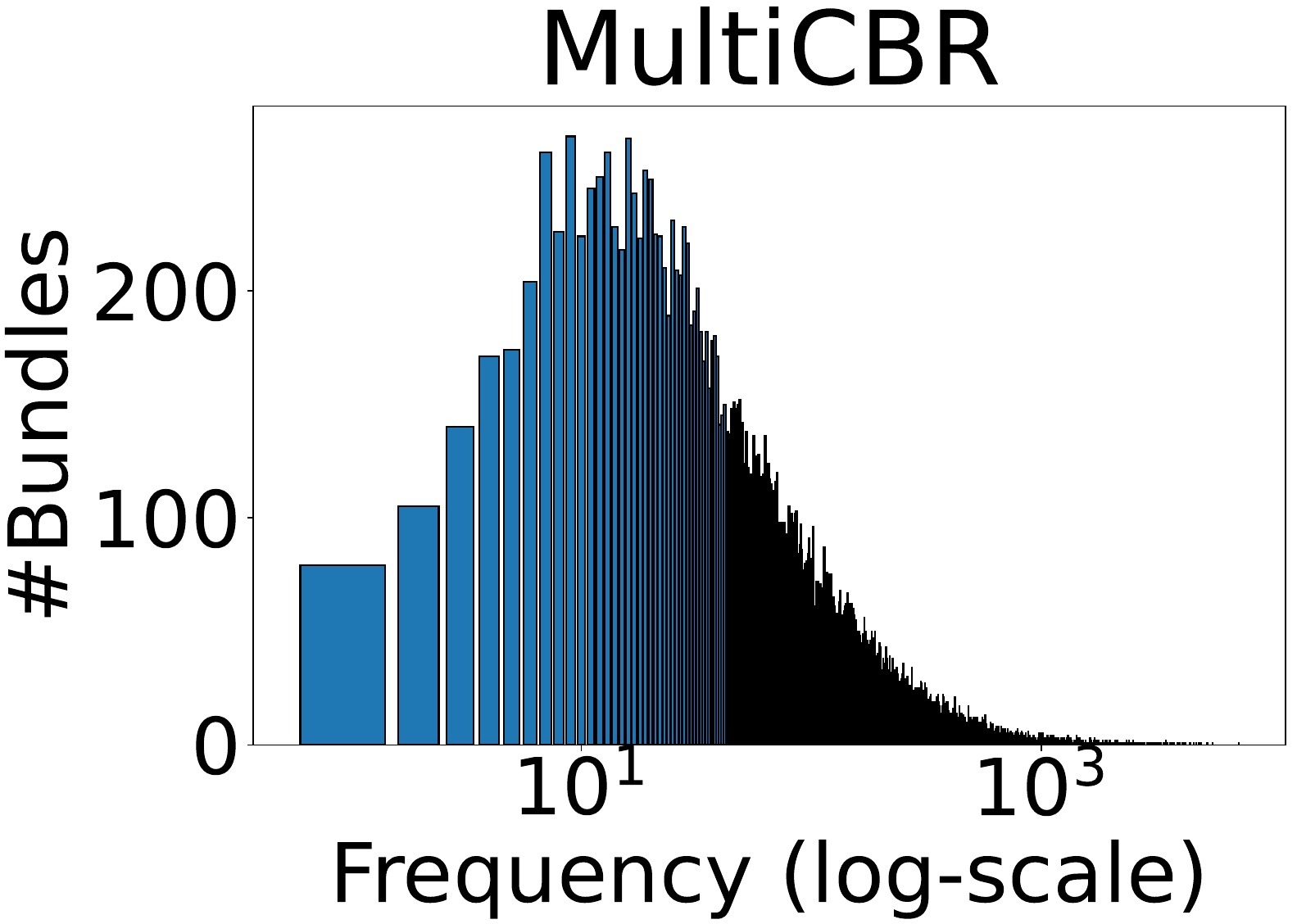}
        \includegraphics[width=0.19\textwidth]{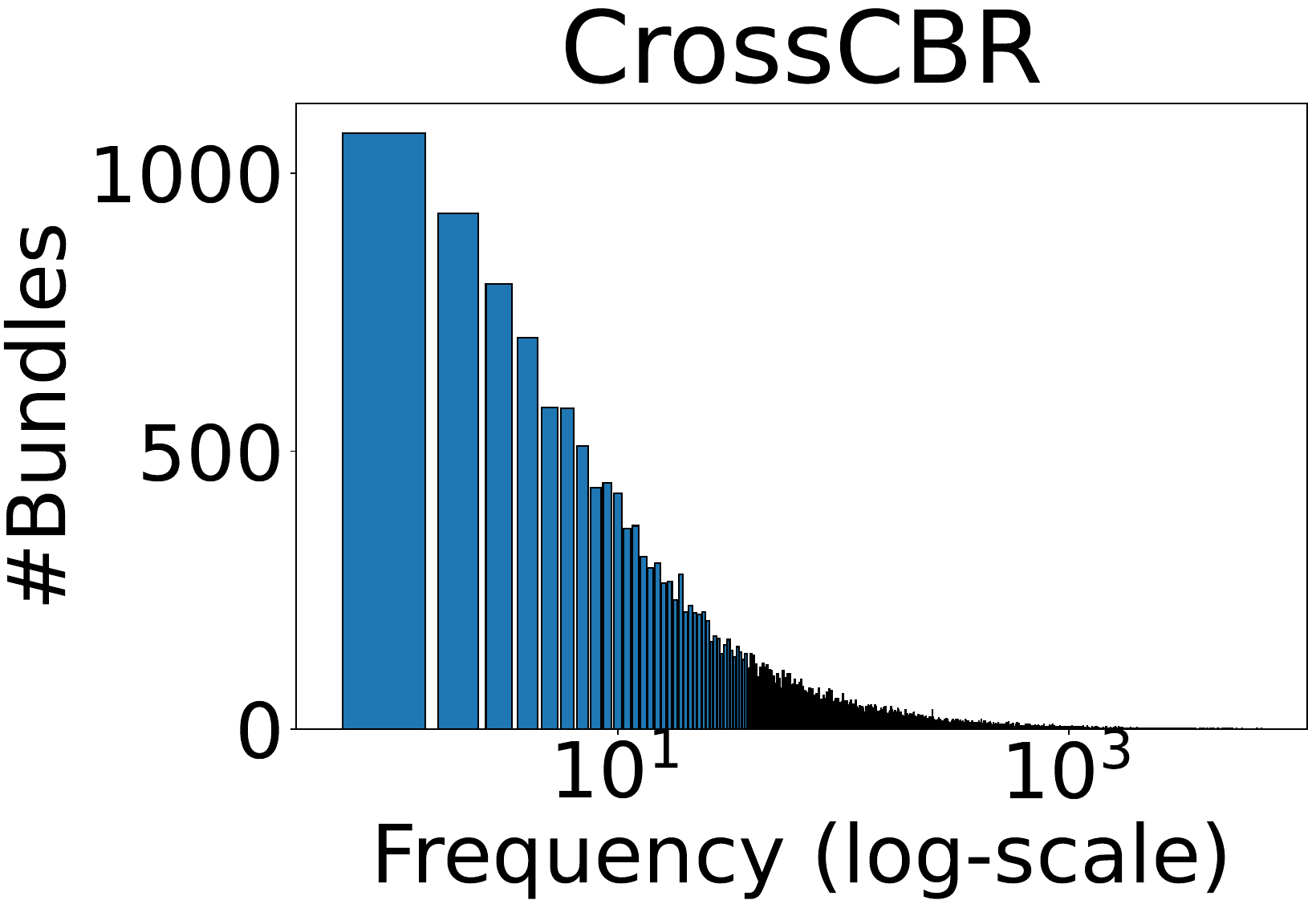}
        \includegraphics[width=0.19\textwidth]{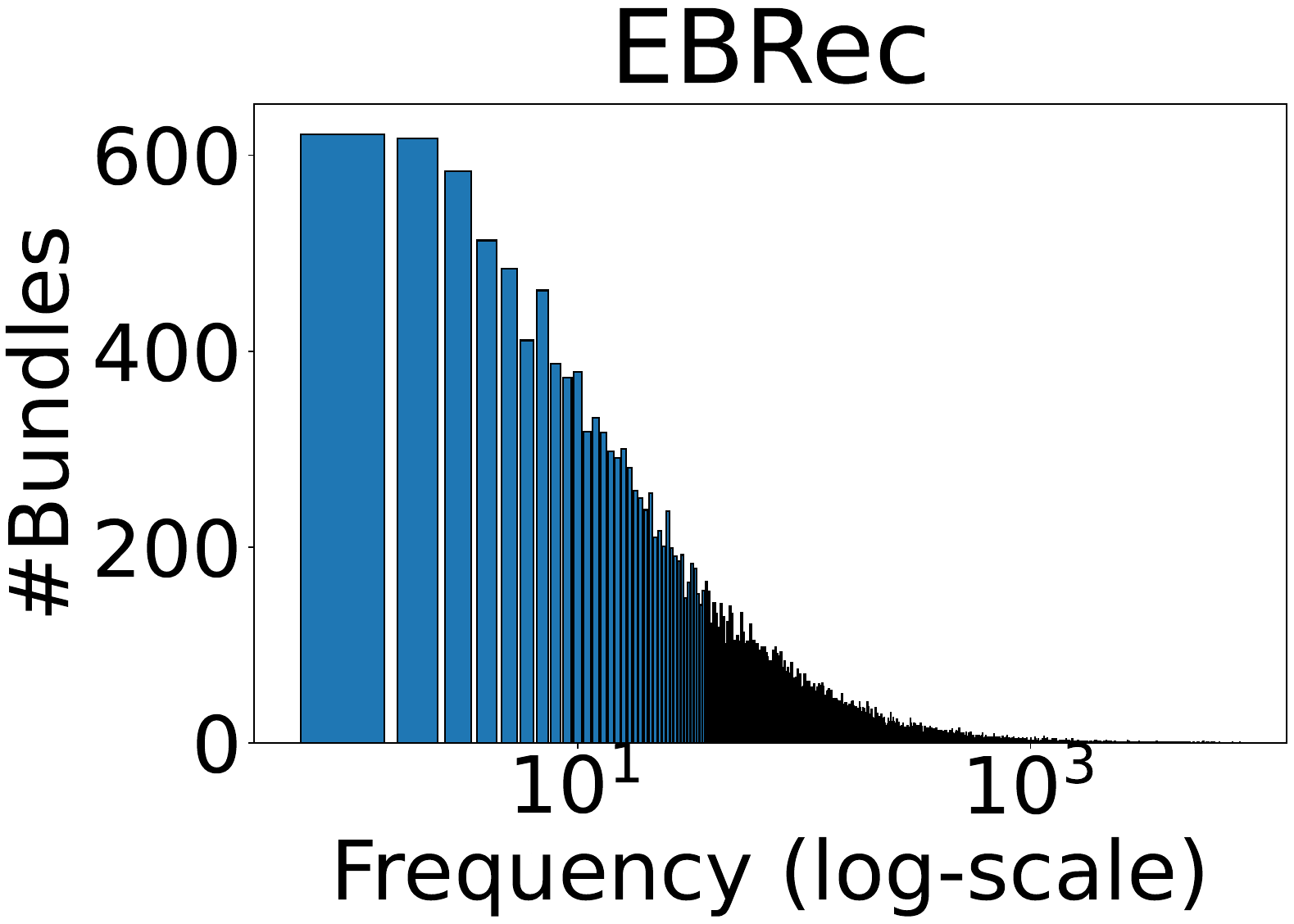}
        \includegraphics[width=0.19\textwidth]{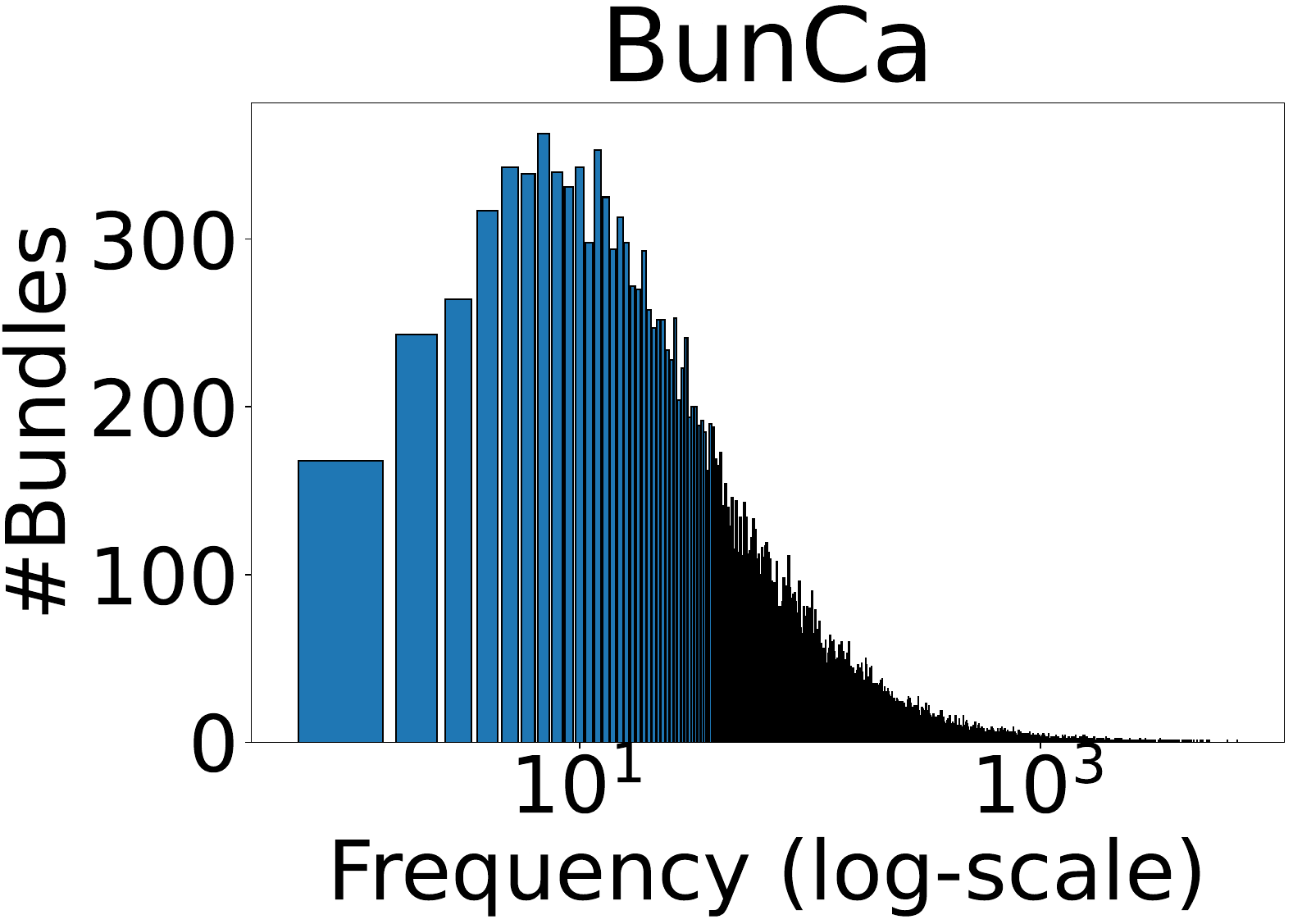}
        \vspace*{-2mm}
        \caption{Bundle distribution}\label{}
    \end{subfigure}
    \begin{subfigure}[b]{.9\textwidth}
        \includegraphics[width=0.19\textwidth]{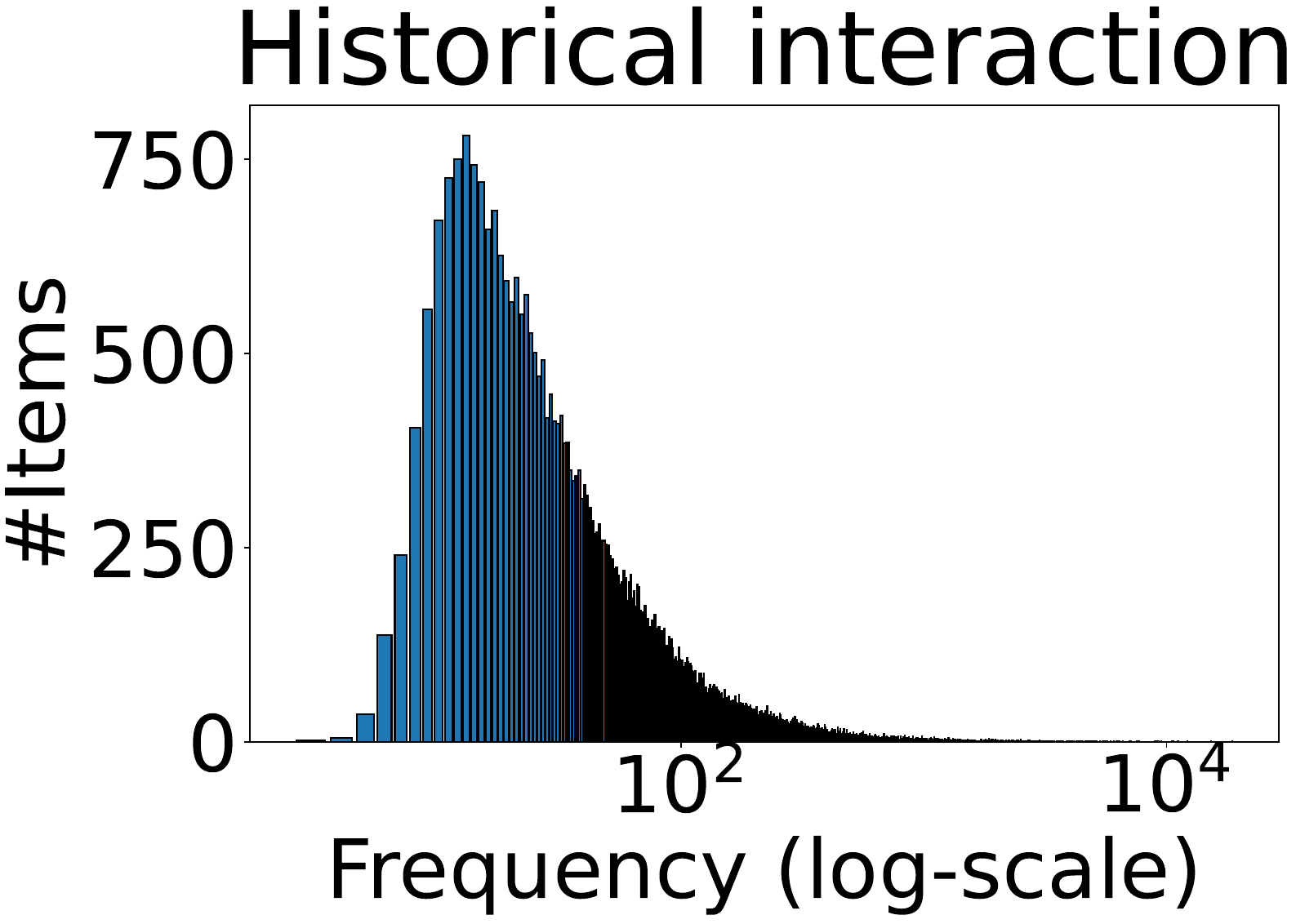}
        \includegraphics[width=0.19\textwidth]{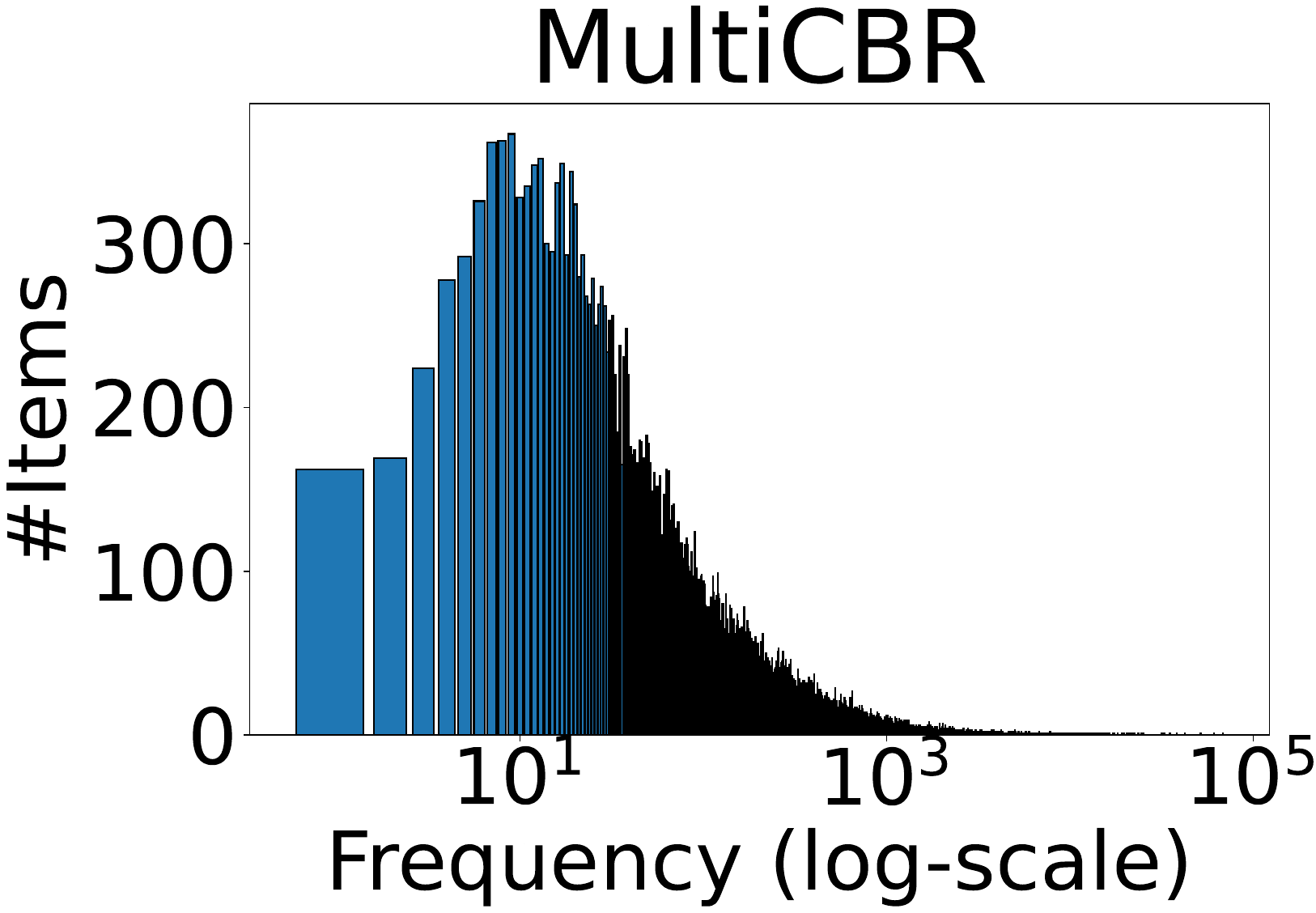}
        \includegraphics[width=0.19\textwidth]{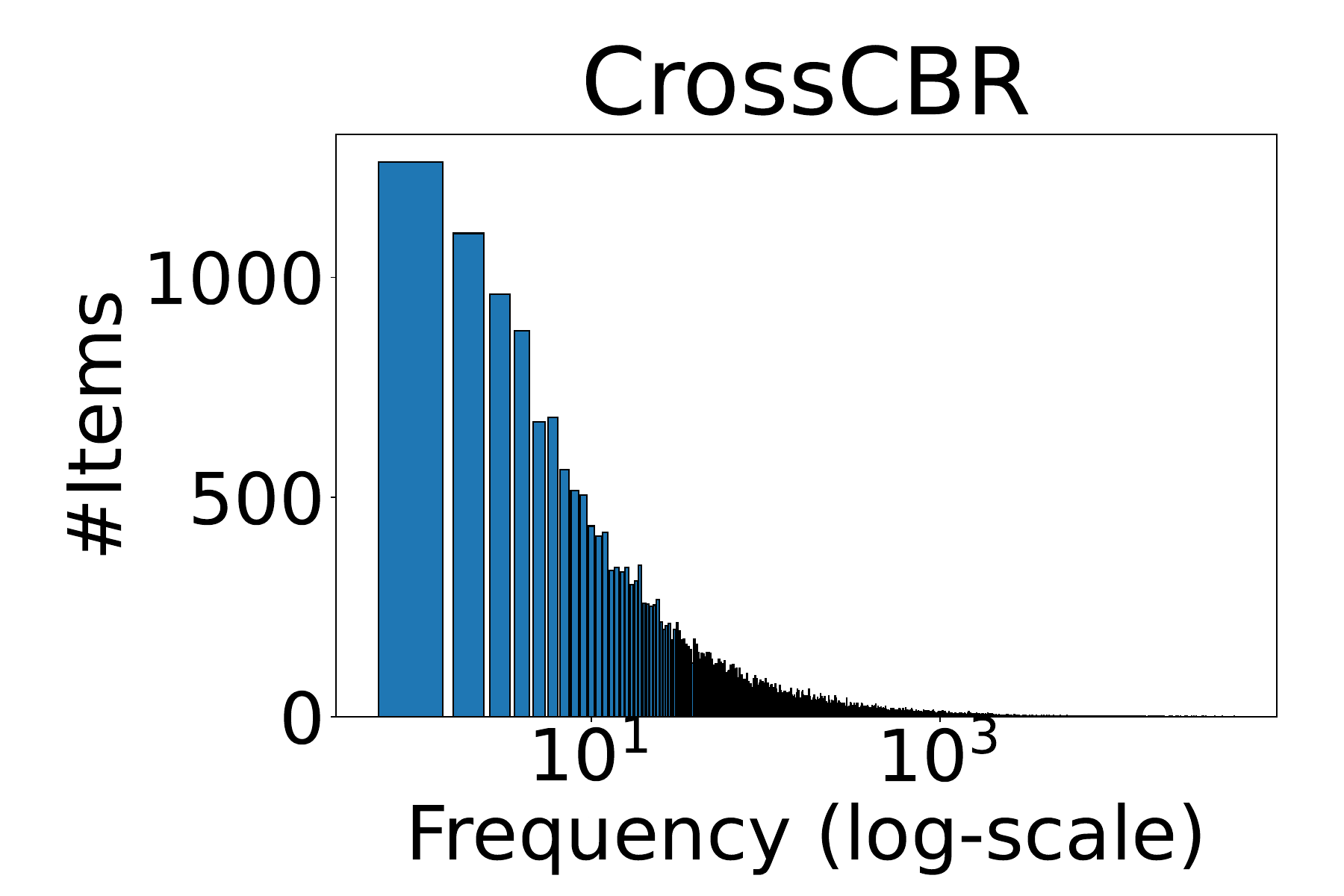}
        \includegraphics[width=0.19\textwidth]{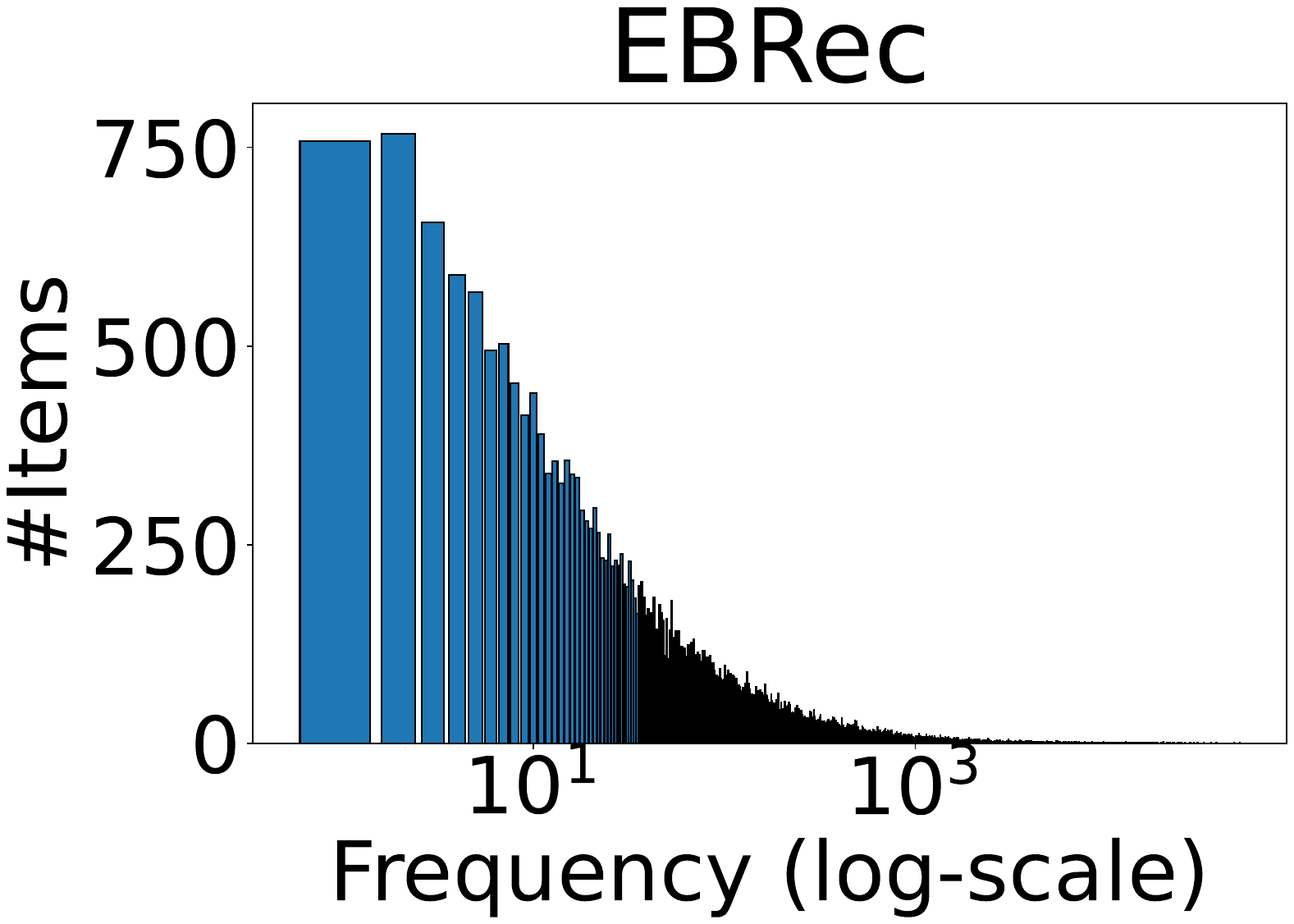}
        \includegraphics[width=0.19\textwidth]{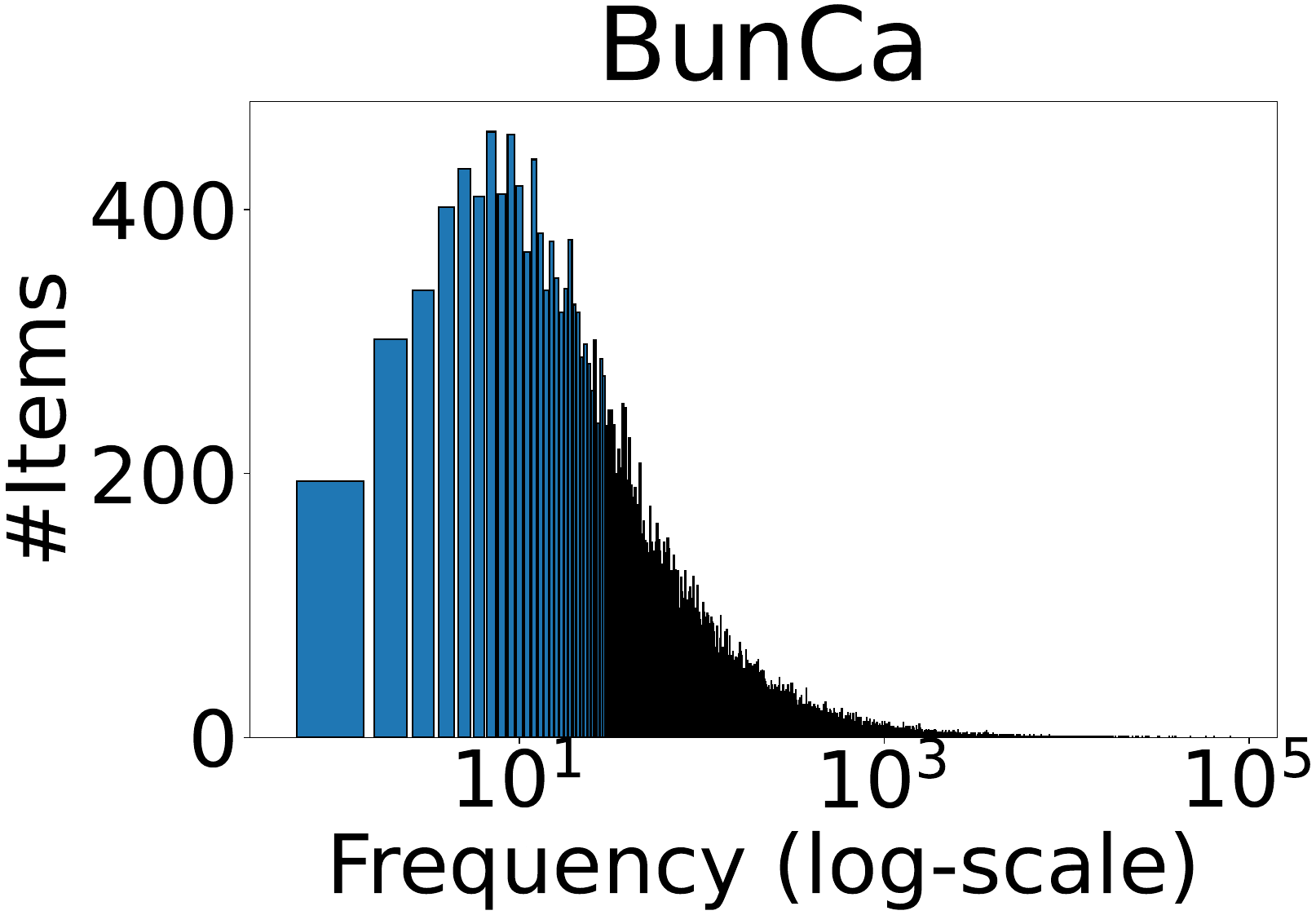}
        \vspace*{-2mm}
        \caption{Item distribution}\label{}
    \end{subfigure}
\caption{Distribution of bundle and item frequency in historical interaction and BR outcomes on iFashion dataset.}
\Description{Distribution of bundle and item frequency on iFashion.}
\label{fig:dist_ifashion}
\vspace*{-1.5mm}
\end{figure*}

To ensure convincing comparisons, our implementation uses settings from state-of-the-art publications on bundle recommendation~\cite{du2023enhancing,ma2022crosscbr,ma2024multicbr,nguyen2024bundle}. 
The initial embedding dimension is set to $64$, Xavier~\cite{glorot2010understanding} is used to initialize the trainable parameters, and the Adam optimizer~\cite{KingBa15} is applied to ensure stable and efficient training. 
Regarding validating hyperparameters, we follow the optimal configurations provided in the original papers.
To facilitate performance validation, the utility metrics, including R@$K$ and N@$K$, are employed with $K=20$. In the Geometric browsing model, $\gamma$ is set to 0.5.
For the training and test phases, our computational environment is built on NVIDIA P100 and T4 GPUs.Our repository is available on Github via \url{https://github.com/Rec4Fun/Fairness_Bundle_RecSys25}. 

\section{Experimental Results}

In this section, we present the results from our experiments to answer our three research questions from the introduction.

\vspace{-2mm}
\subsection{Popularity and Exposure Analysis (RQ1)}

RQ1 investigates the relationship between the distribution of user interactions in the input data and the distribution of exposure in recommendation results, analyzed at both the bundle and item levels. Fig.~\ref{fig:dist_youshu}–\ref{fig:dist_ifashion} visualize the distributions of bundles and items in the user interaction data and the outputs of four BR methods across three datasets. These results reveal key differences in distributional patterns (i.e., degrees of popularity bias) in user interaction with bundles and items, as well as how such biases propagate through recommendation models into their output exposure.

In the leftmost plots of each figure, corresponding to the input interaction data, it becomes evident that user interactions with bundles and with individual items do not always follow the same distributional trend. These histograms show the number of bundles/items on the y-axis and the (log-transformed) interaction frequency on the x-axis. This transformation helps to visualize distributional differences more clearly across varying frequency scales.

On the Youshu and iFashion datasets (Figs.~\ref{fig:dist_youshu} and~\ref{fig:dist_ifashion}, respectively), bundle and item interactions follow similar overall shapes: a long-tail distribution for the Youshu dataset and an approximately normal distribution for the iFashion dataset. However, on the NetEase dataset (Fig.~\ref{fig:dist_netease}), this alignment does not hold. Bundle interactions show a long-tail distribution, whereas item interactions are closer to a normal distribution. This discrepancy aligns with NetEase's low \textit{c-score} (shown in Table~\ref{tab:data}), suggesting that user preferences over bundles and individual items are less consistent in this dataset.

\begin{figure}[t!]
    \centering
    \begin{subfigure}[b]{\textwidth}
        \includegraphics[width=0.16\textwidth]{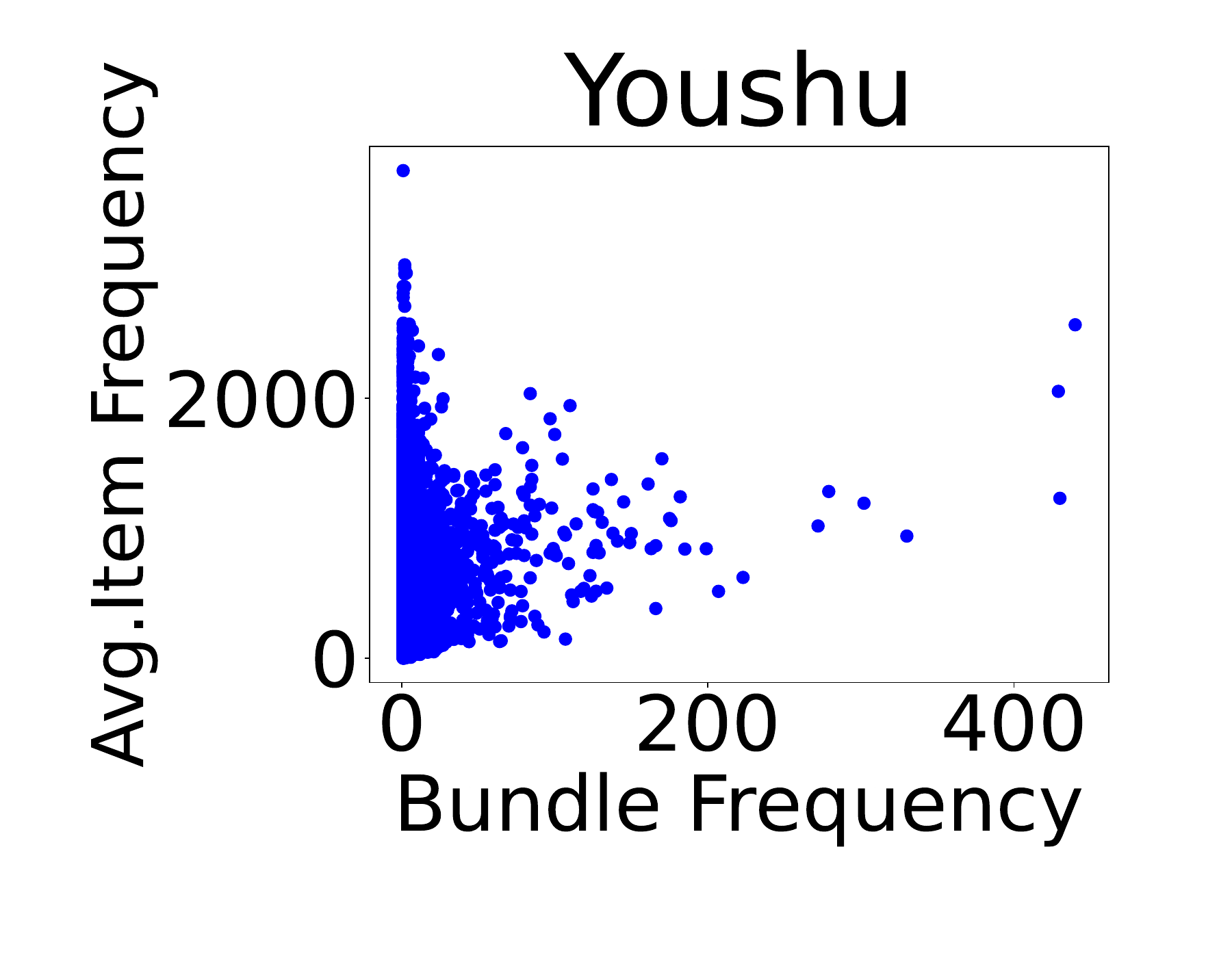}
        \includegraphics[width=0.16\textwidth]{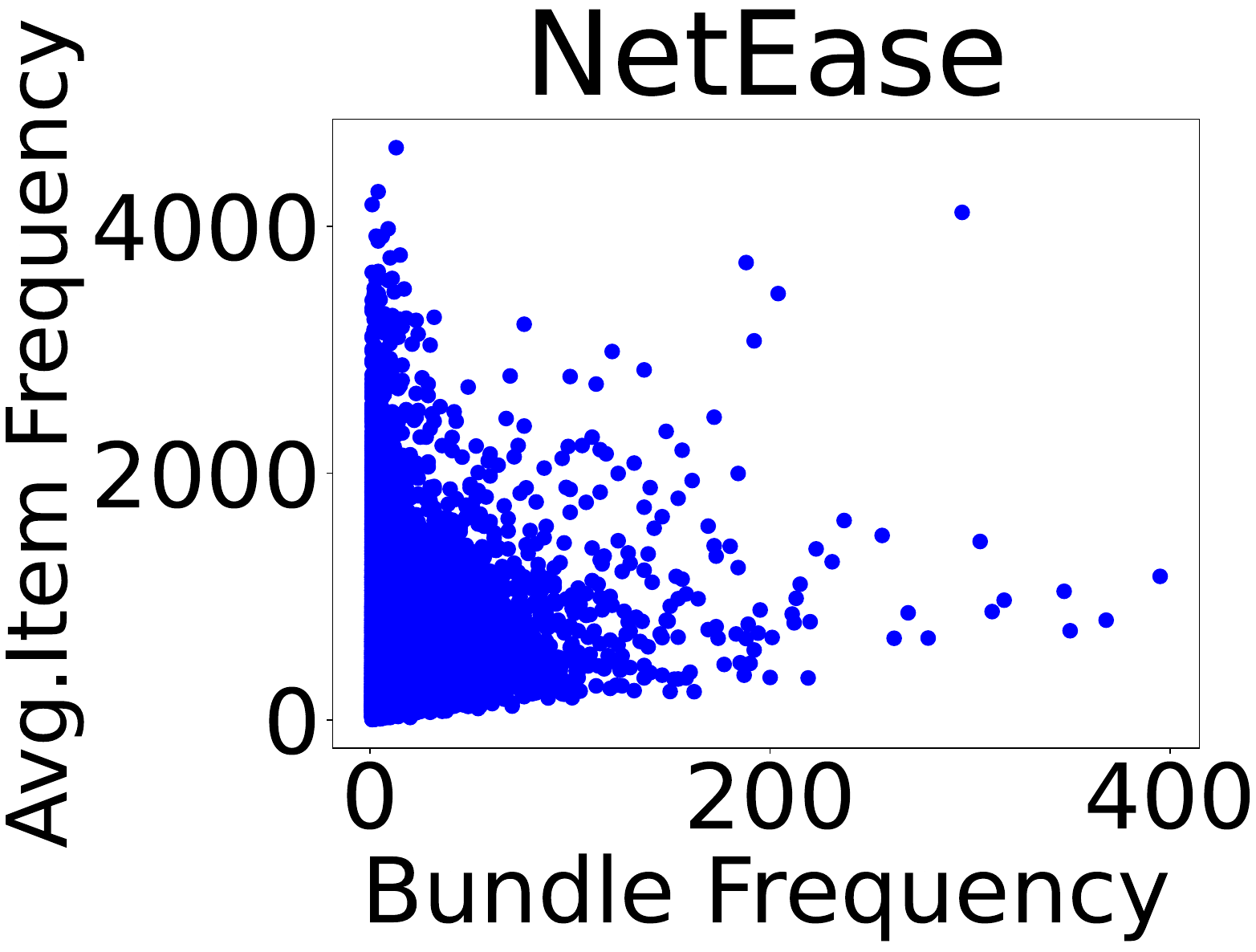}
        \includegraphics[width=0.16\textwidth]{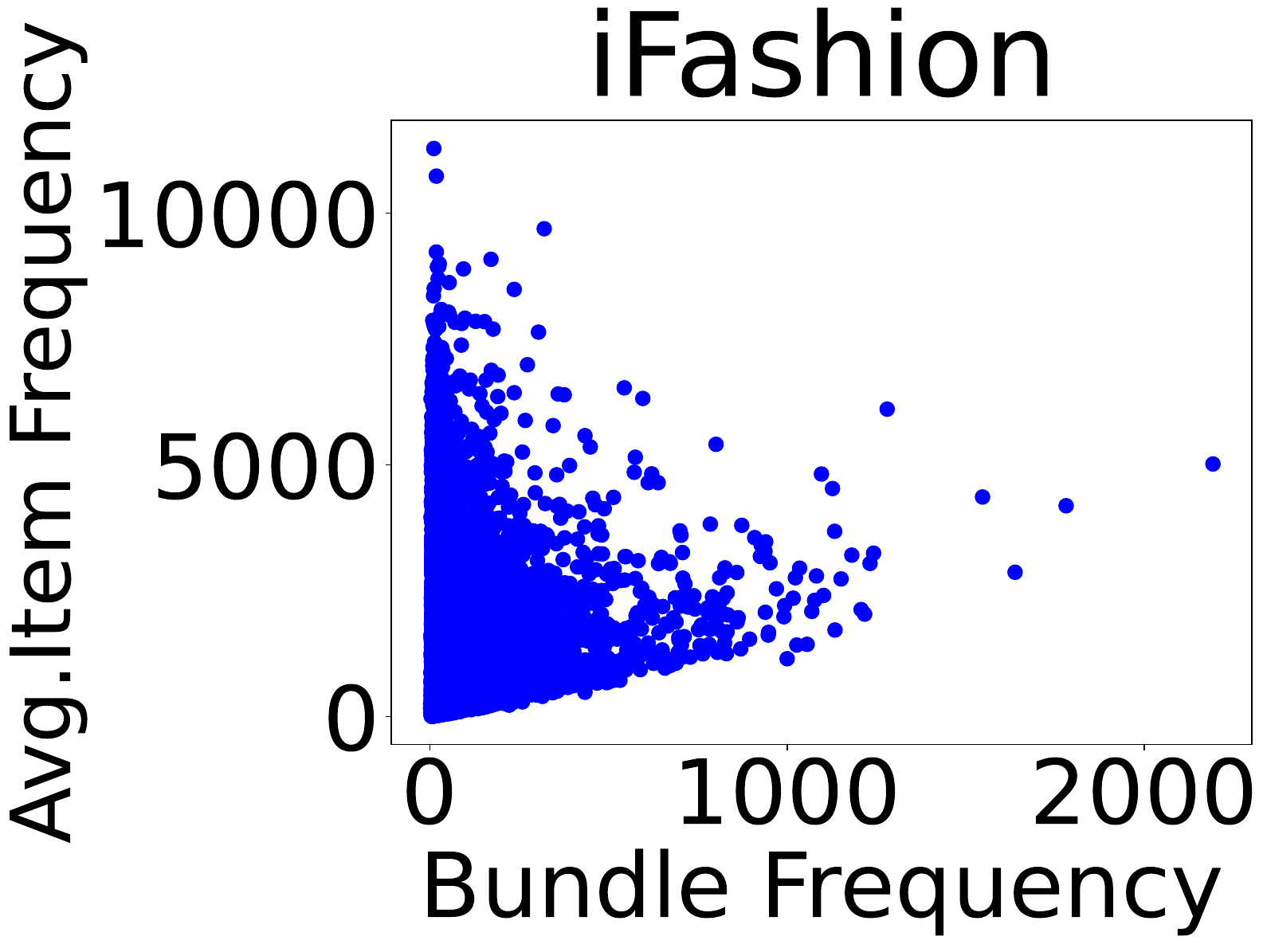}
    \end{subfigure}
\caption{The relationship between each bundle popularity and average popularity of its items.}\label{fig:bundle_avg_item}
\Description{Relation between bundle popularity and its item's average popularity.}
\vspace*{-4mm}
\end{figure}
\begin{figure}[t!]
    \centering
    \begin{subfigure}[b]{.7\linewidth}
        \includegraphics[width=\textwidth]{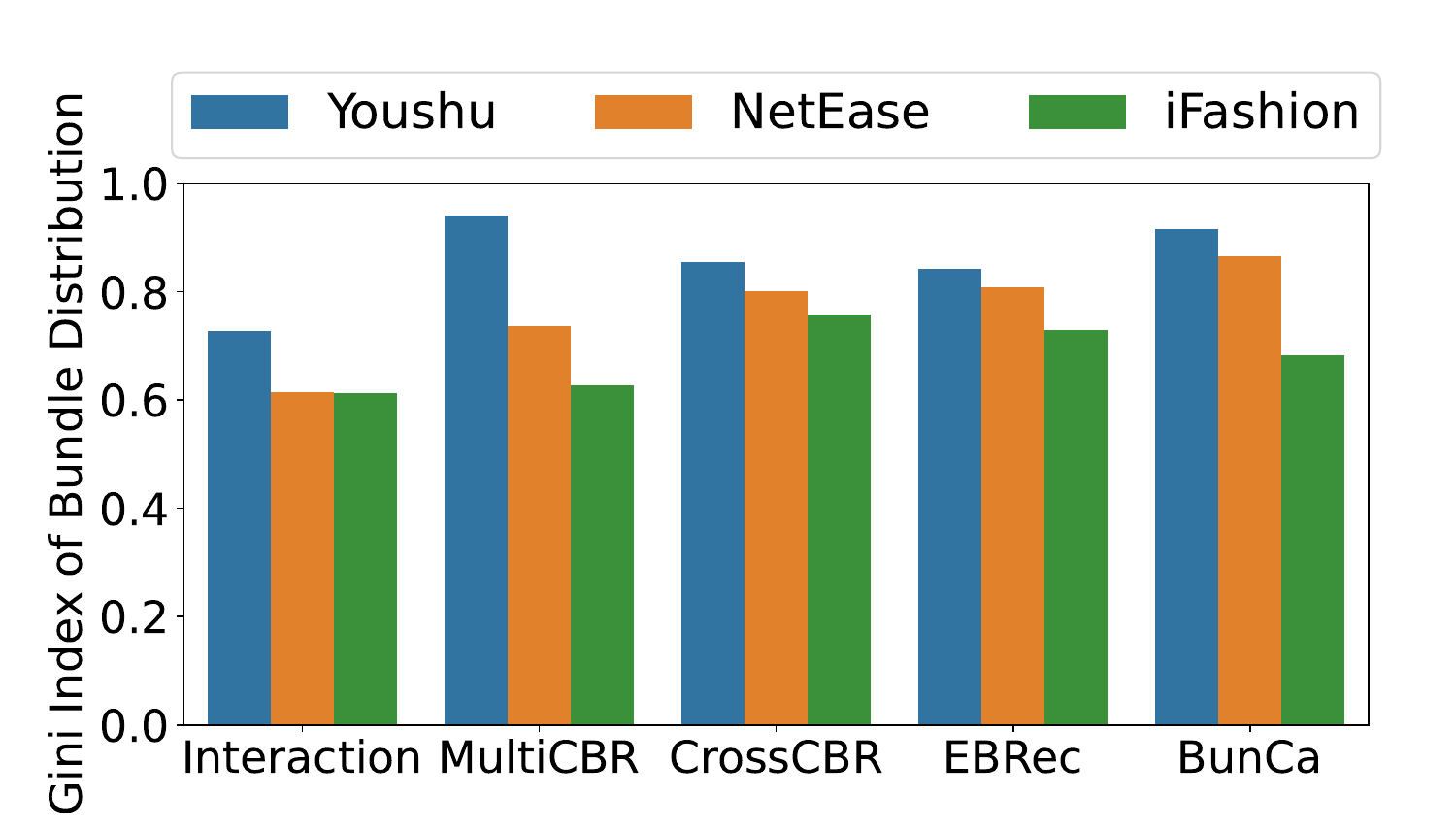}
        \vspace*{-7mm}
        \caption{Uniformity of bundle distribution}
        \Description{Bundle distribution uniformity.}
    \end{subfigure}
    \begin{subfigure}[b]{.7\linewidth}
        \includegraphics[width=\textwidth]{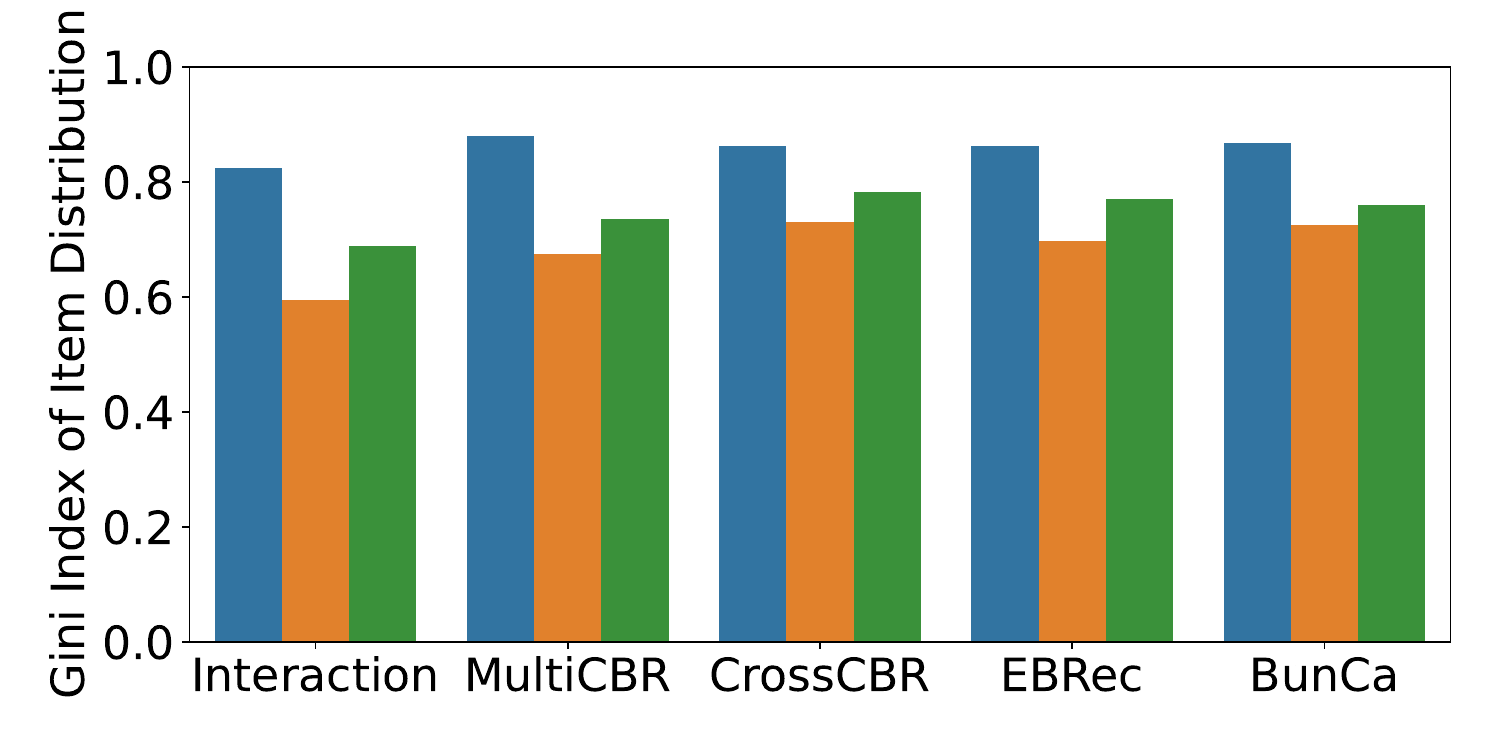}
        \vspace*{-5mm}
        \caption{Uniformity of item distribution}
        \Description{Item distribution uniformity.}
    \end{subfigure}
\caption{Uniformity of (a) bundle and (b) item distribution measured by Gini Index in user interaction data and the recommendation results obtained from four BR methods.}\label{fig:gini}
\Description{Bundle and Item distribution uniformity.}
\vspace*{-4mm}
\end{figure}

To better understand this discrepancy, Fig.~\ref{fig:bundle_avg_item} presents the relationship between each bundle’s interaction frequency and the average interaction frequency of the items it contains. In these plots, the x-axis represents bundle frequency (computed from matrix $X$ in Eq.~\ref{eq:y_prime}), and the y-axis represents the average frequency of items within each bundle (computed from matrix $Y^\prime$ in Eq.~\ref{eq:y_prime}). This analysis confirms that user behavior toward bundles and items can differ significantly. For instance, a frequently interacted bundle may include items that are themselves infrequently interacted with, and vice versa--a highly popular item may appear in bundles that are rarely engaged with. This suggests that bundle-level interaction patterns do not necessarily follow item-level interaction patterns.

Beyond the input data, we also observe how BR methods influence exposure at both levels. Although BR systems recommend only bundles, we infer item-level exposure by distributing each bundle’s exposure score equally among its constituent items as described in Section~\ref{sss:fair_metric}. For example, if a bundle receives an exposure score of $0.9$ and contains $10$ items, each item is assigned $0.09$ exposure. 

The results show method- and dataset-specific patterns. On the Youshu dataset, item and bundle exposure distributions across BR methods closely resemble those in the interaction data. On the NetEase dataset, bundle distributions remain consistent with interaction data, but item-level exposure varies across methods. MultiCBR and EBRec produce item distributions resembling the normal distribution seen in the input data, while CrossCBR and BunCa exhibit long-tail patterns. On the iFashion dataset, MultiCBR and BunCa match the normal distribution observed in the input data, whereas CrossCBR and EBRec tend toward long-tail exposure.


\begin{table*}[ht]
\centering
\small
\setlength{\tabcolsep}{1.2mm}
\caption{Overall performance of bundle recommendation methods in terms of utility and fairness metrics. Bold values indicate the best performance for each metric, while underlined values represent the second-best. Arrows denote metric direction: $\uparrow$ means higher is better, $\downarrow$ means lower is better, and $\circ$ indicates optimal fairness when values are closer to 0.}\label{tab:results}
\begin{tabular}{ll cc cccccc cccccccc}
\toprule
\multirow{2}{*}{Dataset} & \multirow{2}{*}{Method} & \multicolumn{2}{c}{Accuracy} && \multicolumn{6}{c}{Bundle-level Fairness} && \multicolumn{6}{c}{Item-level Fairness} \\ \cmidrule{3-4}\cmidrule{6-11}\cmidrule{13-18}

 && $R@20$$\uparrow$ & $N@20$$\uparrow$ && logEUR$\circ$ & logRUR$\circ$ & EEL$\downarrow$ & EER$\uparrow$ & EED$\downarrow$ & logDP$\circ$ && logEUR$\circ$ & logRUR$\circ$  & EEL$\downarrow$ & EER$\uparrow$ & EED$\downarrow$ & logDP$\circ$ \\
\midrule

\multirow{4}{*}{Youshu} & MultiCBR & \underline{0.286} & \underline{0.168} && 2.667 & 2.249 & 0.476 & 0.486 & 0.660 & 6.024 && 0.565 & \underline{0.083}  & \underline{0.049} & 0.826 & \textbf{0.564} & 4.152 \\
& CrossCBR & 0.276 & 0.166 && \textbf{2.094} & \textbf{2.101}   & \textbf{0.289} & \textbf{0.569} & \textbf{0.556} & \textbf{5.444} && \textbf{0.437} & \textbf{0.079}   & \underline{0.049} & \textbf{0.848} & 0.585 & \textbf{4.025} \\
& EBRec & 0.281 & \underline{0.168} && \underline{2.153} & 2.187 & \underline{0.308} & \underline{0.560} & \underline{0.565} & \underline{5.504} && \underline{0.486} & 0.093   & \textbf{0.048} & \underline{0.840} & 0.577 & \underline{4.074} \\
& BunCa & \textbf{0.295} & \textbf{0.172} && 2.390 & \underline{2.131}  & 0.385 & 0.524 & 0.606 & 5.743 && 0.523 & 0.084  & \textbf{0.048} & 0.834 & \underline{0.571} & 4.111 \\

\midrule

\multirow{4}{*}{Netease} & MultiCBR & \textbf{0.090} & \textbf{0.050} && \underline{2.173} & \underline{0.690}  & \underline{0.387} & \underline{0.560} & \underline{0.589} & \underline{4.744} && 0.458 & \textbf{0.001}  & 0.035 & 0.958 & \underline{0.614} & 2.835 \\
& CrossCBR & 0.081 & 0.043 && \textbf{1.908} & \textbf{0.644}  & \textbf{0.288} & \textbf{0.613} & \textbf{0.544} & \textbf{4.464} && 0.505 & \underline{0.002}  & 0.035 & 0.949 & \textbf{0.605} & 2.872 \\
& EBRec & 0.088 & \underline{0.048} && 2.241 & 0.693  & 0.413 & 0.547 & 0.603 & 4.817 && \textbf{0.443} & \textbf{0.001}  & 0.035 & \textbf{0.961} & 0.617 & \textbf{2.811} \\
& BunCa & \underline{0.089} & \underline{0.048} && 2.515 & 0.755  & 0.519 & 0.500 & 0.661 & 5.110 && \underline{0.452} & \textbf{0.001}  & 0.035 & \underline{0.959} & 0.615 & \underline{2.820} \\

\midrule

\multirow{4}{*}{iFashion} & MultiCBR & \textbf{0.150} & \textbf{0.120} && 1.464 & 0.604  & 0.335 & 0.825 & 0.529 & 4.525 && 0.547 & 0.076  & \underline{0.039} & 1.087 & \underline{0.529} & \underline{3.868} \\
& CrossCBR & 0.115 & 0.088 && \underline{1.383} & \textbf{0.458}  & \underline{0.293} & \underline{0.853} & \underline{0.518} & \underline{4.418} && \textbf{0.350} & \textbf{0.041}  & \textbf{0.014} & \textbf{1.146} & 0.563 & \textbf{3.624} \\
& EBRec & 0.133 & 0.105 && \textbf{1.314} & \underline{0.472}  & \textbf{0.259} & \textbf{0.882} & \textbf{0.510} & \textbf{4.328} && \underline{0.356} & \underline{0.043}  & \textbf{0.014} & \underline{1.144} & 0.561 & \textbf{3.632} \\
& BunCa & \underline{0.143} & \underline{0.112} && 1.578 & 0.594  & 0.399 & 0.781 & 0.550 & 4.680 && 0.601 & 0.076  & 0.048 & 1.070 & \textbf{0.522} & 3.935 \\
\bottomrule
\end{tabular}
\label{tab:sample10x5}
\vspace*{-2mm}
\end{table*}

Fig.~\ref{fig:gini} quantifies distributional uniformity using Gini Index~\cite{vargas2014improving}, computed for both bundles and items. A lower Gini Index indicates a more uniform distribution (fairer representation). While NetEase and iFashion datasets exhibit similar levels of uniformity in the input interaction data, the uniformity in recommendation results varies across BR methods. Two trends emerge: (1)~BR methods typically decrease the uniformity of exposure compared to the input data—indicating an amplification of popularity bias, and (2)~this reduction in uniformity is consistent with the original bias level in the input data: datasets with less uniform interaction distributions, such as the Youshu dataset, tend to result in less uniform exposure distributions in both bundles and items. These findings align with prior work~\cite{liu2023mitigating,abdollahpouri2020multi} and highlight the critical role of interaction bias in shaping fairness outcomes in BR systems.

\subsection{Fairness Evaluation (RQ2)}

To address RQ2, the fairness of BR methods is evaluated by fairness metrics described in Section~\ref{sss:fair_metric}. 
Table~\ref{tab:results} presents the results.

In terms of accuracy, MultiCBR and BunCa consistently perform well across datasets. However, when assessing fairness at the bundle level, CrossCBR (on the  Youshu and NetEase dataset) and EBRec tend to outperform the other methods, reflecting a trade-off between fairness and accuracy similar to findings in previous studies~\cite{mehrotra2018towards,ge2022toward}. At the item level, no single method consistently performs best, and the fairness metrics often disagree on which BR method is superior.

Overall, these results suggest that fairness in bundle recommendation is multi-faceted. Achieving fairness at the bundle level does not necessarily translate to fairness at the item level. This highlights the need for fairness-aware approaches that explicitly consider exposure disparities at both the bundle and item levels.

\subsection{Fairness Analysis on User Tendency (RQ3)}
\label{sec:rq3}

To understand how different user behaviors affect fairness outcomes, we group users based on their interaction tendencies: group $g_1$ tends to purchase bundles rather than individual items; group $g_2$ combines neutral users who treat bundles and items equally; group $g_3$ prefers to buy single items. We quantify this behavior using a \textit{tendency score} that measures the disparity ratio between bundle-based and item-based interaction for each user:
\begin{equation}
r_u = \frac{\sum_{b \in \mathcal{B}} X_{ub}}{\sum_{i \in \mathcal{I}} Y_{ui}}
\end{equation}
where $r_u$ represents the proportion of bundle-based interactions to item-based interactions for each user $u$.
A higher value of $r_u$ depicts a stronger tendency of user $u$ toward bundle-oriented purchases.\footnote{In our experiment, users with $r_u > 1.1$ are bundle-oriented ($g_1$), those with $r_u < 0.9$ are item-oriented ($g_3$), and users with $r_u$ in between are considered neutral ($g_2$). These thresholds were selected based on the observed distribution of $r_u$, though alternative values may also be appropriate depending on dataset characteristics.}



\begin{figure*}[!t]
    \centering
    \vspace{-2mm}
    \begin{subfigure}{0.85\textwidth}
        \includegraphics[width=\linewidth]{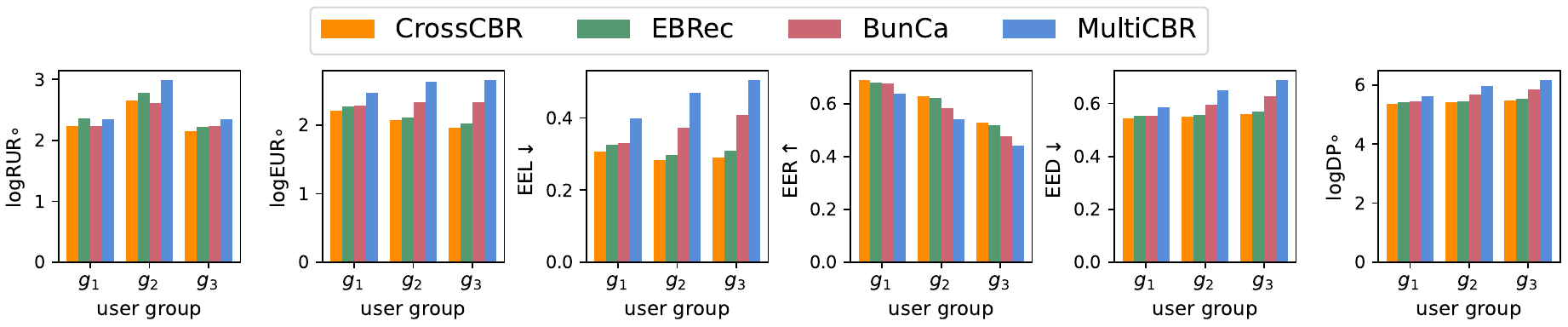}
       \vspace*{-6mm} \caption{Bundle-level fairness}
    \end{subfigure}
    
     \begin{subfigure}{0.85\textwidth}
         \includegraphics[width=\linewidth]{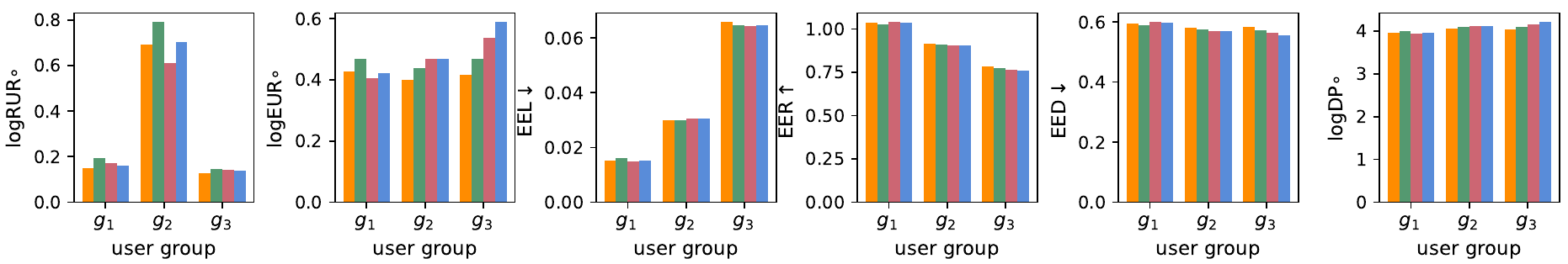}
         \vspace*{-6mm}         \caption{Item-level fairness}
     \end{subfigure}
    \caption{Product fairness evaluation of BR methods for different user groups on Youshu dataset.}
    \Description{Fairness evaluation on Youshu.}
    \label{fig:fairness_group_youshu}
\end{figure*}
\begin{figure*}[!t]
    \centering
    \vspace{-2mm}
    \begin{subfigure}{0.85\textwidth}
        \includegraphics[width=\linewidth]{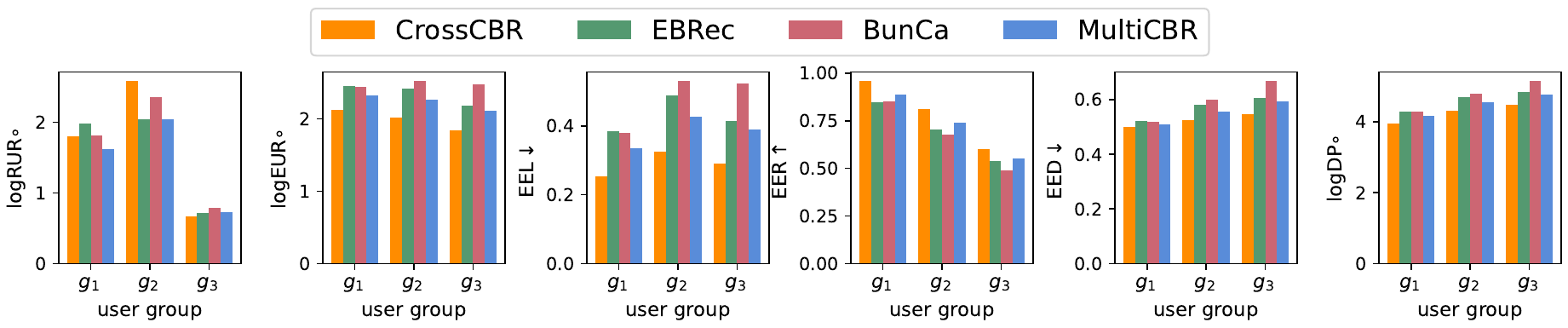}
       \vspace*{-6mm} \caption{Bundle-level fairness}
    \end{subfigure}
    
     \begin{subfigure}{0.85\textwidth}
         \includegraphics[width=\linewidth]{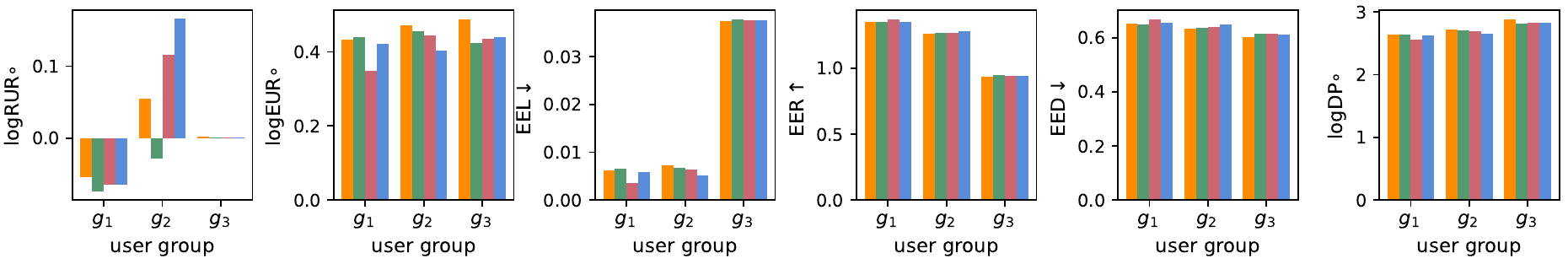}
         \vspace*{-6mm}         \caption{Item-level fairness}
     \end{subfigure}
    \caption{Product fairness evaluation of BR methods for different user groups on NetEase dataset.}
    \Description{{Fairness evaluation on NetEase.}}
    \label{fig:fairness_group_netease}
\end{figure*}
\begin{figure*}[!t]
    \centering
    \vspace{-2mm}
    \begin{subfigure}{0.85\textwidth}
        \includegraphics[width=\linewidth]{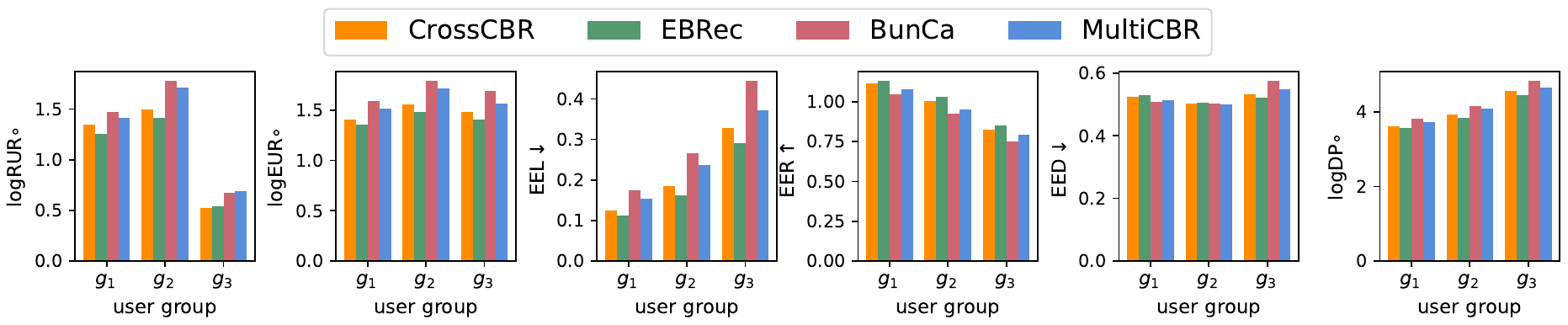}
       \vspace*{-6mm} \caption{Bundle-level fairness}
    \end{subfigure}
    
     \begin{subfigure}{0.85\textwidth}
         \includegraphics[width=\linewidth]{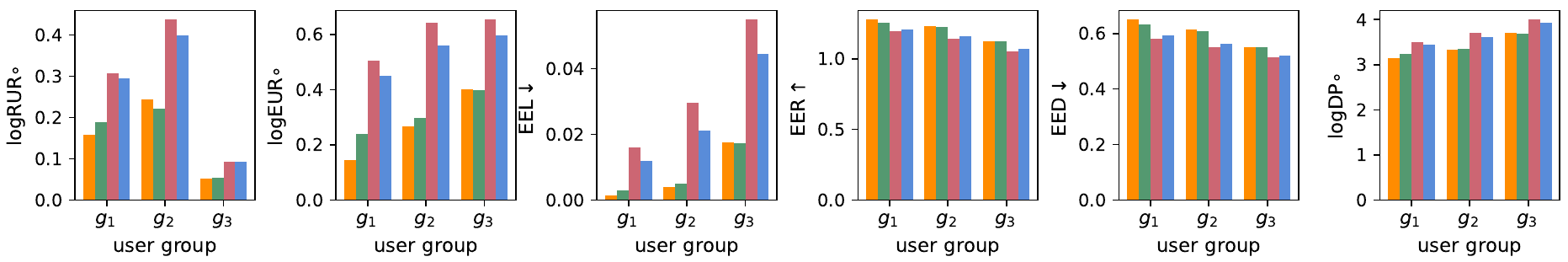}
         \vspace*{-6mm}         \caption{Item-level fairness}
     \end{subfigure}
    \caption{Product fairness evaluation of BR methods for different user groups on iFashion dataset.}
    \Description{Fairness evaluation on iFashion.}
    \label{fig:fairness_group_fashion}
    \vspace*{-4mm}
\end{figure*}

Fig.~\ref{fig:fairness_group_youshu}–\ref{fig:fairness_group_fashion} present the fairness outcomes of four BR methods across these user groups and datasets. Fairness varies meaningfully across user types, confirming that fairness is not solely a property of the recommendation algorithm or dataset, but also a function of user behavior or tendency in interacting with bundles or items. 

For the iFashion dataset, consistency between fairness performances across groups is achieved at both item-level and bundle-level, while it is more confusing on other datasets. This is reasonably expected by the disparity issues between learning item-level and bundle-level user preference~\cite{ren2023distillation}, the meaning of \textit{c-score}~\cite{ren2023distillation}, and the influence of high-frequency items within bundles on BR models~\cite{nguyen2024bundle} demonstrated in related work.
The NetEase dataset exhibits the most inconsistent patterns. The limited accuracy of models and data properties (low \textit{c-score}, large bundle-sizes, and high item-frequency variance) likely introduce noise in fairness evaluations.
Moreover, bundles in the iFashion dataset are outfits, i.e., constructed more elaborately, based on the providers' strategy. Meanwhile, bundles in the Youshu and NetEase datasets are simply defined by grouping items based user sessions.

BR methods generally achieve fairer exposure distributions when serving users who primarily interact with bundles, especially according to the EEL and EER metrics. This may be due to the relatively low overlap among bundles, which promotes greater exposure diversity. EED and logDP have opposite item-level fairness rankings of the three user groups on all datasets. This means that recommendations for $g_3$ have the closest overall exposure between popular and unpopular item groups, while recommendations for $g_1$ have the closest average exposure between two item groups.

\vspace{-2mm}
\section{Conclusion}

We conducted the first in-depth study of product-side fairness in the bundle recommendation task. Unlike conventional recommendation settings where individual items are recommended, BR systems deliver bundles—each consisting of multiple items. While exposure is explicitly given to bundles, items within them also receive indirect exposure. This dual-layer exposure requires fairness assessments at both the bundle-level and item-level aspects.

We found that the distribution of exposure in historical interaction data and in the outputs of BR methods differs notably between bundles and items. Hence, fairness interventions cannot rely solely on bundle-level assumptions and must consider item-specific dynamics. Moreover, consistent with prior research, we observed that different fairness metrics often disagree in their assessments, reinforcing the need for multi-perspective evaluation. Finally, we showed that user behavior plays a key role: when users interact more with bundles than individual items, BR methods tend to yield fairer exposure distributions at both the bundle and item levels.

These findings highlight the complex nature of fairness in bundle recommendation and the need for more targeted strategies that address its unique challenges. To the best of our knowledge, this is the first comprehensive exploration of fairness in the BR scenario.

In future work, we plan to expand our analysis beyond the popularity of bundles in recommender systems. In this study, we defined fairness subgroups based on popularity. However, other grouping strategies—such as item categories, supplier identities, or sensitive user and item attributes—could reveal additional fairness issues. Investigating these dimensions should lead to the development of more holistic and inclusive fairness-aware BR methods.

\begin{acks}
This research was (partially) supported by Ahold Delhaize, through AIRLab Amsterdam, the Dutch Research Council (NWO), under project numbers 024.004.022, NWA.1389.20.\-183, and KICH3.\-LTP.\-20.006, and the European Union's Horizon Europe program under grant agreement No 101070212, and the China Scholarship Council under grant nrs. 202206290080. 
All content represents the opinion of the authors, which is not necessarily shared or endorsed by their respective employers and/or sponsors.
\end{acks}

\bibliographystyle{ACM-Reference-Format}
\balance
\bibliography{ref}


\begin{thebibliography}{65}


\ifx \showCODEN    \undefined \def \showCODEN     #1{\unskip}     \fi
\ifx \showISBNx    \undefined \def \showISBNx     #1{\unskip}     \fi
\ifx \showISBNxiii \undefined \def \showISBNxiii  #1{\unskip}     \fi
\ifx \showISSN     \undefined \def \showISSN      #1{\unskip}     \fi
\ifx \showLCCN     \undefined \def \showLCCN      #1{\unskip}     \fi
\ifx \shownote     \undefined \def \shownote      #1{#1}          \fi
\ifx \showarticletitle \undefined \def \showarticletitle #1{#1}   \fi
\ifx \showURL      \undefined \def \showURL       {\relax}        \fi
\providecommand\bibfield[2]{#2}
\providecommand\bibinfo[2]{#2}
\providecommand\natexlab[1]{#1}
\providecommand\showeprint[2][]{arXiv:#2}

\bibitem[Abdollahpouri et~al\mbox{.}(2020)]%
        {abdollahpouri2020multistakeholder}
\bibfield{author}{\bibinfo{person}{Himan Abdollahpouri}, \bibinfo{person}{Gediminas Adomavicius}, \bibinfo{person}{Robin Burke}, \bibinfo{person}{Ido Guy}, \bibinfo{person}{Dietmar Jannach}, \bibinfo{person}{Toshihiro Kamishima}, \bibinfo{person}{Jan Krasnodebski}, {and} \bibinfo{person}{Luiz Pizzato}.} \bibinfo{year}{2020}\natexlab{}.
\newblock \showarticletitle{Multistakeholder Recommendation: Survey and Research Directions}.
\newblock \bibinfo{journal}{\emph{User Modeling and User-Adapted Interaction}}  \bibinfo{volume}{30} (\bibinfo{year}{2020}), \bibinfo{pages}{127--158}.
\newblock


\bibitem[Abdollahpouri and Mansoury(2020)]%
        {abdollahpouri2020multi}
\bibfield{author}{\bibinfo{person}{Himan Abdollahpouri} {and} \bibinfo{person}{Masoud Mansoury}.} \bibinfo{year}{2020}\natexlab{}.
\newblock \showarticletitle{Multi-sided Exposure Bias in Recommendation}.
\newblock \bibinfo{journal}{\emph{ACM KDD Workshop on Industrial Recommendation Systems}} (\bibinfo{year}{2020}).
\newblock


\bibitem[Beutel et~al\mbox{.}(2019)]%
        {beutel2019fairness}
\bibfield{author}{\bibinfo{person}{Alex Beutel}, \bibinfo{person}{Jilin Chen}, \bibinfo{person}{Tulsee Doshi}, \bibinfo{person}{Hai Qian}, \bibinfo{person}{Li Wei}, \bibinfo{person}{Yi Wu}, \bibinfo{person}{Lukasz Heldt}, \bibinfo{person}{Zhe Zhao}, \bibinfo{person}{Lichan Hong}, \bibinfo{person}{Ed~H Chi}, {et~al\mbox{.}}} \bibinfo{year}{2019}\natexlab{}.
\newblock \showarticletitle{Fairness in Recommendation Ranking through Pairwise Comparisons}. In \bibinfo{booktitle}{\emph{Proceedings of the 25th ACM SIGKDD International Conference on Knowledge Discovery \& Data Mining}}. \bibinfo{pages}{2212--2220}.
\newblock


\bibitem[Biega et~al\mbox{.}(2018)]%
        {biega2018equity}
\bibfield{author}{\bibinfo{person}{Asia~J Biega}, \bibinfo{person}{Krishna~P Gummadi}, {and} \bibinfo{person}{Gerhard Weikum}.} \bibinfo{year}{2018}\natexlab{}.
\newblock \showarticletitle{Equity of Attention: Amortizing Individual Fairness in Rankings}. In \bibinfo{booktitle}{\emph{The 41st international ACM SIGIR Conference on Research \& Development in Information Retrieval}}. \bibinfo{pages}{405--414}.
\newblock


\bibitem[Bui et~al\mbox{.}(2025)]%
        {bui2025personalized}
\bibfield{author}{\bibinfo{person}{Tuan-Nghia Bui}, \bibinfo{person}{Huy-Son Nguyen}, \bibinfo{person}{Cam-Van~Thi Nguyen}, \bibinfo{person}{Hoang-Quynh Le}, {and} \bibinfo{person}{Duc-Trong Le}.} \bibinfo{year}{2025}\natexlab{}.
\newblock \showarticletitle{Personalized Diffusion Model Reshapes Cold-Start Bundle Recommendation}. In \bibinfo{booktitle}{\emph{Companion Proceedings of the ACM on Web Conference 2025}}. \bibinfo{pages}{3088--3091}.
\newblock


\bibitem[Bui et~al\mbox{.}(2024)]%
        {bui2024bridge}
\bibfield{author}{\bibinfo{person}{Tuan-Nghia Bui}, \bibinfo{person}{Huy-Son Nguyen}, \bibinfo{person}{Cam-Van~Nguyen Thi}, \bibinfo{person}{Hoang-Quynh Le}, {and} \bibinfo{person}{Duc-Trong Le}.} \bibinfo{year}{2024}\natexlab{}.
\newblock \showarticletitle{BRIDGE: Bundle Recommendation via Instruction-Driven Generation}.
\newblock \bibinfo{journal}{\emph{arXiv preprint arXiv:2412.18092}} (\bibinfo{year}{2024}).
\newblock


\bibitem[Chang et~al\mbox{.}(2020)]%
        {chang2020bundle}
\bibfield{author}{\bibinfo{person}{Jianxin Chang}, \bibinfo{person}{Chen Gao}, \bibinfo{person}{Xiangnan He}, \bibinfo{person}{Depeng Jin}, {and} \bibinfo{person}{Yong Li}.} \bibinfo{year}{2020}\natexlab{}.
\newblock \showarticletitle{Bundle Recommendation with Graph Convolutional Networks}. In \bibinfo{booktitle}{\emph{Proceedings of the 43rd international ACM SIGIR Conference on Research and Development in Information Retrieval}}. \bibinfo{pages}{1673--1676}.
\newblock


\bibitem[Chang et~al\mbox{.}(2021)]%
        {chang2021bundle}
\bibfield{author}{\bibinfo{person}{Jianxin Chang}, \bibinfo{person}{Chen Gao}, \bibinfo{person}{Xiangnan He}, \bibinfo{person}{Depeng Jin}, {and} \bibinfo{person}{Yong Li}.} \bibinfo{year}{2021}\natexlab{}.
\newblock \showarticletitle{Bundle recommendation and generation with graph neural networks}.
\newblock \bibinfo{journal}{\emph{IEEE Transactions on Knowledge and Data Engineering}} \bibinfo{volume}{35}, \bibinfo{number}{3} (\bibinfo{year}{2021}), \bibinfo{pages}{2326--2340}.
\newblock


\bibitem[Chen et~al\mbox{.}(2019b)]%
        {chen2019matching}
\bibfield{author}{\bibinfo{person}{Liang Chen}, \bibinfo{person}{Yang Liu}, \bibinfo{person}{Xiangnan He}, \bibinfo{person}{Lianli Gao}, {and} \bibinfo{person}{Zibin Zheng}.} \bibinfo{year}{2019}\natexlab{b}.
\newblock \showarticletitle{Matching User with Item Set: Collaborative Bundle Recommendation with Deep Attention Network}. In \bibinfo{booktitle}{\emph{IJCAI}}. \bibinfo{pages}{2095--2101}.
\newblock


\bibitem[Chen et~al\mbox{.}(2019a)]%
        {chen2019pog}
\bibfield{author}{\bibinfo{person}{Wen Chen}, \bibinfo{person}{Pipei Huang}, \bibinfo{person}{Jiaming Xu}, \bibinfo{person}{Xin Guo}, \bibinfo{person}{Cheng Guo}, \bibinfo{person}{Fei Sun}, \bibinfo{person}{Chao Li}, \bibinfo{person}{Andreas Pfadler}, \bibinfo{person}{Huan Zhao}, {and} \bibinfo{person}{Binqiang Zhao}.} \bibinfo{year}{2019}\natexlab{a}.
\newblock \showarticletitle{POG: Personalized Outfit Generation for Fashion Recommendation at Alibaba iFashion}. In \bibinfo{booktitle}{\emph{Proceedings of the 25th ACM SIGKDD International Conference on Knowledge Discovery \& Data Mining}}. \bibinfo{pages}{2662--2670}.
\newblock


\bibitem[Dash et~al\mbox{.}(2022)]%
        {dash2022fairir}
\bibfield{author}{\bibinfo{person}{Abhisek Dash}, \bibinfo{person}{Abhijnan Chakraborty}, \bibinfo{person}{Saptarshi Ghosh}, \bibinfo{person}{Animesh Mukherjee}, {and} \bibinfo{person}{Krishna~P Gummadi}.} \bibinfo{year}{2022}\natexlab{}.
\newblock \showarticletitle{FaiRIR: Mitigating Exposure Bias from Related Item Recommendations in Two-sided Platforms}.
\newblock \bibinfo{journal}{\emph{IEEE Transactions on Computational Social Systems}} \bibinfo{volume}{10}, \bibinfo{number}{3} (\bibinfo{year}{2022}), \bibinfo{pages}{1301--1313}.
\newblock


\bibitem[Deldjoo et~al\mbox{.}(2024)]%
        {deldjoo2024fairness}
\bibfield{author}{\bibinfo{person}{Yashar Deldjoo}, \bibinfo{person}{Dietmar Jannach}, \bibinfo{person}{Alejandro Bellogin}, \bibinfo{person}{Alessandro Difonzo}, {and} \bibinfo{person}{Dario Zanzonelli}.} \bibinfo{year}{2024}\natexlab{}.
\newblock \showarticletitle{Fairness in Recommender Systems: Research Landscape and Future Directions}.
\newblock \bibinfo{journal}{\emph{User Modeling and User-Adapted Interaction}} \bibinfo{volume}{34}, \bibinfo{number}{1} (\bibinfo{year}{2024}), \bibinfo{pages}{59--108}.
\newblock


\bibitem[Deng et~al\mbox{.}(2020)]%
        {deng2020personalized}
\bibfield{author}{\bibinfo{person}{Qilin Deng}, \bibinfo{person}{Kai Wang}, \bibinfo{person}{Minghao Zhao}, \bibinfo{person}{Zhene Zou}, \bibinfo{person}{Runze Wu}, \bibinfo{person}{Jianrong Tao}, \bibinfo{person}{Changjie Fan}, {and} \bibinfo{person}{Liang Chen}.} \bibinfo{year}{2020}\natexlab{}.
\newblock \showarticletitle{Personalized Bundle Recommendation in Online Games}. In \bibinfo{booktitle}{\emph{Proceedings of the 29th ACM International Conference on Information \& Knowledge Management}}. \bibinfo{pages}{2381--2388}.
\newblock


\bibitem[Diaz et~al\mbox{.}(2020)]%
        {diaz2020evaluating}
\bibfield{author}{\bibinfo{person}{Fernando Diaz}, \bibinfo{person}{Bhaskar Mitra}, \bibinfo{person}{Michael~D Ekstrand}, \bibinfo{person}{Asia~J Biega}, {and} \bibinfo{person}{Ben Carterette}.} \bibinfo{year}{2020}\natexlab{}.
\newblock \showarticletitle{Evaluating Stochastic Rankings with Expected Exposure}. In \bibinfo{booktitle}{\emph{Proceedings of the 29th ACM International Conference on Information \& Knowledge Management}}. \bibinfo{pages}{275--284}.
\newblock


\bibitem[Ding et~al\mbox{.}(2023)]%
        {ding2023computational}
\bibfield{author}{\bibinfo{person}{Yujuan Ding}, \bibinfo{person}{Zhihui Lai}, \bibinfo{person}{PY Mok}, {and} \bibinfo{person}{Tat-Seng Chua}.} \bibinfo{year}{2023}\natexlab{}.
\newblock \showarticletitle{Computational Technologies for Fashion Recommendation: A Survey}.
\newblock \bibinfo{journal}{\emph{Comput. Surveys}} \bibinfo{volume}{56}, \bibinfo{number}{5} (\bibinfo{year}{2023}), \bibinfo{pages}{1--45}.
\newblock


\bibitem[Du et~al\mbox{.}(2023)]%
        {du2023enhancing}
\bibfield{author}{\bibinfo{person}{Xiaoyu Du}, \bibinfo{person}{Kun Qian}, \bibinfo{person}{Yunshan Ma}, {and} \bibinfo{person}{Xinguang Xiang}.} \bibinfo{year}{2023}\natexlab{}.
\newblock \showarticletitle{Enhancing Item-level Bundle Representation for Bundle Recommendation}.
\newblock \bibinfo{journal}{\emph{ACM Transactions on Recommender Systems}} (\bibinfo{year}{2023}).
\newblock


\bibitem[Ekstrand et~al\mbox{.}(2022)]%
        {ekstrand2022fairness}
\bibfield{author}{\bibinfo{person}{Michael~D Ekstrand}, \bibinfo{person}{Anubrata Das}, \bibinfo{person}{Robin Burke}, \bibinfo{person}{Fernando Diaz}, {et~al\mbox{.}}} \bibinfo{year}{2022}\natexlab{}.
\newblock \showarticletitle{Fairness in Information Access Systems}.
\newblock \bibinfo{journal}{\emph{Foundations and Trends in Information Retrieval}} \bibinfo{volume}{16}, \bibinfo{number}{1-2} (\bibinfo{year}{2022}), \bibinfo{pages}{1--177}.
\newblock


\bibitem[Gao et~al\mbox{.}(2023)]%
        {gao2023survey}
\bibfield{author}{\bibinfo{person}{Chen Gao}, \bibinfo{person}{Yu Zheng}, \bibinfo{person}{Nian Li}, \bibinfo{person}{Yinfeng Li}, \bibinfo{person}{Yingrong Qin}, \bibinfo{person}{Jinghua Piao}, \bibinfo{person}{Yuhan Quan}, \bibinfo{person}{Jianxin Chang}, \bibinfo{person}{Depeng Jin}, \bibinfo{person}{Xiangnan He}, {et~al\mbox{.}}} \bibinfo{year}{2023}\natexlab{}.
\newblock \showarticletitle{A Survey of Graph Neural Networks for Recommender Systems: Challenges, Methods, and Directions}.
\newblock \bibinfo{journal}{\emph{ACM Transactions on Recommender Systems}} \bibinfo{volume}{1}, \bibinfo{number}{1} (\bibinfo{year}{2023}), \bibinfo{pages}{1--51}.
\newblock


\bibitem[Ge et~al\mbox{.}(2021)]%
        {ge2021towards}
\bibfield{author}{\bibinfo{person}{Yingqiang Ge}, \bibinfo{person}{Shuchang Liu}, \bibinfo{person}{Ruoyuan Gao}, \bibinfo{person}{Yikun Xian}, \bibinfo{person}{Yunqi Li}, \bibinfo{person}{Xiangyu Zhao}, \bibinfo{person}{Changhua Pei}, \bibinfo{person}{Fei Sun}, \bibinfo{person}{Junfeng Ge}, \bibinfo{person}{Wenwu Ou}, {et~al\mbox{.}}} \bibinfo{year}{2021}\natexlab{}.
\newblock \showarticletitle{Towards long-term fairness in recommendation}. In \bibinfo{booktitle}{\emph{Proceedings of the 14th ACM international conference on web search and data mining}}. \bibinfo{pages}{445--453}.
\newblock


\bibitem[Ge et~al\mbox{.}(2022)]%
        {ge2022toward}
\bibfield{author}{\bibinfo{person}{Yingqiang Ge}, \bibinfo{person}{Xiaoting Zhao}, \bibinfo{person}{Lucia Yu}, \bibinfo{person}{Saurabh Paul}, \bibinfo{person}{Diane Hu}, \bibinfo{person}{Chu-Cheng Hsieh}, {and} \bibinfo{person}{Yongfeng Zhang}.} \bibinfo{year}{2022}\natexlab{}.
\newblock \showarticletitle{Toward Pareto Efficient Fairness-utility Trade-off in Recommendation through Reinforcement Learning}. In \bibinfo{booktitle}{\emph{Proceedings of the fifteenth ACM international Conference on Web Search and Data Mining}}. \bibinfo{pages}{316--324}.
\newblock


\bibitem[Glorot and Bengio(2010)]%
        {glorot2010understanding}
\bibfield{author}{\bibinfo{person}{Xavier Glorot} {and} \bibinfo{person}{Yoshua Bengio}.} \bibinfo{year}{2010}\natexlab{}.
\newblock \showarticletitle{Understanding the Difficulty of Training Deep Feedforward Neural Networks}. In \bibinfo{booktitle}{\emph{Proceedings of the Thirteenth International Conference on Artificial Intelligence and Statistics}}. JMLR Workshop and Conference Proceedings, \bibinfo{pages}{249--256}.
\newblock


\bibitem[He et~al\mbox{.}(2020)]%
        {he2020lightgcn}
\bibfield{author}{\bibinfo{person}{Xiangnan He}, \bibinfo{person}{Kuan Deng}, \bibinfo{person}{Xiang Wang}, \bibinfo{person}{Yan Li}, \bibinfo{person}{Yongdong Zhang}, {and} \bibinfo{person}{Meng Wang}.} \bibinfo{year}{2020}\natexlab{}.
\newblock \showarticletitle{LightGCN: Simplifying and Powering Graph Convolution Network for Recommendation}. In \bibinfo{booktitle}{\emph{Proceedings of the 43rd International ACM SIGIR Conference on Research and Development in Information Retrieval}}. \bibinfo{pages}{639--648}.
\newblock


\bibitem[Islam et~al\mbox{.}(2021)]%
        {islam2021debiasing}
\bibfield{author}{\bibinfo{person}{Rashidul Islam}, \bibinfo{person}{Kamrun~Naher Keya}, \bibinfo{person}{Ziqian Zeng}, \bibinfo{person}{Shimei Pan}, {and} \bibinfo{person}{James Foulds}.} \bibinfo{year}{2021}\natexlab{}.
\newblock \showarticletitle{Debiasing Career Recommendations with Neural Fair Collaborative Filtering}. In \bibinfo{booktitle}{\emph{Proceedings of the Web Conference 2021}}. \bibinfo{pages}{3779--3790}.
\newblock


\bibitem[Jin et~al\mbox{.}(2023)]%
        {jin2023survey}
\bibfield{author}{\bibinfo{person}{Di Jin}, \bibinfo{person}{Luzhi Wang}, \bibinfo{person}{He Zhang}, \bibinfo{person}{Yizhen Zheng}, \bibinfo{person}{Weiping Ding}, \bibinfo{person}{Feng Xia}, {and} \bibinfo{person}{Shirui Pan}.} \bibinfo{year}{2023}\natexlab{}.
\newblock \showarticletitle{A Survey on Fairness-aware Recommender Systems}.
\newblock \bibinfo{journal}{\emph{Information Fusion}}  \bibinfo{volume}{100} (\bibinfo{year}{2023}), \bibinfo{pages}{101906}.
\newblock


\bibitem[Kim et~al\mbox{.}(2024)]%
        {kim2024towards}
\bibfield{author}{\bibinfo{person}{Kyungho Kim}, \bibinfo{person}{Sunwoo Kim}, \bibinfo{person}{Geon Lee}, {and} \bibinfo{person}{Kijung Shin}.} \bibinfo{year}{2024}\natexlab{}.
\newblock \showarticletitle{Towards Better Utilization of Multiple Views for Bundle Recommendation}. In \bibinfo{booktitle}{\emph{Proceedings of the 33rd ACM International Conference on Information and Knowledge Management}}. \bibinfo{pages}{3827--3831}.
\newblock


\bibitem[Kingma and Ba(2015)]%
        {KingBa15}
\bibfield{author}{\bibinfo{person}{Diederik Kingma} {and} \bibinfo{person}{Jimmy Ba}.} \bibinfo{year}{2015}\natexlab{}.
\newblock \showarticletitle{Adam: A Method for Stochastic Optimization}. In \bibinfo{booktitle}{\emph{International Conference on Learning Representations (ICLR)}}. \bibinfo{address}{San Diego, CA, USA}.
\newblock


\bibitem[Kipf and Welling(2017)]%
        {KipfW17}
\bibfield{author}{\bibinfo{person}{Thomas~N. Kipf} {and} \bibinfo{person}{Max Welling}.} \bibinfo{year}{2017}\natexlab{}.
\newblock \showarticletitle{Semi-Supervised Classification with Graph Convolutional Networks}. In \bibinfo{booktitle}{\emph{5th International Conference on Learning Representations, 2017, Conference Track Proceedings}}.
\newblock


\bibitem[Li et~al\mbox{.}(2023b)]%
        {li2023next}
\bibfield{author}{\bibinfo{person}{Ming Li}, \bibinfo{person}{Sami Jullien}, \bibinfo{person}{Mozhdeh Ariannezhad}, {and} \bibinfo{person}{Maarten de Rijke}.} \bibinfo{year}{2023}\natexlab{b}.
\newblock \showarticletitle{A Next Basket Recommendation Reality Check}.
\newblock \bibinfo{journal}{\emph{ACM Transactions on Information Systems}} \bibinfo{volume}{41}, \bibinfo{number}{4} (\bibinfo{year}{2023}), \bibinfo{pages}{1--29}.
\newblock


\bibitem[Li et~al\mbox{.}(2024a)]%
        {li2024mealrec+}
\bibfield{author}{\bibinfo{person}{Ming Li}, \bibinfo{person}{Lin Li}, \bibinfo{person}{Xiaohui Tao}, {and} \bibinfo{person}{Jimmy~Xiangji Huang}.} \bibinfo{year}{2024}\natexlab{a}.
\newblock \showarticletitle{MealRec+: A Meal Recommendation Dataset with Meal-Course Affiliation for Personalization and Healthiness}. In \bibinfo{booktitle}{\emph{Proceedings of the 47th International ACM SIGIR Conference on Research and Development in Information Retrieval}}. \bibinfo{pages}{564--574}.
\newblock


\bibitem[Li et~al\mbox{.}(2024b)]%
        {li2024we}
\bibfield{author}{\bibinfo{person}{Ming Li}, \bibinfo{person}{Yuanna Liu}, \bibinfo{person}{Sami Jullien}, \bibinfo{person}{Mozhdeh Ariannezhad}, \bibinfo{person}{Andrew Yates}, \bibinfo{person}{Mohammad Aliannejadi}, {and} \bibinfo{person}{Maarten de Rijke}.} \bibinfo{year}{2024}\natexlab{b}.
\newblock \showarticletitle{Are We Really Achieving Better Beyond-Accuracy Performance in Next Basket Recommendation?}. In \bibinfo{booktitle}{\emph{Proceedings of the 47th International ACM SIGIR Conference on Research and Development in Information Retrieval}}. \bibinfo{pages}{924--934}.
\newblock


\bibitem[Li et~al\mbox{.}(2023a)]%
        {li2023fairness}
\bibfield{author}{\bibinfo{person}{Yunqi Li}, \bibinfo{person}{Hanxiong Chen}, \bibinfo{person}{Shuyuan Xu}, \bibinfo{person}{Yingqiang Ge}, \bibinfo{person}{Juntao Tan}, \bibinfo{person}{Shuchang Liu}, {and} \bibinfo{person}{Yongfeng Zhang}.} \bibinfo{year}{2023}\natexlab{a}.
\newblock \showarticletitle{Fairness in Recommendation: Foundations, Methods, and Applications}.
\newblock \bibinfo{journal}{\emph{ACM Transactions on Intelligent Systems and Technology}} \bibinfo{volume}{14}, \bibinfo{number}{5} (\bibinfo{year}{2023}), \bibinfo{pages}{1--48}.
\newblock


\bibitem[Li et~al\mbox{.}(2023c)]%
        {li2023afairness}
\bibfield{author}{\bibinfo{person}{Yunqi Li}, \bibinfo{person}{Michiharu Yamashita}, \bibinfo{person}{Hanxiong Chen}, \bibinfo{person}{Dongwon Lee}, {and} \bibinfo{person}{Yongfeng Zhang}.} \bibinfo{year}{2023}\natexlab{c}.
\newblock \showarticletitle{Fairness in Job Recommendation under Quantity Constraints}. In \bibinfo{booktitle}{\emph{AAAI-23 Workshop on AI for Web Advertising}}.
\newblock


\bibitem[Liu et~al\mbox{.}(2025)]%
        {liu2025repeat}
\bibfield{author}{\bibinfo{person}{Yuanna Liu}, \bibinfo{person}{Ming Li}, \bibinfo{person}{Mohammad Aliannejadi}, {and} \bibinfo{person}{Maarten de Rijke}.} \bibinfo{year}{2025}\natexlab{}.
\newblock \showarticletitle{Repeat-Bias-Aware Optimization of Beyond-Accuracy Metrics for Next Basket Recommendation}. In \bibinfo{booktitle}{\emph{European Conference on Information Retrieval}}. Springer, \bibinfo{pages}{214--229}.
\newblock


\bibitem[Liu et~al\mbox{.}(2024)]%
        {liu2024measuring}
\bibfield{author}{\bibinfo{person}{Yuanna Liu}, \bibinfo{person}{Ming Li}, \bibinfo{person}{Mozhdeh Ariannezhad}, \bibinfo{person}{Masoud Mansoury}, \bibinfo{person}{Mohammad Aliannejadi}, {and} \bibinfo{person}{Maarten de Rijke}.} \bibinfo{year}{2024}\natexlab{}.
\newblock \showarticletitle{Measuring item fairness in next basket recommendation: A reproducibility study}. In \bibinfo{booktitle}{\emph{European Conference on Information Retrieval}}. Springer, \bibinfo{pages}{210--225}.
\newblock


\bibitem[Liu et~al\mbox{.}(2023)]%
        {liu2023mitigating}
\bibfield{author}{\bibinfo{person}{Zhongzhou Liu}, \bibinfo{person}{Yuan Fang}, {and} \bibinfo{person}{Min Wu}.} \bibinfo{year}{2023}\natexlab{}.
\newblock \showarticletitle{Mitigating Popularity Bias for Users and Items with Fairness-centric Adaptive Recommendation}.
\newblock \bibinfo{journal}{\emph{ACM Transactions on Information Systems}} \bibinfo{volume}{41}, \bibinfo{number}{3} (\bibinfo{year}{2023}), \bibinfo{pages}{1--27}.
\newblock


\bibitem[Ma et~al\mbox{.}(2024a)]%
        {ma2024multicbr}
\bibfield{author}{\bibinfo{person}{Yunshan Ma}, \bibinfo{person}{Yingzhi He}, \bibinfo{person}{Xiang Wang}, \bibinfo{person}{Yinwei Wei}, \bibinfo{person}{Xiaoyu Du}, \bibinfo{person}{Yuyangzi Fu}, {and} \bibinfo{person}{Tat-Seng Chua}.} \bibinfo{year}{2024}\natexlab{a}.
\newblock \showarticletitle{MultiCBR: Multi-view Contrastive Learning for Bundle Recommendation}.
\newblock \bibinfo{journal}{\emph{ACM Transactions on Information Systems}} \bibinfo{volume}{42}, \bibinfo{number}{4} (\bibinfo{year}{2024}), \bibinfo{pages}{1--23}.
\newblock


\bibitem[Ma et~al\mbox{.}(2022)]%
        {ma2022crosscbr}
\bibfield{author}{\bibinfo{person}{Yunshan Ma}, \bibinfo{person}{Yingzhi He}, \bibinfo{person}{An Zhang}, \bibinfo{person}{Xiang Wang}, {and} \bibinfo{person}{Tat-Seng Chua}.} \bibinfo{year}{2022}\natexlab{}.
\newblock \showarticletitle{CrossCBR: Cross-view Contrastive Learning for Bundle Recommendation}. In \bibinfo{booktitle}{\emph{Proceedings of the 28th ACM SIGKDD Conference on Knowledge Discovery and Data Mining}}. \bibinfo{pages}{1233--1241}.
\newblock


\bibitem[Ma et~al\mbox{.}(2024b)]%
        {ma2024leveraging}
\bibfield{author}{\bibinfo{person}{Yunshan Ma}, \bibinfo{person}{Xiaohao Liu}, \bibinfo{person}{Yinwei Wei}, \bibinfo{person}{Zhulin Tao}, \bibinfo{person}{Xiang Wang}, {and} \bibinfo{person}{Tat-Seng Chua}.} \bibinfo{year}{2024}\natexlab{b}.
\newblock \showarticletitle{Leveraging multimodal features and item-level user feedback for bundle construction}. In \bibinfo{booktitle}{\emph{Proceedings of the 17th ACM International Conference on Web Search and Data Mining}}. \bibinfo{pages}{510--519}.
\newblock


\bibitem[Mansoury(2022)]%
        {mansoury2022understanding}
\bibfield{author}{\bibinfo{person}{Masoud Mansoury}.} \bibinfo{year}{2022}\natexlab{}.
\newblock \showarticletitle{Understanding and Mitigating Multi-sided Exposure Bias in Recommender Systems}.
\newblock \bibinfo{journal}{\emph{ACM SIGWEB Newsletter}} \bibinfo{volume}{2022}, \bibinfo{number}{Autumn} (\bibinfo{year}{2022}), \bibinfo{pages}{1--4}.
\newblock


\bibitem[Mansoury et~al\mbox{.}(2020)]%
        {mansoury2020feedback}
\bibfield{author}{\bibinfo{person}{Masoud Mansoury}, \bibinfo{person}{Himan Abdollahpouri}, \bibinfo{person}{Mykola Pechenizkiy}, \bibinfo{person}{Bamshad Mobasher}, {and} \bibinfo{person}{Robin Burke}.} \bibinfo{year}{2020}\natexlab{}.
\newblock \showarticletitle{Feedback loop and bias amplification in recommender systems}. In \bibinfo{booktitle}{\emph{Proceedings of the 29th ACM international conference on information \& knowledge management}}. \bibinfo{pages}{2145--2148}.
\newblock


\bibitem[Mansoury et~al\mbox{.}(2024)]%
        {mansoury2024mitigating}
\bibfield{author}{\bibinfo{person}{Masoud Mansoury}, \bibinfo{person}{Bamshad Mobasher}, {and} \bibinfo{person}{Herke van Hoof}.} \bibinfo{year}{2024}\natexlab{}.
\newblock \showarticletitle{Mitigating exposure bias in online learning to rank recommendation: A novel reward model for cascading bandits}. In \bibinfo{booktitle}{\emph{Proceedings of the 33rd ACM International Conference on Information and Knowledge Management}}. \bibinfo{pages}{1638--1648}.
\newblock


\bibitem[Mehrotra et~al\mbox{.}(2018)]%
        {mehrotra2018towards}
\bibfield{author}{\bibinfo{person}{Rishabh Mehrotra}, \bibinfo{person}{James McInerney}, \bibinfo{person}{Hugues Bouchard}, \bibinfo{person}{Mounia Lalmas}, {and} \bibinfo{person}{Fernando Diaz}.} \bibinfo{year}{2018}\natexlab{}.
\newblock \showarticletitle{Towards a Fair Marketplace: Counterfactual Evaluation of the Trade-off between Relevance, Fairness \& Satisfaction in Recommendation Systems}. In \bibinfo{booktitle}{\emph{Proceedings of the 27th ACM International Conference on Information and Knowledge Management}}. \bibinfo{pages}{2243--2251}.
\newblock


\bibitem[Naghiaei et~al\mbox{.}(2022)]%
        {naghiaei2022cpfair}
\bibfield{author}{\bibinfo{person}{Mohammadmehdi Naghiaei}, \bibinfo{person}{Hossein~A Rahmani}, {and} \bibinfo{person}{Yashar Deldjoo}.} \bibinfo{year}{2022}\natexlab{}.
\newblock \showarticletitle{CPFair: Personalized Consumer and Producer Fairness Re-ranking for Recommender Systems}. In \bibinfo{booktitle}{\emph{Proceedings of the 45th International ACM SIGIR Conference on Research and Development in Information Retrieval}}. \bibinfo{pages}{770--779}.
\newblock


\bibitem[Nguyen et~al\mbox{.}(2023)]%
        {nguyen2023hhmc}
\bibfield{author}{\bibinfo{person}{Huy-Son Nguyen}, \bibinfo{person}{Tuan-Nghia Bui}, \bibinfo{person}{Long-Hai Nguyen}, \bibinfo{person}{Duy-Cat Can}, \bibinfo{person}{Cam-Van~Thi Nguyen}, \bibinfo{person}{Duc-Trong Le}, {and} \bibinfo{person}{Hoang-Quynh Le}.} \bibinfo{year}{2023}\natexlab{}.
\newblock \showarticletitle{HHMC: a heterogeneous x homogeneous graph-based network for multimodal cross-selling recommendation}. In \bibinfo{booktitle}{\emph{2023 15th International Conference on Knowledge and Systems Engineering (KSE)}}. IEEE, \bibinfo{pages}{1--6}.
\newblock


\bibitem[Nguyen et~al\mbox{.}(2024)]%
        {nguyen2024bundle}
\bibfield{author}{\bibinfo{person}{Huy-Son Nguyen}, \bibinfo{person}{Tuan-Nghia Bui}, \bibinfo{person}{Long-Hai Nguyen}, \bibinfo{person}{Hung Hoang}, \bibinfo{person}{Cam-Van Thi~Nguyen}, \bibinfo{person}{Hoang-Quynh Le}, {and} \bibinfo{person}{Duc-Trong Le}.} \bibinfo{year}{2024}\natexlab{}.
\newblock \showarticletitle{Bundle Recommendation with Item-Level Causation-Enhanced Multi-view Learning}. In \bibinfo{booktitle}{\emph{Joint European Conference on Machine Learning and Knowledge Discovery in Databases}}. Springer, \bibinfo{pages}{324--341}.
\newblock


\bibitem[Qi et~al\mbox{.}(2022)]%
        {qi2022profairrec}
\bibfield{author}{\bibinfo{person}{Tao Qi}, \bibinfo{person}{Fangzhao Wu}, \bibinfo{person}{Chuhan Wu}, \bibinfo{person}{Peijie Sun}, \bibinfo{person}{Le Wu}, \bibinfo{person}{Xiting Wang}, \bibinfo{person}{Yongfeng Huang}, {and} \bibinfo{person}{Xing Xie}.} \bibinfo{year}{2022}\natexlab{}.
\newblock \showarticletitle{ProFairRec: Provider Fairness-aware News Recommendation}. In \bibinfo{booktitle}{\emph{Proceedings of the 45th International ACM SIGIR Conference on Research and Development in Information Retrieval}}. \bibinfo{pages}{1164--1173}.
\newblock


\bibitem[Rahmani et~al\mbox{.}(2022)]%
        {rahmani2022experiments}
\bibfield{author}{\bibinfo{person}{Hossein~A Rahmani}, \bibinfo{person}{Mohammadmehdi Naghiaei}, \bibinfo{person}{Mahdi Dehghan}, {and} \bibinfo{person}{Mohammad Aliannejadi}.} \bibinfo{year}{2022}\natexlab{}.
\newblock \showarticletitle{Experiments on Generalizability of User-oriented Fairness in Recommender Systems}. In \bibinfo{booktitle}{\emph{Proceedings of the 45th International ACM SIGIR Conference on Research and Development in Information Retrieval}}. \bibinfo{pages}{2755--2764}.
\newblock


\bibitem[Raj and Ekstrand(2022)]%
        {raj2022measuring}
\bibfield{author}{\bibinfo{person}{Amifa Raj} {and} \bibinfo{person}{Michael~D. Ekstrand}.} \bibinfo{year}{2022}\natexlab{}.
\newblock \showarticletitle{Measuring Fairness in Ranked Results: An Analytical and Empirical Comparison}. In \bibinfo{booktitle}{\emph{Proceedings of the 45th International ACM SIGIR Conference on Research and Development in Information Retrieval}}. \bibinfo{pages}{726--736}.
\newblock


\bibitem[Ren et~al\mbox{.}(2023)]%
        {ren2023distillation}
\bibfield{author}{\bibinfo{person}{Yuyang Ren}, \bibinfo{person}{Zhang Haonan}, \bibinfo{person}{Luoyi Fu}, \bibinfo{person}{Xinbing Wang}, {and} \bibinfo{person}{Chenghu Zhou}.} \bibinfo{year}{2023}\natexlab{}.
\newblock \showarticletitle{Distillation-enhanced Graph Masked Autoencoders for Bundle Recommendation}. In \bibinfo{booktitle}{\emph{Proceedings of the 46th International ACM SIGIR Conference on Research and Development in Information Retrieval}}. \bibinfo{pages}{1660--1669}.
\newblock


\bibitem[Sapiezynski et~al\mbox{.}(2019)]%
        {sapiezynski2019quantifying}
\bibfield{author}{\bibinfo{person}{Piotr Sapiezynski}, \bibinfo{person}{Wesley Zeng}, \bibinfo{person}{Ronald E~Robertson}, \bibinfo{person}{Alan Mislove}, {and} \bibinfo{person}{Christo Wilson}.} \bibinfo{year}{2019}\natexlab{}.
\newblock \showarticletitle{Quantifying the Impact of User Attentionon fair Group Representation in Ranked Lists}. In \bibinfo{booktitle}{\emph{Companion proceedings of the 2019 world wide web conference}}. \bibinfo{pages}{553--562}.
\newblock


\bibitem[Singh and Joachims(2018)]%
        {singh2018fairness}
\bibfield{author}{\bibinfo{person}{Ashudeep Singh} {and} \bibinfo{person}{Thorsten Joachims}.} \bibinfo{year}{2018}\natexlab{}.
\newblock \showarticletitle{Fairness of Exposure in Rankings}. In \bibinfo{booktitle}{\emph{Proceedings of the 24th ACM SIGKDD International Conference on Knowledge Discovery \& Data Mining}}. \bibinfo{pages}{2219--2228}.
\newblock


\bibitem[Sun et~al\mbox{.}(2024c)]%
        {sun2024survey}
\bibfield{author}{\bibinfo{person}{Meng Sun}, \bibinfo{person}{Lin Li}, \bibinfo{person}{Ming Li}, \bibinfo{person}{Xiaohui Tao}, \bibinfo{person}{Dong Zhang}, \bibinfo{person}{Peipei Wang}, {and} \bibinfo{person}{Jimmy~Xiangji Huang}.} \bibinfo{year}{2024}\natexlab{c}.
\newblock \showarticletitle{A Survey on Bundle Recommendation: Methods, Applications, and Challenges}.
\newblock \bibinfo{journal}{\emph{arXiv preprint arXiv:2411.00341}} (\bibinfo{year}{2024}).
\newblock


\bibitem[Sun et~al\mbox{.}(2024a)]%
        {sun2024revisiting}
\bibfield{author}{\bibinfo{person}{Zhu Sun}, \bibinfo{person}{Kaidong Feng}, \bibinfo{person}{Jie Yang}, \bibinfo{person}{Hui Fang}, \bibinfo{person}{Xinghua Qu}, \bibinfo{person}{Yew-Soon Ong}, {and} \bibinfo{person}{Wenyuan Liu}.} \bibinfo{year}{2024}\natexlab{a}.
\newblock \showarticletitle{Revisiting Bundle Recommendation for Intent-aware Product Bundling}.
\newblock \bibinfo{journal}{\emph{ACM Transactions on Recommender Systems}} \bibinfo{volume}{2}, \bibinfo{number}{3} (\bibinfo{year}{2024}), \bibinfo{pages}{1--34}.
\newblock


\bibitem[Sun et~al\mbox{.}(2024b)]%
        {sun2024adaptive}
\bibfield{author}{\bibinfo{person}{Zhu Sun}, \bibinfo{person}{Kaidong Feng}, \bibinfo{person}{Jie Yang}, \bibinfo{person}{Xinghua Qu}, \bibinfo{person}{Hui Fang}, \bibinfo{person}{Yew-Soon Ong}, {and} \bibinfo{person}{Wenyuan Liu}.} \bibinfo{year}{2024}\natexlab{b}.
\newblock \showarticletitle{Adaptive In-Context Learning with Large Language Models for Bundle Generation}. In \bibinfo{booktitle}{\emph{Proceedings of the 47th International ACM SIGIR Conference on Research and Development in Information Retrieval}}. \bibinfo{pages}{966--976}.
\newblock


\bibitem[Vargas and Castells(2014)]%
        {vargas2014improving}
\bibfield{author}{\bibinfo{person}{Sa{\'u}l Vargas} {and} \bibinfo{person}{Pablo Castells}.} \bibinfo{year}{2014}\natexlab{}.
\newblock \showarticletitle{Improving Sales Diversity by Recommending Users to Items}. In \bibinfo{booktitle}{\emph{Proceedings of the 8th ACM Conference on Recommender systems}}. \bibinfo{pages}{145--152}.
\newblock


\bibitem[Wang and Joachims(2021)]%
        {wang2021user}
\bibfield{author}{\bibinfo{person}{Lequn Wang} {and} \bibinfo{person}{Thorsten Joachims}.} \bibinfo{year}{2021}\natexlab{}.
\newblock \showarticletitle{User Fairness, Item Fairness, and Diversity for Rankings in Two-sided Markets}. In \bibinfo{booktitle}{\emph{Proceedings of the 2021 ACM SIGIR International Conference on Theory of Information Retrieval}}. \bibinfo{pages}{23--41}.
\newblock


\bibitem[Wang et~al\mbox{.}(2023)]%
        {wang2023survey}
\bibfield{author}{\bibinfo{person}{Yifan Wang}, \bibinfo{person}{Weizhi Ma}, \bibinfo{person}{Min Zhang}, \bibinfo{person}{Yiqun Liu}, {and} \bibinfo{person}{Shaoping Ma}.} \bibinfo{year}{2023}\natexlab{}.
\newblock \showarticletitle{A Survey on the Fairness of Recommender Systems}.
\newblock \bibinfo{journal}{\emph{ACM Transactions on Information Systems}} \bibinfo{volume}{41}, \bibinfo{number}{3} (\bibinfo{year}{2023}), \bibinfo{pages}{1--43}.
\newblock


\bibitem[Wei et~al\mbox{.}(2023)]%
        {wei2023strategy}
\bibfield{author}{\bibinfo{person}{Yinwei Wei}, \bibinfo{person}{Xiaohao Liu}, \bibinfo{person}{Yunshan Ma}, \bibinfo{person}{Xiang Wang}, \bibinfo{person}{Liqiang Nie}, {and} \bibinfo{person}{Tat-Seng Chua}.} \bibinfo{year}{2023}\natexlab{}.
\newblock \showarticletitle{Strategy-aware Bundle Recommender System}. In \bibinfo{booktitle}{\emph{Proceedings of the 46th International ACM SIGIR Conference on Research and Development in Information Retrieval}}. \bibinfo{pages}{1198--1207}.
\newblock


\bibitem[Wu et~al\mbox{.}(2021)]%
        {wu2021fairness}
\bibfield{author}{\bibinfo{person}{Chuhan Wu}, \bibinfo{person}{Fangzhao Wu}, \bibinfo{person}{Xiting Wang}, \bibinfo{person}{Yongfeng Huang}, {and} \bibinfo{person}{Xing Xie}.} \bibinfo{year}{2021}\natexlab{}.
\newblock \showarticletitle{Fairness-aware News Recommendation with Decomposed Adversarial Learning}. In \bibinfo{booktitle}{\emph{Proceedings of the AAAI Conference on Artificial Intelligence}}, Vol.~\bibinfo{volume}{35}. \bibinfo{pages}{4462--4469}.
\newblock


\bibitem[Xu et~al\mbox{.}(2025a)]%
        {xu2025fairdiverse}
\bibfield{author}{\bibinfo{person}{Chen Xu}, \bibinfo{person}{Zhirui Deng}, \bibinfo{person}{Clara Rus}, \bibinfo{person}{Xiaopeng Ye}, \bibinfo{person}{Yuanna Liu}, \bibinfo{person}{Jun Xu}, \bibinfo{person}{Zhicheng Dou}, \bibinfo{person}{Ji-Rong Wen}, {and} \bibinfo{person}{Maarten de Rijke}.} \bibinfo{year}{2025}\natexlab{a}.
\newblock \showarticletitle{FairDiverse: A Comprehensive Toolkit for Fair and Diverse Information Retrieval Algorithms}.
\newblock \bibinfo{journal}{\emph{arXiv preprint arXiv:2502.11883}} (\bibinfo{year}{2025}).
\newblock


\bibitem[Xu et~al\mbox{.}(2025b)]%
        {xu2025bridging}
\bibfield{author}{\bibinfo{person}{Chen Xu}, \bibinfo{person}{Yuxin Li}, \bibinfo{person}{Wenjie Wang}, \bibinfo{person}{Liang Pang}, \bibinfo{person}{Jun Xu}, {and} \bibinfo{person}{Tat-Seng Chua}.} \bibinfo{year}{2025}\natexlab{b}.
\newblock \showarticletitle{Bridging Jensen Gap for Max-Min Group Fairness Optimization in Recommendation}.
\newblock \bibinfo{journal}{\emph{arXiv preprint arXiv:2502.09319}} (\bibinfo{year}{2025}).
\newblock


\bibitem[Yang and Stoyanovich(2017)]%
        {yang2017measuring}
\bibfield{author}{\bibinfo{person}{Ke Yang} {and} \bibinfo{person}{Julia Stoyanovich}.} \bibinfo{year}{2017}\natexlab{}.
\newblock \showarticletitle{Measuring Fairness in Ranked Outputs}. In \bibinfo{booktitle}{\emph{Proceedings of the 29th International Conference on Scientific and Statistical Database Management}}. \bibinfo{pages}{1--6}.
\newblock


\bibitem[Zehlike et~al\mbox{.}(2017)]%
        {zehlike2017fa}
\bibfield{author}{\bibinfo{person}{Meike Zehlike}, \bibinfo{person}{Francesco Bonchi}, \bibinfo{person}{Carlos Castillo}, \bibinfo{person}{Sara Hajian}, \bibinfo{person}{Mohamed Megahed}, {and} \bibinfo{person}{Ricardo Baeza-Yates}.} \bibinfo{year}{2017}\natexlab{}.
\newblock \showarticletitle{FA*IR: A Fair Top-k Ranking Algorithm}. In \bibinfo{booktitle}{\emph{Proceedings of the 2017 ACM on Conference on Information and Knowledge Management}}. \bibinfo{pages}{1569--1578}.
\newblock


\bibitem[Zhao et~al\mbox{.}(2022)]%
        {zhao2022multi}
\bibfield{author}{\bibinfo{person}{Sen Zhao}, \bibinfo{person}{Wei Wei}, \bibinfo{person}{Ding Zou}, {and} \bibinfo{person}{Xianling Mao}.} \bibinfo{year}{2022}\natexlab{}.
\newblock \showarticletitle{Multi-view Intent Disentangle Graph Networks for Bundle Recommendation}. In \bibinfo{booktitle}{\emph{Proceedings of the AAAI Conference on Artificial Intelligence}}, Vol.~\bibinfo{volume}{36}. \bibinfo{pages}{4379--4387}.
\newblock


\bibitem[Zheng et~al\mbox{.}(2023)]%
        {zheng2023interaction}
\bibfield{author}{\bibinfo{person}{Zhi Zheng}, \bibinfo{person}{Chao Wang}, \bibinfo{person}{Tong Xu}, \bibinfo{person}{Dazhong Shen}, \bibinfo{person}{Penggang Qin}, \bibinfo{person}{Xiangyu Zhao}, \bibinfo{person}{Baoxing Huai}, \bibinfo{person}{Xian Wu}, {and} \bibinfo{person}{Enhong Chen}.} \bibinfo{year}{2023}\natexlab{}.
\newblock \showarticletitle{Interaction-aware Drug Package Recommendation via Policy Gradient}.
\newblock \bibinfo{journal}{\emph{ACM Transactions on Information Systems}} \bibinfo{volume}{41}, \bibinfo{number}{1} (\bibinfo{year}{2023}), \bibinfo{pages}{1--32}.
\newblock


\end{thebibliography}

\end{document}